\documentclass[fleqn,usenatbib]{mnras} 

\usepackage{mathptmx}

\usepackage[T1]{fontenc}
\usepackage{ae,aecompl}

\usepackage[pdftex]{graphicx}	
\usepackage{epstopdf}
\usepackage{amsmath}	
\usepackage{amssymb}	
\usepackage{subfigure}
\usepackage{natbib}
\usepackage{threeparttable}
\usepackage{hyperref}
\usepackage{color}

\newcommand{\lfgas}{$\log f_\mathrm{gas}\,$}
\newcommand{\lfwise}{$\log f_\mathrm{12}/f_\mathrm{4.6}\,$}
\newcommand{\lfwiseh}{$\log f_\mathrm{4.6}/f_\mathrm{3.4}\,$}

\newcommand{\alphacom}{$\alpha_\mathrm{CO} = 4.3 \,\mathrm{ M_\odot (K \,km \,s^{-1} pc^{2})^{-1}}\,$}
\newcommand{\degree}{$^{\circ}$}
\newcommand{\arcsecond}{$^{\prime\prime}$}

\title[]{Molecular Gas during the Post-Starburst Phase: Low Gas Fractions in Green Valley Seyfert Post-Starburst Galaxies}

\author[Yesuf et al.]{Hassen M. Yesuf$^{1}$\thanks{E-mail: myesuf@gmail.com}
 K. Decker French$^{2}$
 S. M. Faber$^{1}$
 David C. Koo$^{1}$
\\
$^{1}$University of California Observatories and the Department of Astronomy \& Astrophysics, University of California, Santa Cruz, \\
CA 95064, USA\\
$^{2}$Steward Observatory, University of Arizona, 933 North Cherry Avenue, Tucson, AZ 85721, USA
}

\begin{document}
\label{firstpage}
\pagerange{\pageref{firstpage}--\pageref{lastpage}}
\maketitle

\begin{abstract}
Post-starbursts (PSBs) are candidate for rapidly transitioning from star-bursting to quiescent galaxies. We study the molecular gas evolution of PSBs at $z \sim 0.03 - 0.2$. We undertook new CO\,(2--1) observations of 22 Seyfert PSBs candidates using the ARO Submillimeter Telescope. This sample complements previous samples of PSBs by including green valley PSBs with Seyfert-like emission, allowing us to analyze for the first time the molecular gas properties of 116 PSBs with a variety of AGN properties. The distribution of molecular gas to stellar mass fractions in PSBs is significantly different than normal star-forming galaxies in the COLD GASS survey. The combined samples of PSBs with Seyfert-like emission line ratios have a gas fraction distribution which is even more significantly different and is broader ($\sim 0.03-0.3$). Most of them have lower gas fractions than normal star-forming galaxies. We find a highly significant correlation between the WISE $12 \mu$m to $4.6\mu$m flux ratios and molecular gas fractions in both PSBs and normal galaxies. We detect molecular gas in 27\% of our Seyfert PSBs. Taking into account the upper limits, the mean and the dispersion of the distribution of the gas fraction in our Seyfert PSB sample are much smaller $(\mu = 0.025, \sigma = 0.018)$ than previous samples of Seyfert PSBs or PSBs in general $(\mu \sim 0.1 - 0.2, \sigma \sim 0.1 - 0.2)$. 
\end{abstract}


\section{Introduction}
 
Despite extensive observational and theoretical studies, the physical mechanisms that regulate the star formation rates of galaxies are still poorly understood. Star formation quenching, by yet unknown mechanisms, causes star-forming galaxies to migrate to the ``red sequence" \citep{Gladders+98,Faber+07}. One likely such formation of mechanism of the red-sequence is the transformation of star forming, disk-dominated, gas-rich galaxies into early types via mergers and their associated feedback \citep{Toomre+72,Hopkins+06}. 

Post-starburst (PSB) galaxies are candidate post-merger objects, rapidly transitioning from the blue-cloud to the red-sequence \citep[e.g.,][]{Dressler+83, Couch+87,Zabludoff+96, Wild+09,Snyder+11,Yesuf+14,Pawlik+16}. In their quiescent phase, their spectra reveal little-to-no current star formation, but a substantial burst of star formation before an abrupt cessation $\sim$1 Gyr ago, long enough for the ionizing O\&B stars to evolve away, but recent enough for A-stars to dominate the stellar light. Recent observational efforts have enlarged the traditional definition of post-starburst to include a more complete and less biased sample of galaxies with ongoing star formation or/and AGN activity \citep{Wild+10,Yesuf+14,Alatalo+14}. With the identification of this sample, which spans the entire starburst to quenched post-starburst evolutionary sequence, better constraints on theoretical models of galaxy evolution have started to emerge. One of the the firm constraints is the observed time delay between the starburst phase and the AGN activity by about 200 Myr \citep{Wild+10,Yesuf+14}. This time delay suggests that AGN are not primarily responsible in the original quenching of starbursts but may be responsible for keeping star formation at a low level by removing gas and dust during the post-starburst phase. The other emerging but contested phenomenon is molecular gas evolution after the starburst has ended \citep{Schawinski+09,Rowlands+15}. This work aims to further examine the molecular gas evolution along the starburst sequence using both new and existing data, by adding molecular gas observations from green valley Seyfert PSBs, which were excluded from other PSB samples. 

In simulations of gas-rich mergers, gas is funneled to galaxy centers, powering intense nuclear starbursts and obscured nuclear AGN activity. At the end of the starburst, the leftover gas and dust are cleared out due to feedback from the AGN \citep[e.g.,][]{Sanders+88,Barnes+91,Silk+98,DiMatteo+05,springel05b,Kaviraj+07,Hopkins+06,Hopkins+08,Wild+09,Snyder+11, Cen12}. For instance, \citet{Narayanan+08} found that galactic winds are a natural consequence of merger-induced star formation and black hole growth. In their simulated galaxies, the galactic winds can entrain molecular gas of $\sim 10^8-10^9$M$_\odot$, which, the authors showed, should be observable in CO emission. The molecular gas entrained in the winds driven by AGN are predicted to be longer-lived than the gas entrained solely in starburst-driven winds.  The wind velocities in the simulated galaxies with AGN-feedback can reach close to 2.5 times the circular velocity. Thus, making AGN-feedback a viable mechanism to get rid off a residual gas and dust at the end of a starburst.

Despite its theoretical appeal, the evidence that connects AGN activity with the end of star-formation in galaxies has been elusive, with evidence both for \citep{Schawinski+09,Alatalo+11,Cicone+14, Garcia-Burillo+14} and against \citep{Fabello+11,French+15,Gereb+15,Rowlands+15,Alatalo+16}. 

Now we review previous works on molecular gas contents of PSBs and AGN. Using IRAM CO observations, \citet{Rowlands+15} investigated the evolution of molecular gas and dust properties in 11 PSBs on the starburst to quenched post-starburst sequence at $z \sim 0.03$. Two of these PSBs are Seyfert galaxies while the rest are either star-forming or composite galaxies of star formation and AGN activity. 10/11 of the PSBs were detected in the CO\,(1--0) transition and 9/11 of the PSBs were detected in CO\,(2--1) transition. The gas and dust contents, the star-formation efficiency, the gas depletion time of majority of these PSBs are similar to those of local star-forming spiral galaxies \citep{Saintonge+11,Boselli+14} and gas-rich elliptical galaxies \citep{Young+11,Davis+14}.  In addition, the authors found a decrease in dust temperature with the starburst age but they did not find evidence for dust heating by AGN at late times. 

Similarly, \citet{Alatalo+16} studied 52 PSBs with shock signatures at $z=0.02-0.2$ using IRAM and CARMA. About of half of these PSBs are at $z > 0.1$, 14/52 are Seyferts, and $90\%$ have CO\,(1--0) detections. The molecular gas properties of these PSBs are also similar to those of normal star-forming galaxies. More than $80\%$ of the PSBs in \citet{Alatalo+16}  and \citet{Rowlands+15} samples are located in the blue-cloud. 

\citet{French+15} studied 32 PSBs in the green valley at $z=0.01-0.12$ using the IRAM 30m and the Sub-millimeter Telescope (SMT). Almost all of these PSBs have signatures mimicking low-ionization nuclear emission line regions (LINERs), and 53\% of them have CO detections. Those detected in CO have gas masses and gas to stellar mass fractions comparable to those of star-forming galaxies while the non-detected PSBs have gas fractions more consistent with those observed in quiescent galaxies. 

The three aforementioned studies on the molecular gas contents of PSBs suggested that the end of starburst in these galaxies cannot be ascribed to a complete exhaustion or removal or destruction of molecular gas. The studies also suggested that multiple episodes of starburst or/and AGN activities may be needed for the eventual migration of these galaxies to the red-sequence and that a transition time longer than 1\,Gyr may be needed for this migration to take place. 

\citet{Saintonge+12} found that, among the gas-rich, disk-dominated galaxy population, those which are ongoing mergers or are morphologically disturbed have the shortest molecular gas depletion times. They found no link between the presence of AGN and the long depletion times observed in bulge-dominated galaxies.  Even though their AGN sample has lower molecular gas fractions than the control sample matched in NUV-r color and stellar mass surface density, the depletion times of the two populations are similar. More than 90\% of the AGN studied by \citet{Saintonge+12} are not Seyferts, and instead are LINERs.

In contrast, \citet{Schawinski+09} presented evidence that AGN are responsible for the destruction of molecular gas in morphologically early-type galaxies at $z \sim 0.05$. Their sample included 10 star-forming galaxies, 10 star-formation and AGN composite galaxies and 4 Seyfert galaxies. The galaxies studied by \citet{Schawinski+09} are not post-starbursts but might have experienced mild recent star-formation \citep[see][]{Schawinski+07}. The authors found that the molecular gas mass drops significantly 200 Myr after a recent star formation in the composite galaxies, and none of their Seyferts have CO detections. The authors interpreted their observations as evidence for a destruction of molecular gas and for a suppression of residual star formation by low-luminosity AGN. Likewise, many studies have reported molecular outflows with high mass-outflow rates as evidence for AGN feedback in non-PSB AGN host galaxies \citep[e.g.,][]{Fischer+10, Feruglio+10,Sturm+11,Spoon+13,Veilleux+13,Cicone+14,Garcia-Burillo+14,Sun+14}.

In this work, we study molecular gas in 22 green-valley, Seyfert PSBs candidates using SMT CO\,(2--1) observations, in combination with an existing sample of 94 PSBs from the literature. Our observations were motivated by \citet{French+15} and \citet{Rowlands+15} and were designed to be complementary to the samples in these two works. Our sample is also complementary to the recently published sample of shocked PSB galaxies  \citep{Alatalo+16}, which contains 14 galaxies with Seyfert-like emission line ratios, which are mainly located in the blue-cloud. Our sample is comparable in number to the existing sample of Seyfert PSBs but represents those in the UV-optical green valley. When combined with other samples, our sample is indispensable in sampling PSBs with a variety of AGN properties.

\subsection{Simple energetic argument for no feedback or delayed AGN feedback after the molecular gas fraction is $\lesssim 4\%$}

In this section, we present the condition required for momentum-driven AGN wind to clear gas in a galactic disk or very close to the galaxy, following \citet{Silk+10}. These authors argued, from the condition they derived, that AGN cannot supply enough momentum in radiation to unbind gas out of halos of galaxies. They theoretically estimated the total gas mass within a dark matter halo assuming a gas to total mass fraction of 10\%. In this work, we measure the molecular gas mass to the stellar mass fraction, $f_{\rm H_2} = \frac{M_{\rm H_2}}{M_{\star}}$, in the disks of our galaxies (or very close to the vicinities of the galaxies). Our estimated condition is local, while \citet{Silk+10} were interested if the AGN-driven winds eventually escape from the halos of galaxies. We recast their derivation in terms of molecular gas fraction to show that AGN may only impart sufficient momentum to clear galactic disks once the molecular gas fraction is below about $\lesssim 4\%$, perhaps explaining the delayed AGN molecular gas destruction hinted in this work, and the minor role of AGN in quenching starbursts \citep[e.g.,][]{Wild+10,Yesuf+14}. Note that starbursts have molecular gas fractions of 20--30\% while normal star-forming galaxies have gas fractions of about 10\%.

Let us assume an isothermal sphere galaxy. Within a halo radius $r$, its total mass is  M$_{\rm gal}\, = \,2 \sigma^2r/G$, where G is Newton's gravitational constant and $\sigma$ is the velocity dispersion. 

The total gas fraction,  $f_{\rm gas}$,  in terms of the molecular hydrogen gas fraction is :

\begin{equation}
 f_{\rm gas}  =  M_{\rm gas}/M_{\star }  =  (M_{\rm HI}+M_{\rm H_2})/M_{\star}  =  (1+M_{\rm HI}/M_{\rm H_2})f_{\rm H_2}
\end{equation}

If $f_{\rm \star}$ is the ratio between stellar mass and the total (halo) mass of the galaxy, $f_{\rm \star}  = \frac{M_{\star}}{M_{\rm gal}}$, then :
\begin{equation}
M_{\rm gas} =  (1+M_{\rm HI}/M_{\rm H_2})f_{\rm H_2}  f_{\star} M_{\rm gal}
\end{equation}

Assuming the AGN luminosity, $L$, is a fraction of the Eddington luminosity, $L = \lambda_{\rm Edd}  L_{\rm Edd}$, where $\lambda_{\rm Edd}$ is the Eddington ratio and the Eddington luminosity is $L_{\rm {Edd}} = 4\pi GM_{\rm BH}m_{\rm p}c/\sigma _{\rm T}$. If the the radiation pressure force balances and gravitational force \citep{Murray+05},
\begin{equation}
 \lambda_{\rm Edd} L_{\rm {Edd}}/c = GM_{\rm gal}M_{\rm gas}/r^2
\end{equation}
\begin{equation}
\implies M_{\rm BH} = (1+M_{\rm HI}/M_{\rm H_2})f_{\rm H_2} \frac{f_{\star}}{\lambda_{\rm Edd}}\frac {\sigma_{\rm T}}{m_p}\frac{\sigma^4}{\pi G^2}
\end{equation}
 which has similar scaling as the observed $M_{\rm BH}-\sigma$ relation. 
 
The total energy radiated by the black hole, $E_{\rm out}$, while driving the gas at radius $R$ near the vicinity of the host galaxy by the radiation pressure is $E_{\rm out}\,= \,L\int_{R_0}^{R}1/v(r)dr > L\cdot(R-R_{0})/v_e(R)$, where $v_e(R)$ is the escape speed at the radius $R$. For an isothermal sphere truncated at $R_{\rm max}$, $v_e(R)\,=\,2\sigma\sqrt{1+{\rm ln\,R_{\rm max}/R}}$. The work done by radiation pressure in moving the gas from $R_0$ to $R$ must be greater than the kinetic energy required for the gas to the escape, $L\cdot (R-R_0)/c > 1/2M_{\rm gas} v_e^2(R)$. Thus, 
\begin{equation}
E_{\rm out}   > c/2M_{\rm gas}v_e(R) > M_{\rm gas}c\sigma \sqrt{1+{\rm ln\,R_{\rm max}/R}}
\end{equation}
\begin{equation}
E_{\rm out}   >  (1+M_{\rm HI}/M_{\rm H_2})f_{\rm H_2}M_{\rm \star} c\sigma  \sqrt{1+{\rm ln\,R_{\rm max}/R}}
\end{equation}
The accretion energy of the black hole must also satisfy :
 \begin{equation*}
 \eta {\rm M}_{\rm BH}c^2 > (1+M_{\rm HI}/M_{\rm H_2})f_{\rm H_2}M_{\rm \star} c\sigma  \sqrt{1+{\rm ln\,R_{\rm max}/R}}
\end{equation*}
 Dividing both sides by $M_{\rm \star}$ leads to the condition
 \begin{equation*}
f_{\rm H_2} < \frac{\eta}{(1+M_{\rm HI}/M_{\rm H_2})}\frac{\rm M_{BH}}{\rm M_{bulge}}\frac{\rm M_{bulge}}{\rm M_{\star}}\frac{c}{\sigma \sqrt{1+{\rm ln\,R_{\rm max}/R}}}
\end{equation*}
Using the observed ratio $M_{\rm BH}/M_{\rm bulge} \sim 3 \times 10^{-3}$ \citep[e.g.,][]{Kormendy+13},  $\eta = 0.1$, $M_{\rm bulge}/M_{\rm \star}=0.5$, $\sigma=150$\,km\,s$^{-1}$, $(1+M_{\rm HI}/M_{\rm H_2})  \sim 4 $ \citep{Saintonge+11} and $R_{\rm max}/R = 10$ gives, 
\begin{eqnarray}
f_{\rm H_2} \lesssim 0.04 \left(\frac{M_{\rm bulge}}{M_{\rm \star}}/0.5\right)\left(\frac{150{\rm \,km\,s^{-1}}}{\sigma}\right)\left(\frac{4}{1+M_{\rm HI}/M_{\rm H_2}}\right)  \nonumber \\
 \times \left(\frac{1.82}{\sqrt{1+{\rm ln\,R_{\rm max}/R}}}\right)
\end{eqnarray}

The above assumed values are reasonable for post-starburst galaxies. Similar assumption of $R_{\rm max}/R = 10$ is also made by \citet{Rupke+02} in studying galactic winds in local ultraluminous infrared  galaxies (ULIRGs). The gas measurements are made within $R\sim 10-20$ kpc of our galaxies while their dark matter halos may extend to few 100 kpc. $R_{\rm max}/R$ may range between 10 and 100. Adopting $R_{\rm max}/R = 100$ instead will only change our $f_{\rm H_2}$ limit by a factor of 1.3. Only few percent of the AGN luminosity is expected to coupled to the gas. Therefore, the above lower limit is very liberal estimate since it assumes 100\% feedback efficiency. 5\% feedback efficiency is often used in AGN feedback models \citep[e.g.,][]{ Scannapieco+04,DiMatteo+05, Zubovas+12}. Therefore, our estimate implies that that AGN feedback may only be effective in late-stage post-starbursts after the molecular gas fraction is below $\lesssim 4\%$. 

The rest of the paper is organized as follows: section~\ref{sec:samp} presents the sample selection. Section~\ref{sec:stat} presents overview of statistical methods used in the paper. Section~\ref{sec:res} presents the main results. Section~\ref{sec:discus} discusses our sample in comparison with other samples of PSBs. Section~\ref{sec:sum} summarizes the main findings of this work. Section~\ref{sec:app} provides ancillary information on how our sample relates to existing samples of PSBs with molecular gas measurements. We assume $(\Omega_m,\Omega_\Lambda,h) = (0.3,0.7,0.7)$ cosmology.

\section{Sample Selection \& Observations}\label{sec:samp}

\subsection{Sample selection}

\begin{figure*}
\includegraphics[width=3in, angle=270]{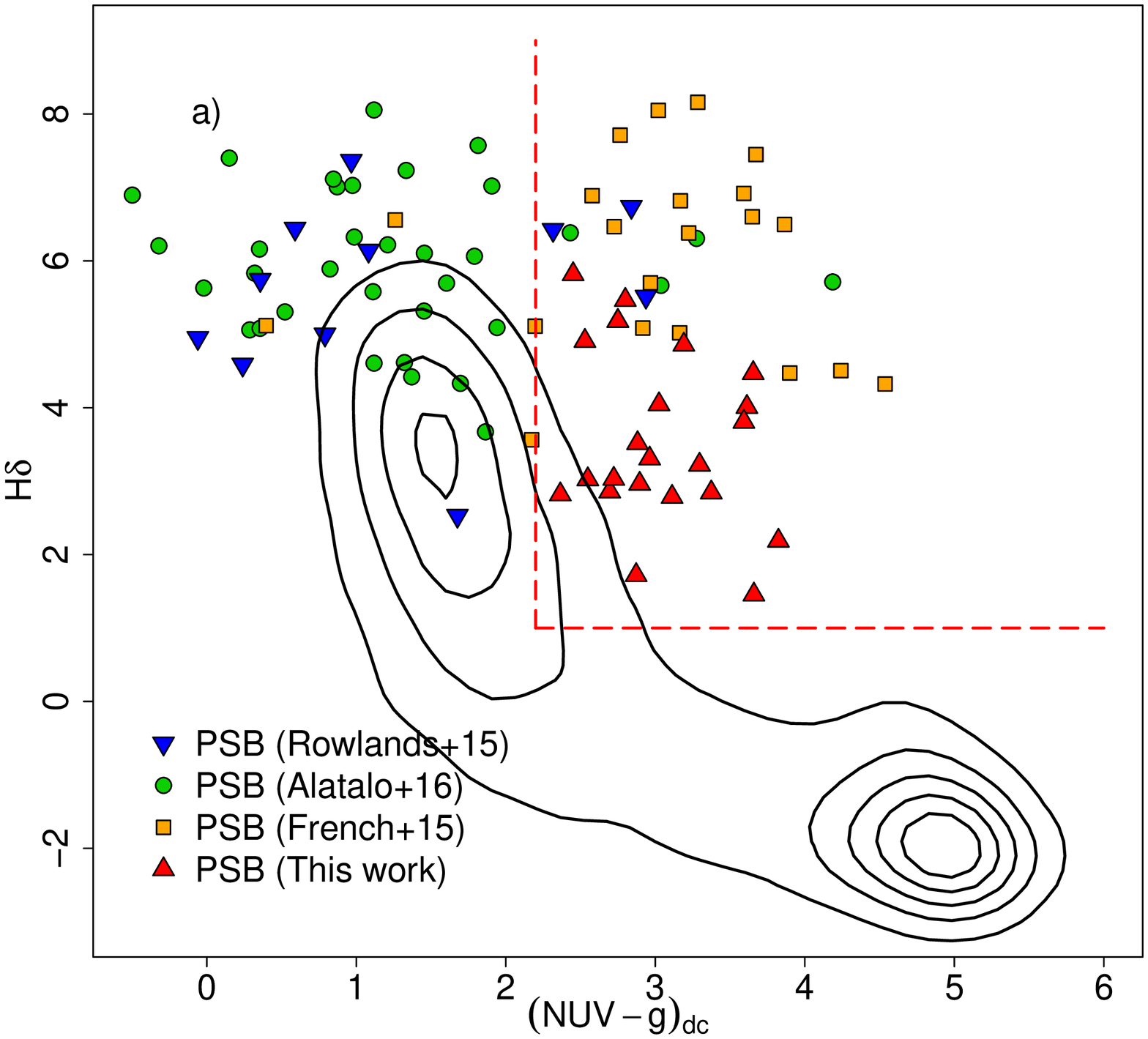}
\includegraphics[width=3in, angle=270]{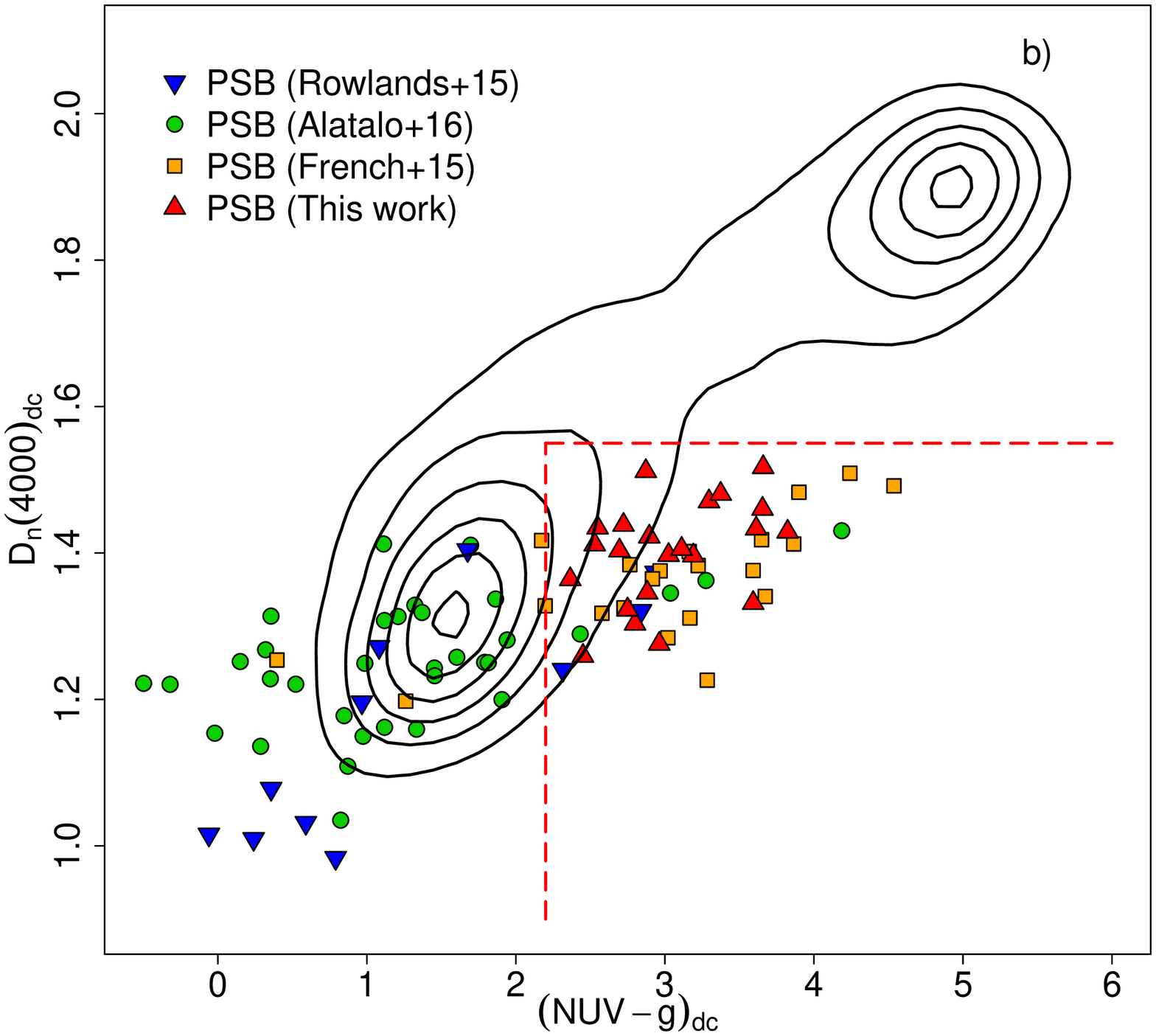}
\includegraphics[width=3in, angle=270]{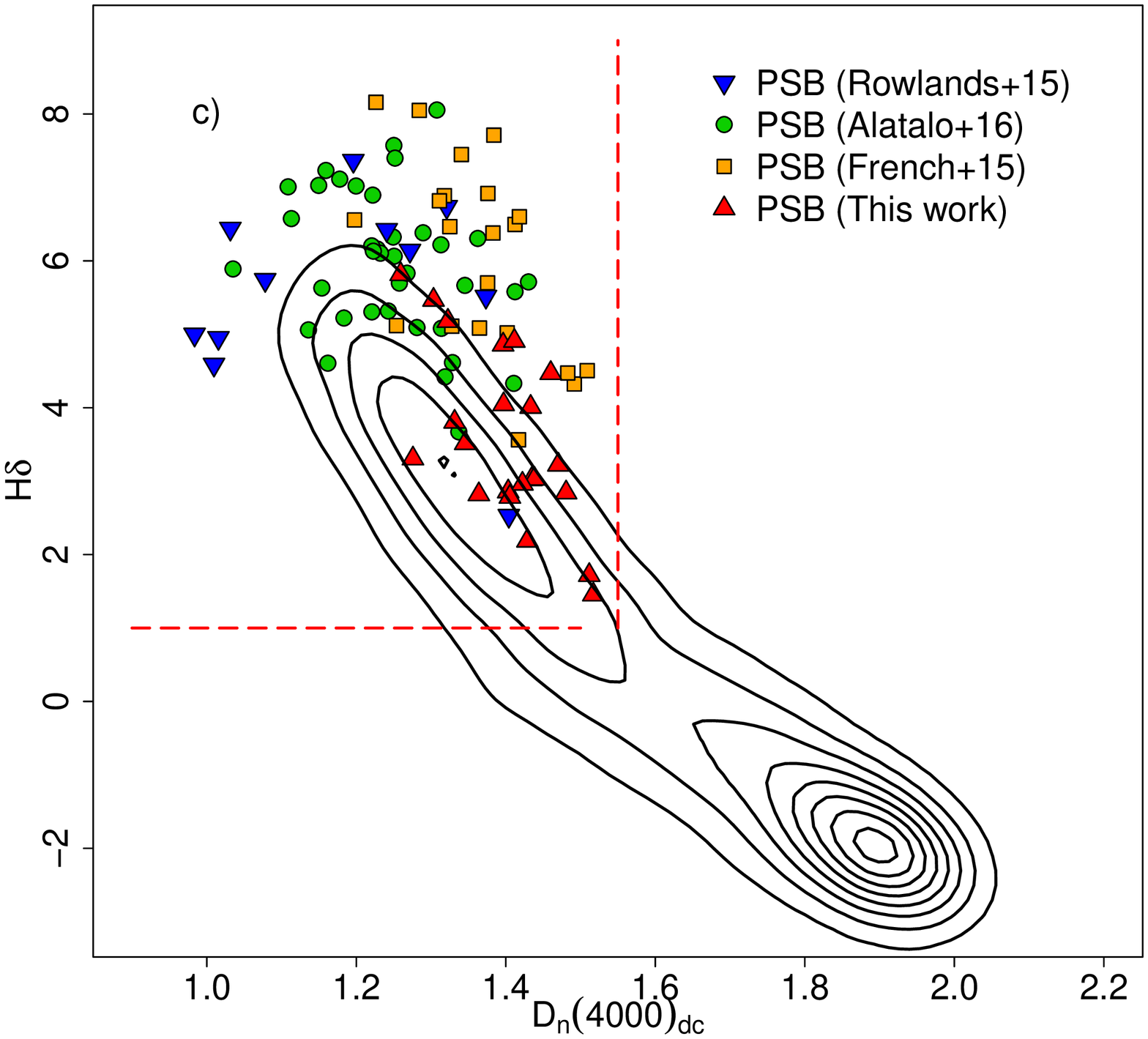}
\includegraphics[width=3in, angle=270]{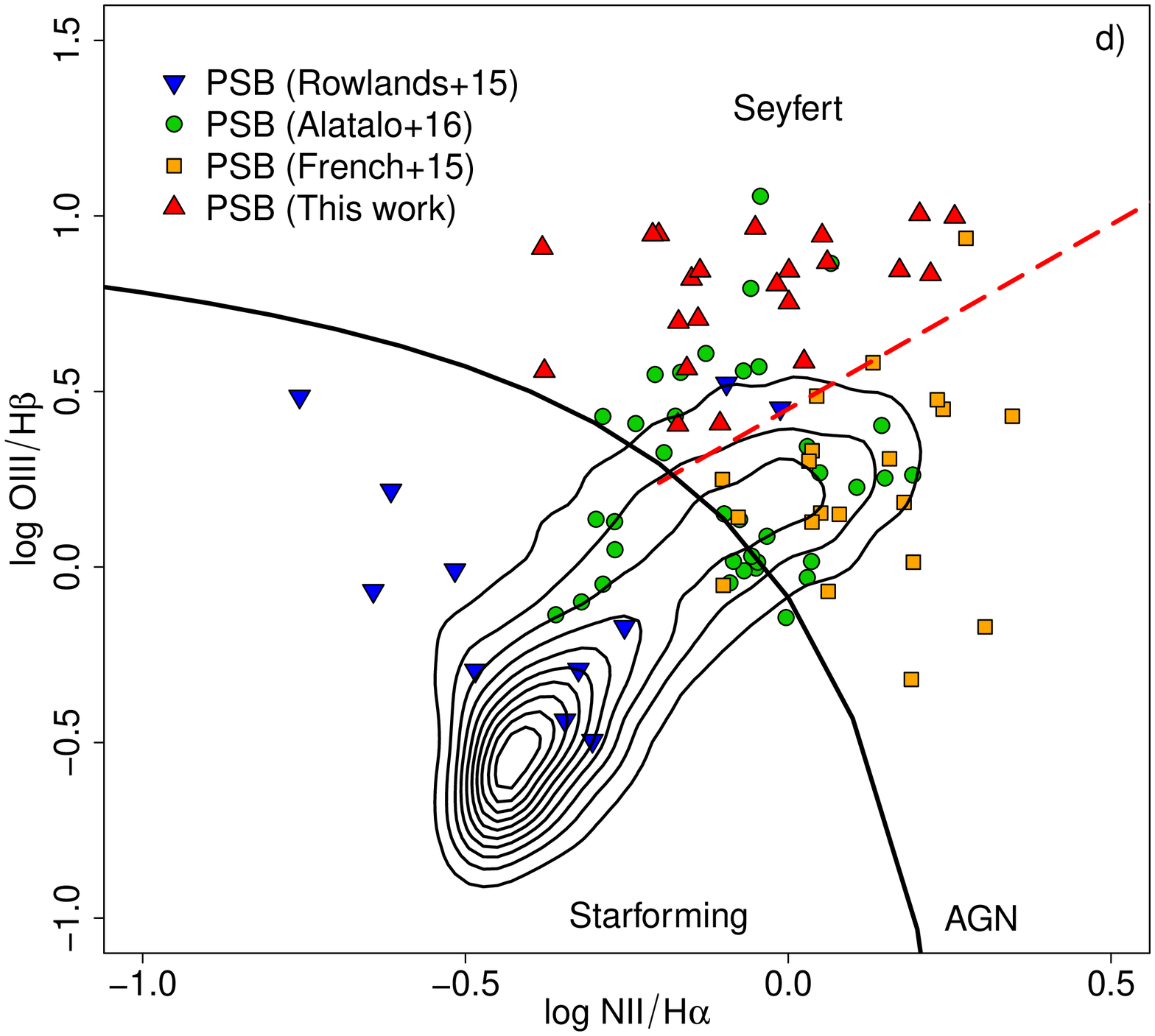}
\caption{Shows the sample selection. The panels clearly show that PSB galaxies as a class differ strongly from normal SDSS galaxies and further that the various types of PSB of galaxies differ from one another depending on how the samples are selected. Panel a) : H$\delta$ absorption equivalent width against dust-corrected NUV-g color.  Panel b) : the $D_n(4000)$ index versus the dust-corrected NUV-g color. Panel c) : the H$\delta$ absorption equivalent width against the $D_n(4000)$ index. Panel d) : The BPT emission-line ratio AGN diagnostic. Seyferts are located in the upper corner of the AGN region \citep{Kewley+01,Schawinski+07}. The contours in all panels represent the number densities of SDSS galaxies at $z=0.02-0.06$ and with stellar mass M$=10^{10}-10^{11}$ M$_\odot$. The colored points are post-starburst galaxies. As detailed in \citet{Yesuf+14}, the combination NUV-g color with H$\delta$ and $D_n(4000)$ is useful in identifying late stage post-starburst galaxies. The red triangles are the new sample of green valley Seyfert PSBs selected in this work. They are selected quantitatively as objects that are 2$\sigma$ outliers from the loci of normal galaxies, at a given $D_n(4000)$, in their dust-corrected NUV - g colors or H$\delta$ \citep[See also Figure 4 of][]{Yesuf+14}. The red dashed lines outline the sample selection for the new Seyfert PSBs. The orange squares are PSBs studied by \citet{French+15}.The green circles and the blue triangle are PSBs studied by \citet{Alatalo+16} and \citet{Rowlands+15} respectively \label{fig:samp_sel}}
\end{figure*}

Using the cross-matched catalog of SDSS, GALEX and WISE \citep{Martin+05,Wright+10,Aihara+11,Yesuf+14}, we select a sample of 22 transition post-starburst Seyfert galaxy candidates based on the evolutionary path that starbursts and post-starbursts follow in the dust-corrected NUV-g color, H$\delta$ absorption equivalent width and the 4000{\AA} break \citep{Yesuf+14}, and based on the BPT line ratio AGN diagnostic \citep{Baldwin+81,Kewley+01,kauffmann03c}, as shown in Figure~\ref{fig:samp_sel}. $D_n(4000)$ probes the average temperature of the stars responsible for the continuum emission and is a good indicator of the mean stellar age \citep{Bruzual83,Balogh+99}. The combination NUV-g color with H$\delta$ and $D_n(4000)$ is useful in identifying late stage post-starburst galaxies, which are outliers at intermediate age from loci of galaxies with regular (continuous) star formation histories. As discussed in \citet{Yesuf+14}, we acquired the measurements for physical parameters such as stellar masses and spectral indices from the publicly available catalogs on SDSS website \footnote{https://www.sdss3.org/dr8/spectro/galspec.php} \citep{Aihara+11}. Our sample is restricted to galaxies with redshift, $z < 0.06$, stellar mass, M $> 10^{10}$ M$_\odot$, (NUV-g)$_{\rm dc}  > 2.2$ and $D_n(4000)_{\rm dc} <1.55$ and H$\delta > 1${\AA}. The subscript ``dc" denotes dust-correction. We use the H$\alpha$/H$\beta$ flux ratio with two-component dust attenuation model of \citet{Charlot+00} to correct for dust attenuation of the nebular emission lines, and the empirical relationship between the emission line and continuum optical depths found in \citet{Wild+11} to correct the continuum fluxes \citep[for more details, see][]{Yesuf+14}. Two candidate PSB galaxies with CO observations are removed from the main analyses because they do not satisfy the above cuts. TPSB12 is removed because it has H$\delta < 1${\AA} while TPSB13 is removed because it has (NUV-g)$_{\rm dc}  < 2$ and it is also an ongoing merger. Including these two PSBs instead does not change the main conclusions. Furthermore, note that in Figure~\ref{fig:samp_sel} we only show galaxies with NUV detections \citep{Yesuf+14} in previous samples of PSBs \citep{French+15,Rowlands+15,Alatalo+16}. In later analyses, this restriction is not required.

The redshift cut, $z < 0.06$, was imposed due to the sensitivity of the SMT to achieve the desired signal-to-noise ratio in 6--8 hours. This severely limited the number of Seyferts available for the observation. In addition, in the second year of the observation, the sample was restricted to be above a declination of 35\degree (away from the sun avoidance zone for the SMT and the bulge of the Galaxy), further limiting the observable sample. Therefore, we did not impose the H$\delta >$ 4 {\AA} absorption cut that was done in the original sample selection of \citet{Yesuf+14}. Most (19/22) of the galaxies in our sample have, within the measurement errors, H$\delta$ absorptions $ \gtrsim 3${\AA}, which is expected in post-starbursts with weak or strong bursts or rapidly truncated star-forming galaxies \citep[e.g.,][]{Poggianti+99}. Our sample on average has lower H$\delta$ than do the aforementioned previous samples. Our PSBs are consistent with being later stage PSBs or PSBs with weaker starbursts compared to most of the PSBs in the other samples considered here. Note that the H$\delta$ equivalent width is $\sim 2${\AA} at 1 Gyr after a starburst with star-formation timescale $\tau = 0.1$ Gyr and burst mass fraction $b_f = $ 20\%. For starburst with $\tau = 0.1$ Gyr and $b_f =$ 3\%, H$\delta \sim 0$ at 1\,Gyr (see Figure~\ref{fig:t_bc03} in the Appendix). Thus, not imposing H$\delta > $ 4{\AA} cut helps select old PSBs. However, it remains uncertain whether these galaxies underwent weak or strong starbursts or rapidly truncated star-formation.

It should be noted that at $z < 0.06$, the SDSS fiber covers only $ \sim 3.5$ kpc of the central region of a galaxy, and the spectroscopic measurements may not reflect the galaxy-wide values. On the other hand, the NUV-g color is an integrated galaxy-wide quantity. By combining the spectroscopic measurements with NUV-g color , we select only (PSB) galaxies that are fading galaxy-wide.

\subsection{SMT CO observations}\label{smt_obs}

The observations were carried out using the ARO Submillimeter Telescope (SMT) on Mount Graham, Arizona. The observing runs were in February 25 -- March 10, 2015 and in March 4 -- 25, 2016. We follow the same instrument set up and observing strategy as \cite{French+15}. Namely, we used the 1 mm ALMA Band 6 dual polarization sideband separating SIS (superconductor-insulator-superconductor) receiver and 1 mHz filterbank to measure the CO\,(2--1) 230.5 GHz emission line. The beam size of the SMT for this line is about 33\arcsecond. Beam switching was done with the secondary at 2.5 Hz switching rate and a throw of 120\arcsecond, in the BSP (beam switching plus position switching) mode.  Calibration using a hot load and the standard chopper wheel method was performed every 6 minutes. Calibration using a cold load was performed at every tuning. The observing times range between 4 -- 9 hours. \citet{French+15} observed a subset of 13 PSBs using both SMT and IRAM, which we use to guide comparisons between SMT- and IRAM- observed samples.

The data reduction is done using CLASS, a program within the GILDAS software package\footnote{http://www.iram.fr/IRAMFR/GILDAS/}. The main beam efficiency is calculated using Jupiter in each polarization. 
A first-order polynomial baseline is subtracted from the spectra using data between [-600, 600] km s$^{-1}$, excluding the central regions of [-300, 300] km s$^{-1}$. The spectra are scaled using the main beam efficiency, and are coadded by weighting with the root-mean square (RMS) noise. The spectra are rebinned to 14 km s$^{-1}$ velocity bins. The typical RMS error per bin is 1--2 mK. Thus, we achieve similar sensitivity as previous works \citep{Saintonge+11,French+15,Rowlands+15,Alatalo+16}.

To calculate the integrated CO line intensity, $I_{\rm CO}$, we fit a Gaussian profile to each line, allowing the peak velocity to differ from the optical systemic velocity by up to 200 km s$^{-1}$. The statistical uncertainty in the line intensity is calculated following \citet{Young+11} as $\sigma_I= (\Delta v)^2\sigma^2N_I \left(1+\frac{N_I}{N_b} \right)$, where $\Delta v$ is the channel velocity width, $\sigma$ is the channel RMS noise, $N_I$ is the number of channels used to integrate over the line, and $N_b$ is the number of channels used to fit the baseline. When the line is not detected, the upper limits of the line intensity is calculated as three times the statistical uncertainty. Following \citet{Solomon+97}, the CO line luminosity in {\rm K km s$^{-1}$ pc$^{2}$} is $L^\prime _{\rm CO} = 23.5\, \Omega_{s*b} D_{\rm L}^2 I_{\rm CO} (1+z)^{-3}$ where $\Omega_{s*b}$ is the solid angle of the source convolved with the beam, $z$ is the redshift from the SDSS optical spectrum, and $D_{\rm L}$ is the luminosity distance in Mpc. If the source is much smaller than the beam, then $\Omega_{s*b} \approx \Omega_{b}$. We do not know the CO emitting sizes of our galaxies. We adopt this approximation as a simplifying assumption. We do not expect this to affect our conclusion. Previous works also adopt the same approximation. The fact that SMT has larger beam-size than IRAM, and CO\,(2--1) is known to be more centrally concentrated than CO\,(1--0), make the effect of this approximation less significant in this work compared to previous works. \citet{French+15} estimated for their sample that the $L^\prime _{\rm CO}$ may be underestimated by about $1.5 \times$ because of this effect.

\begin{figure*}
\subfigure{\includegraphics[width=2.0in]{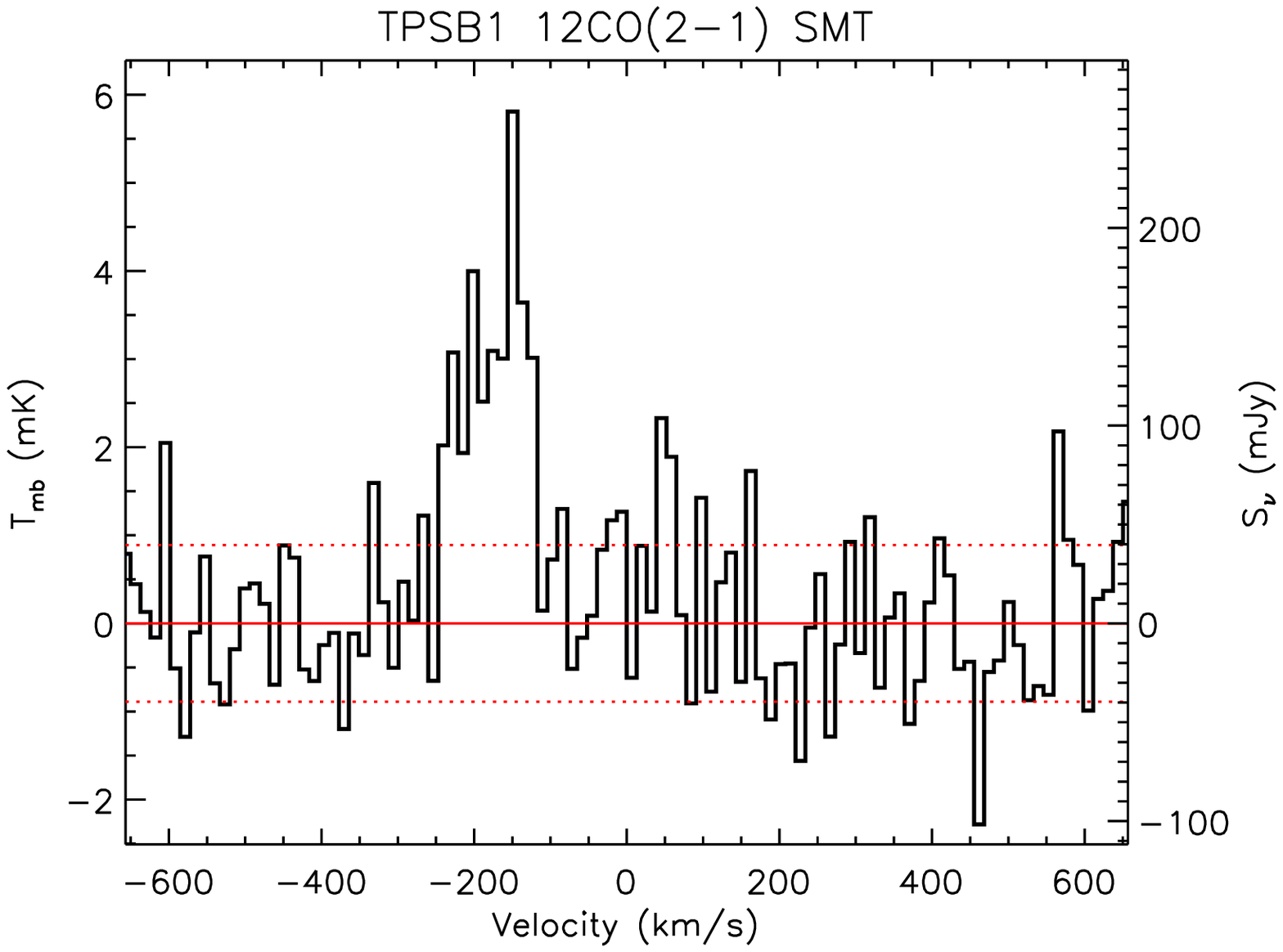}
\includegraphics[width=2.0in]{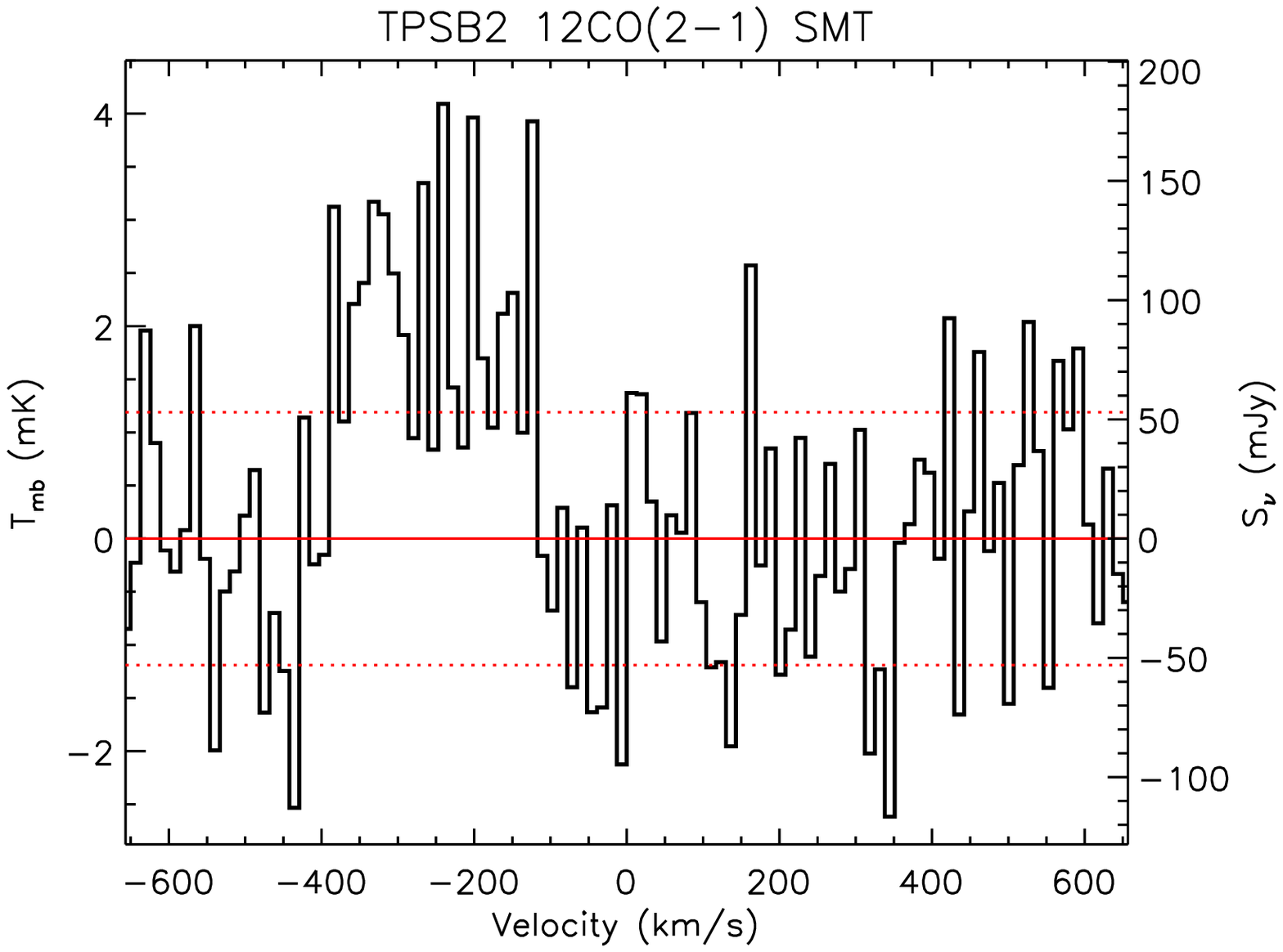}
\includegraphics[width=2.0in]{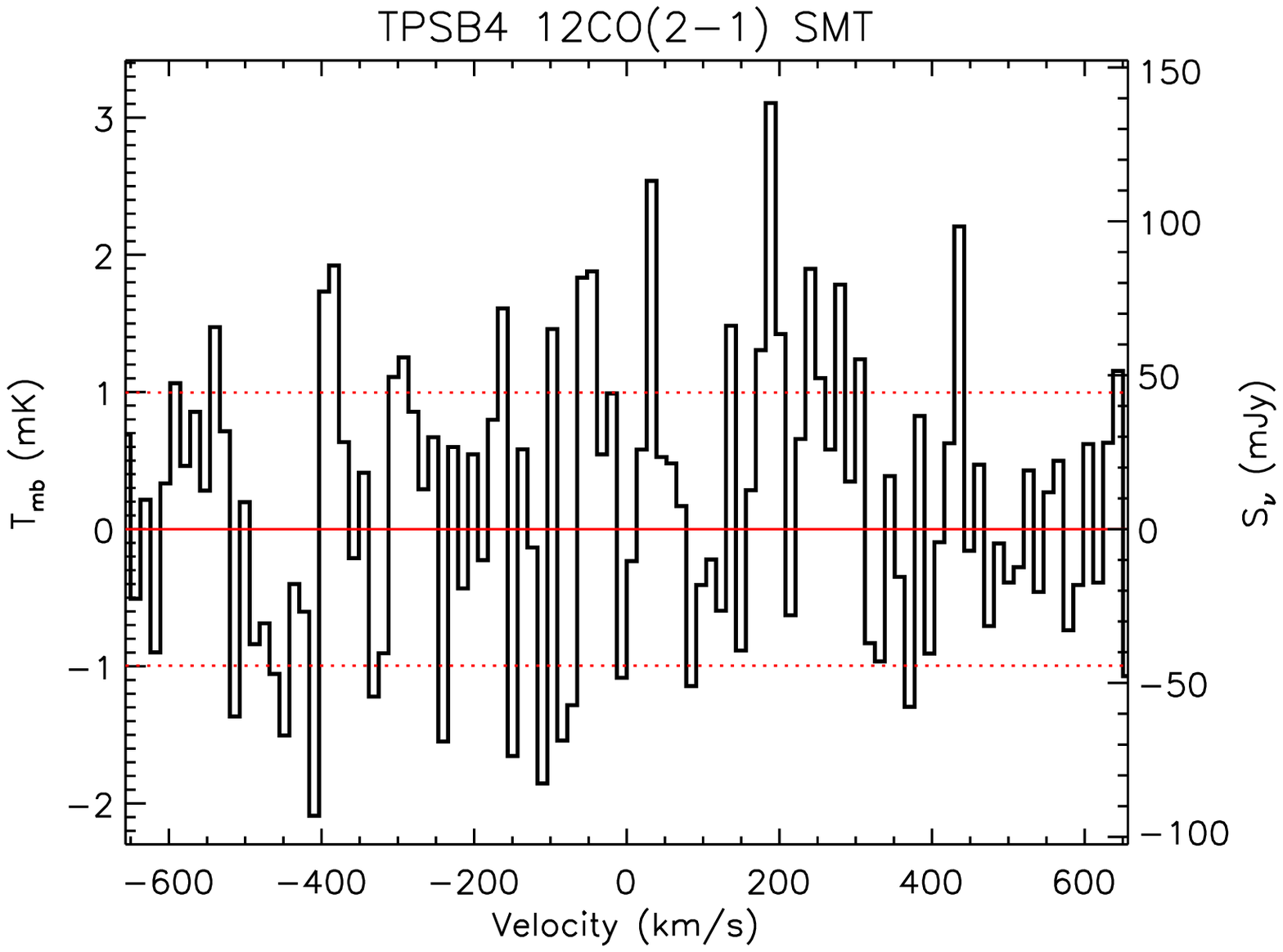}}

\subfigure{\includegraphics[width=2.0in]{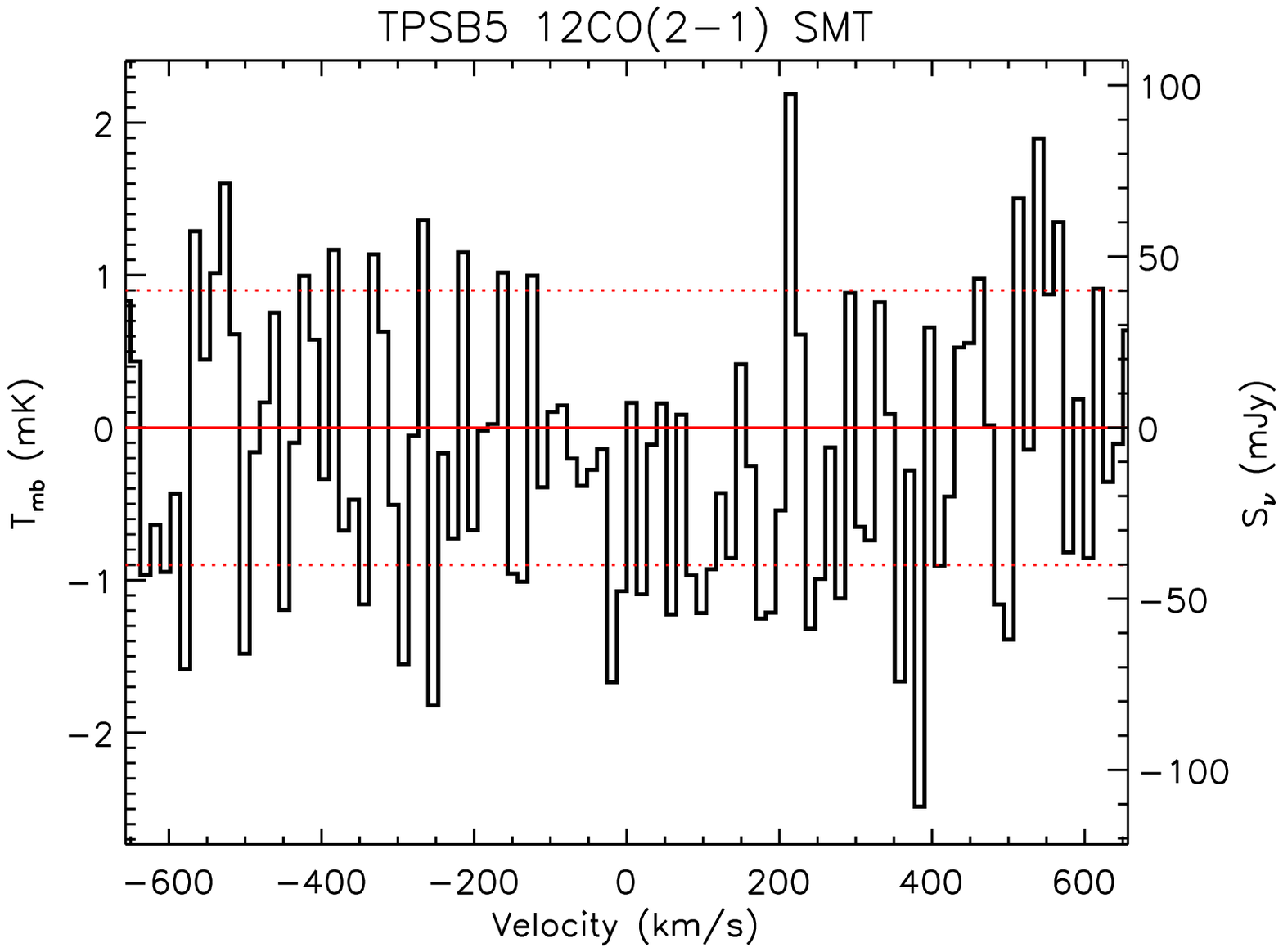}
\includegraphics[width=2.0in]{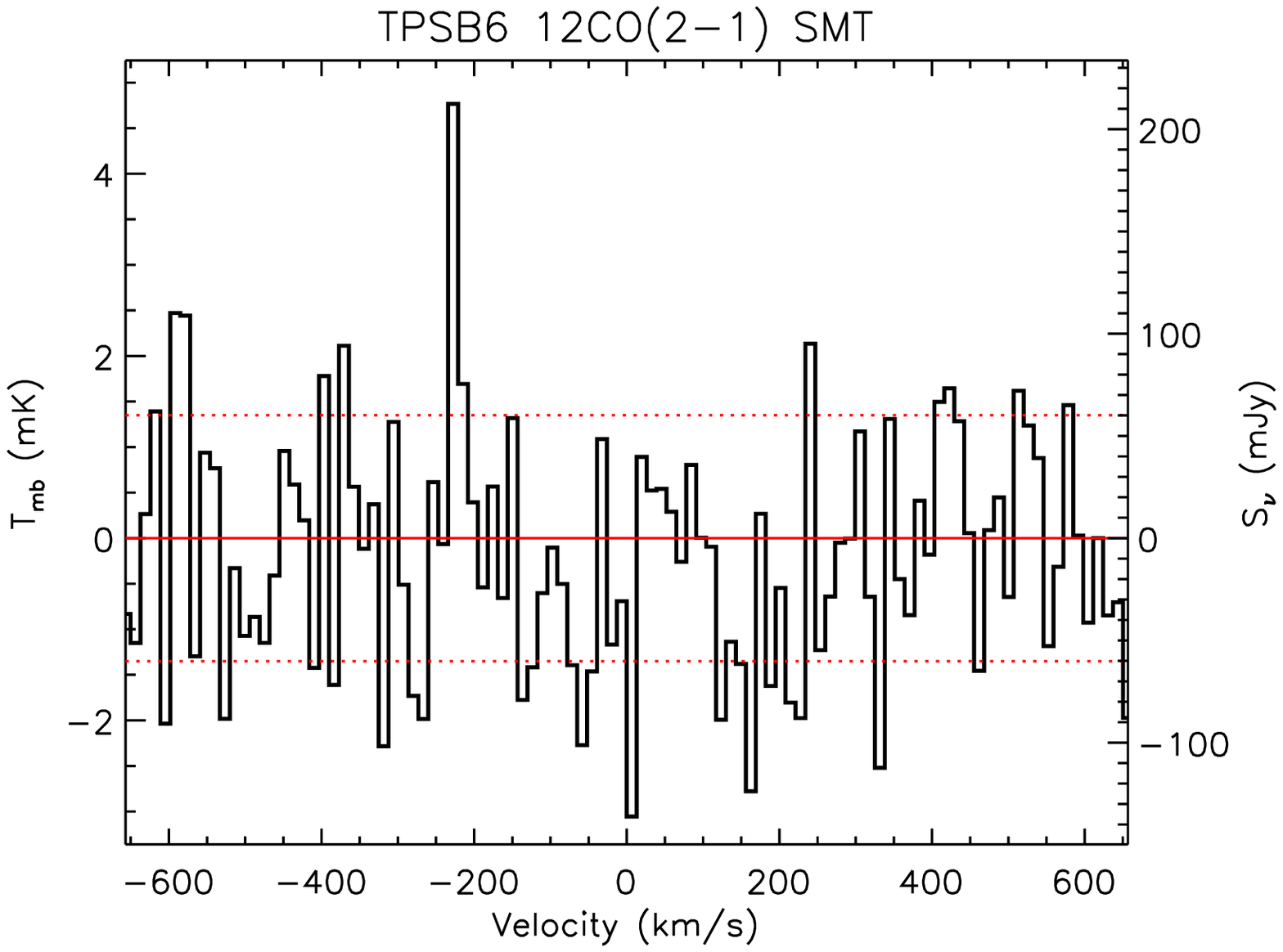}
\includegraphics[width=2.0in]{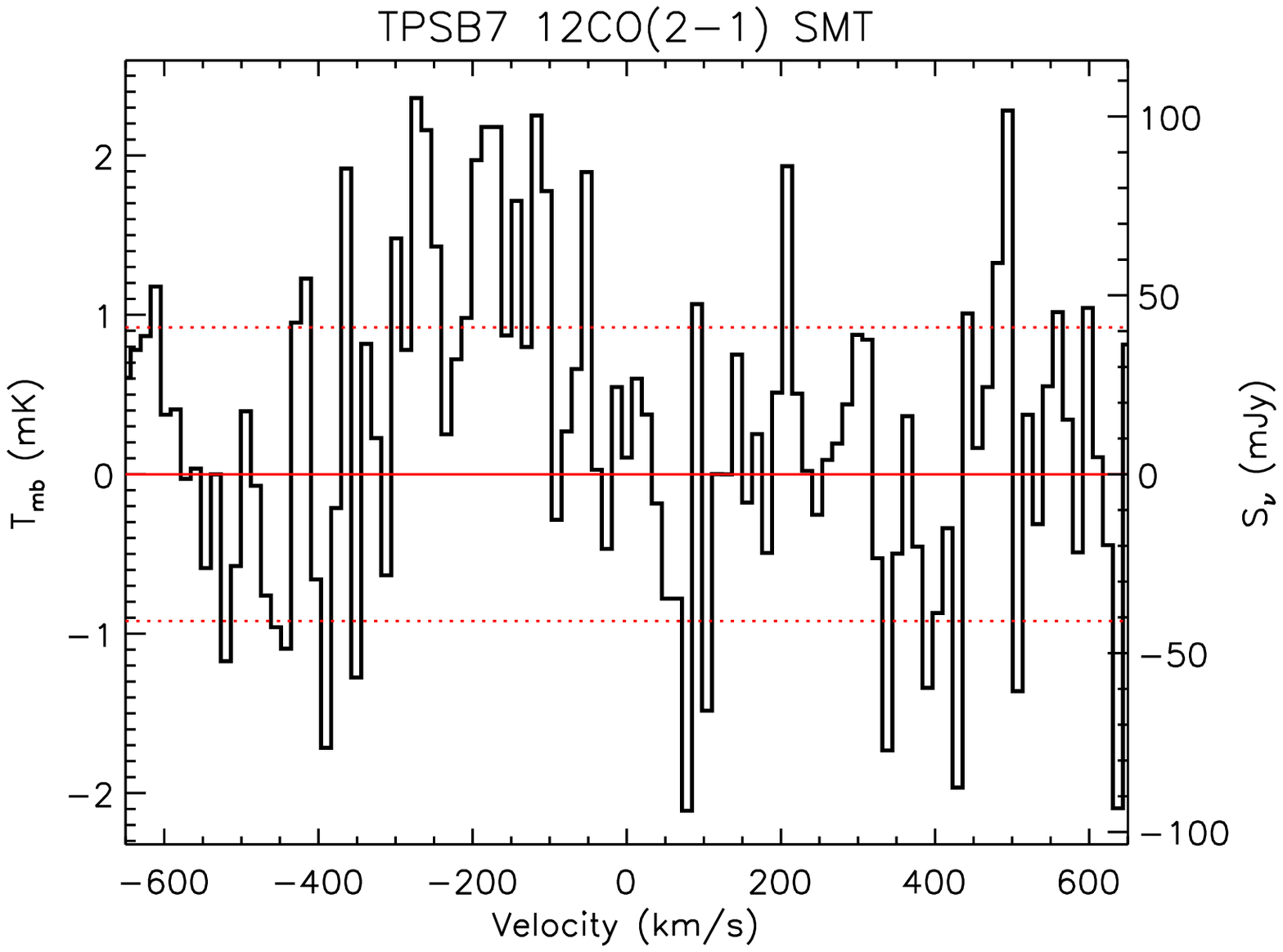}}

\subfigure{\includegraphics[width=2.0in]{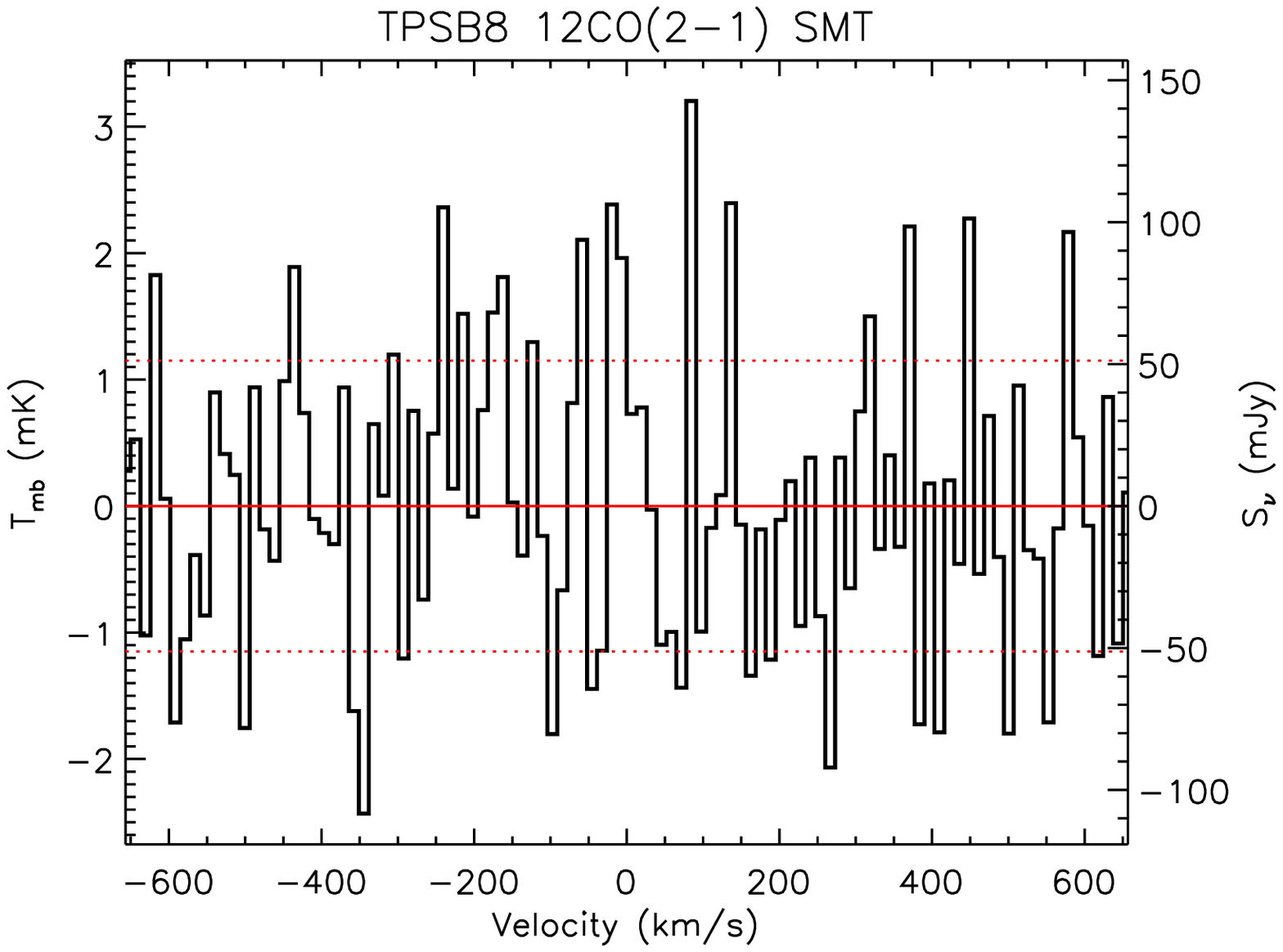}
\includegraphics[width=2.0in]{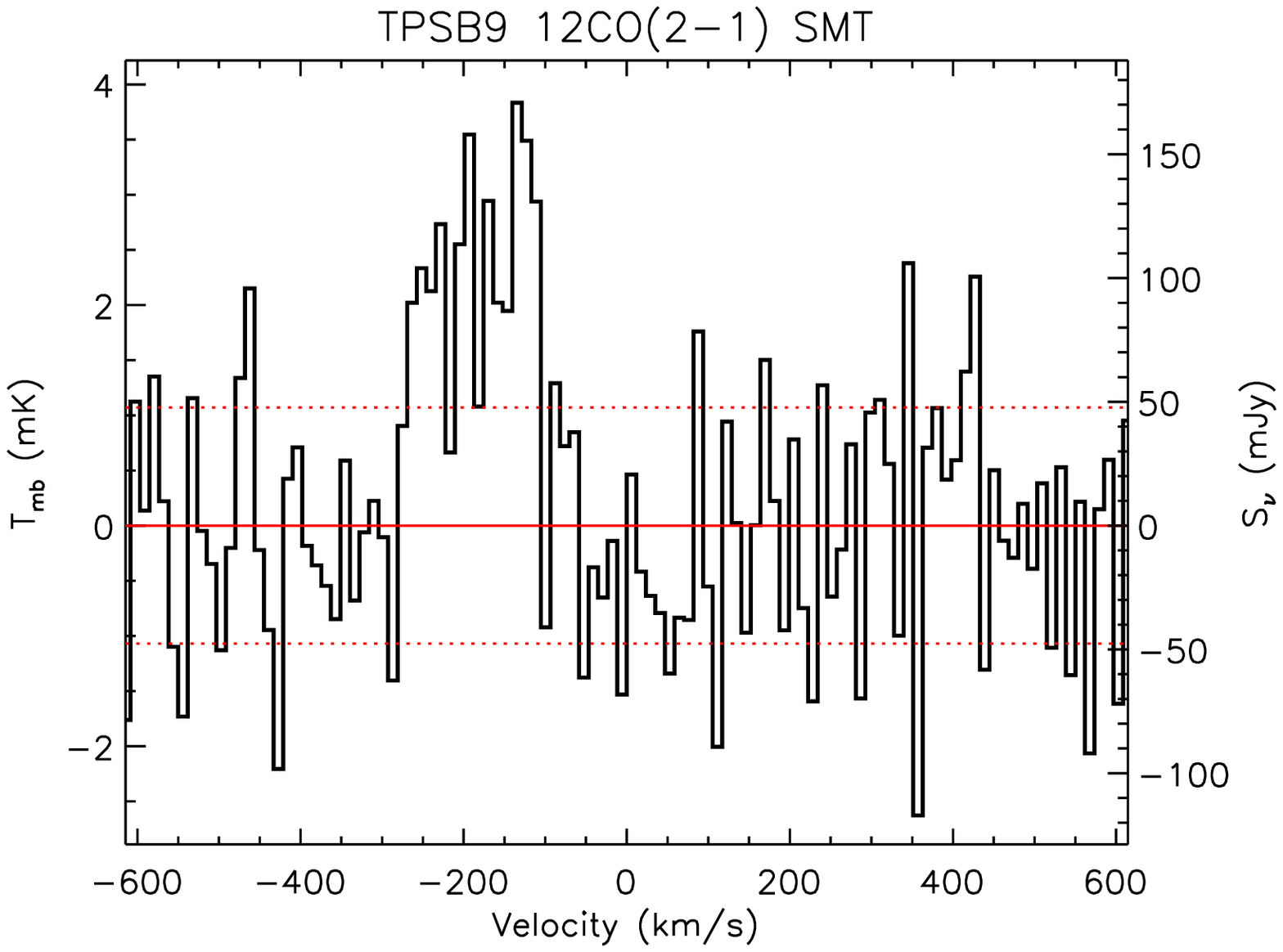}
\includegraphics[width=2.0in]{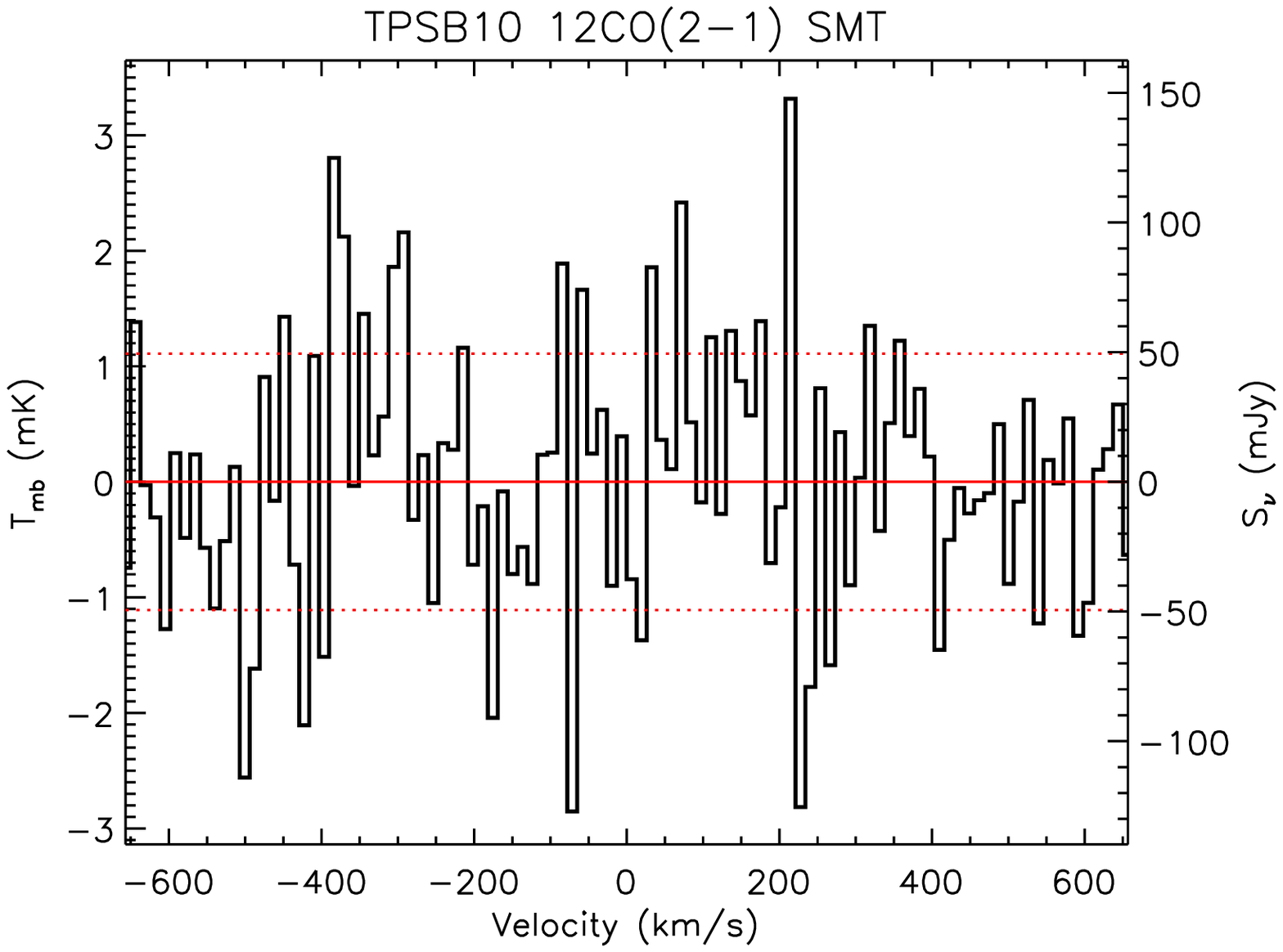}}

\subfigure{\includegraphics[width=2.0in]{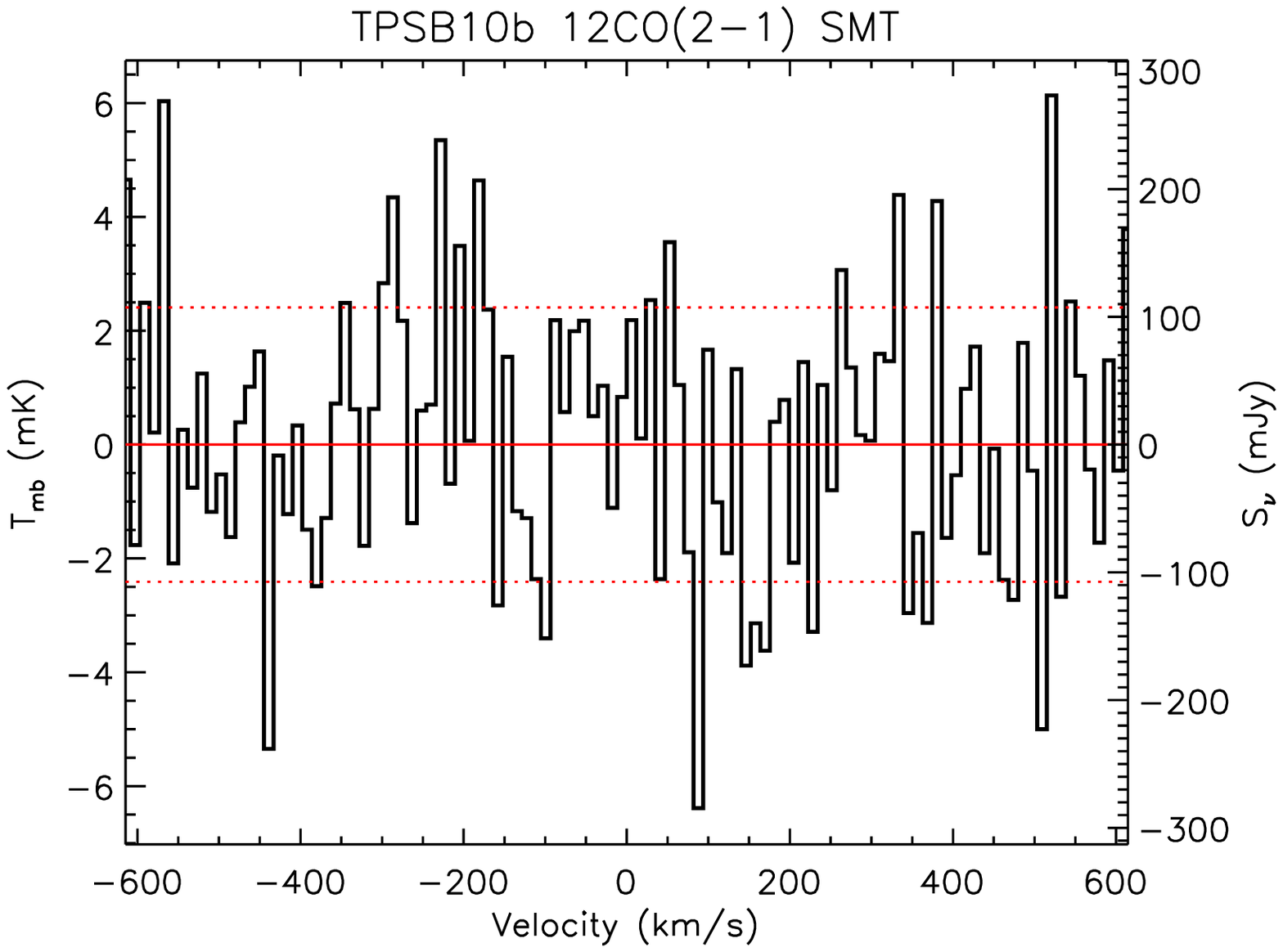}
\includegraphics[width=2.0in]{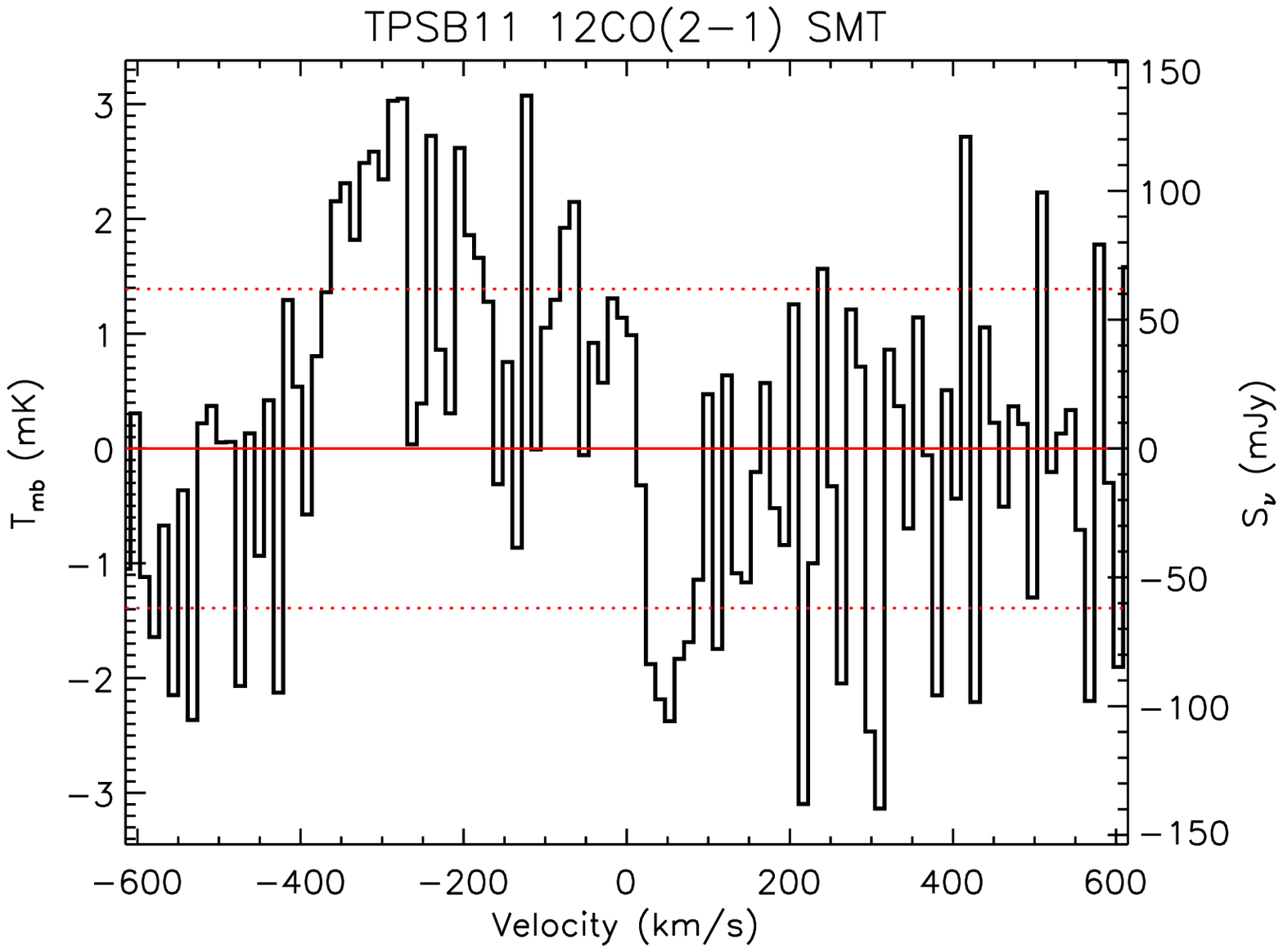}
\includegraphics[width=2.0in]{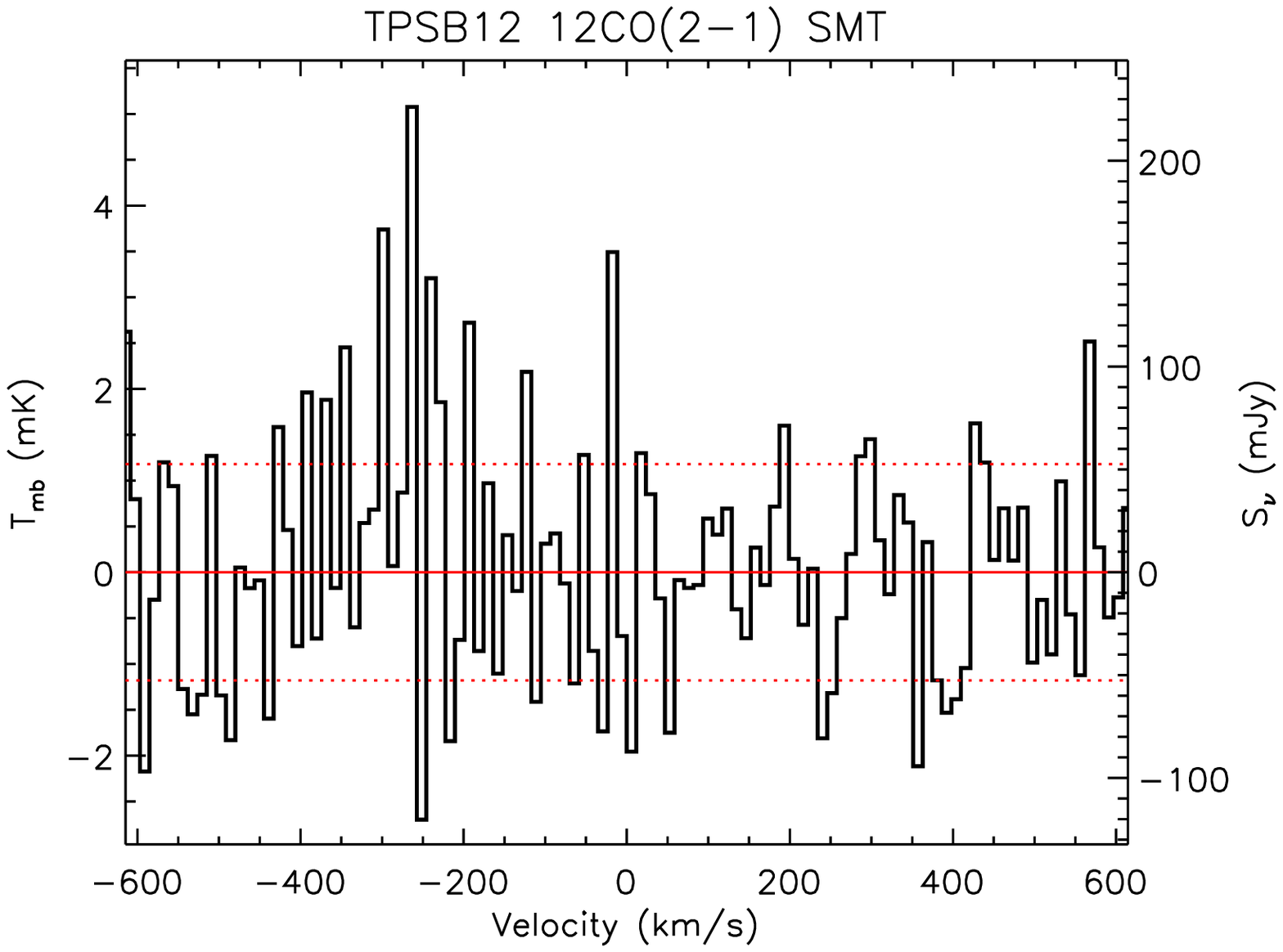}}
\caption{Submillimeter Telescope (SMT) CO\,(2--1) spectra for Seyfert PSBs. Spectra are shown in units of both main beam temperature $T_{\rm mb}$ (mK) and $S_\nu$ (Jy). 
The dashed red lines represent the RMS of the binned data. The data are binned by 14\,km\,s$^{-1}$. \label{fig:cospec1}}
\end{figure*}

\begin{figure*}
\subfigure{\includegraphics[width=2.0in]{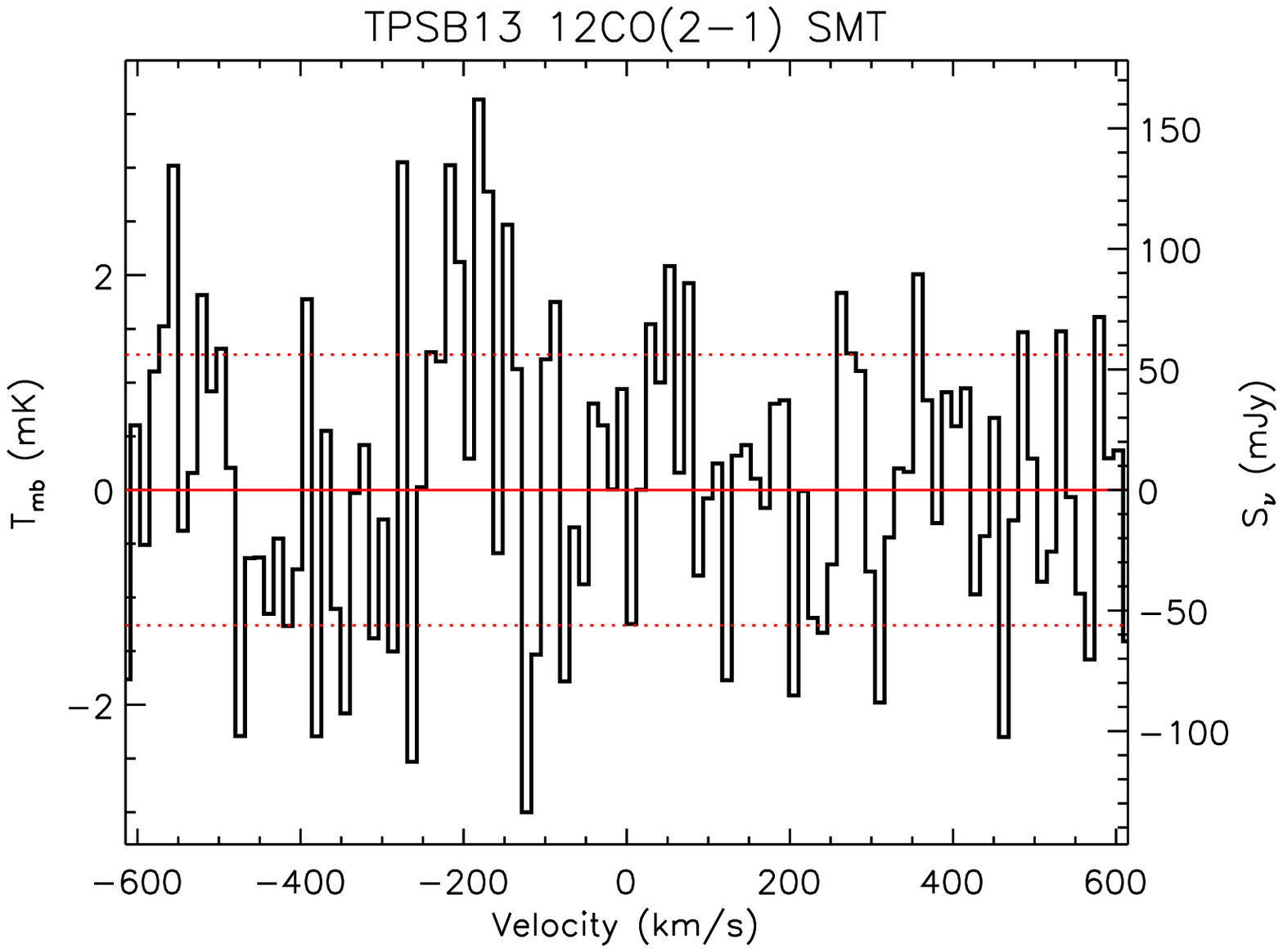}
\includegraphics[width=2.0in]{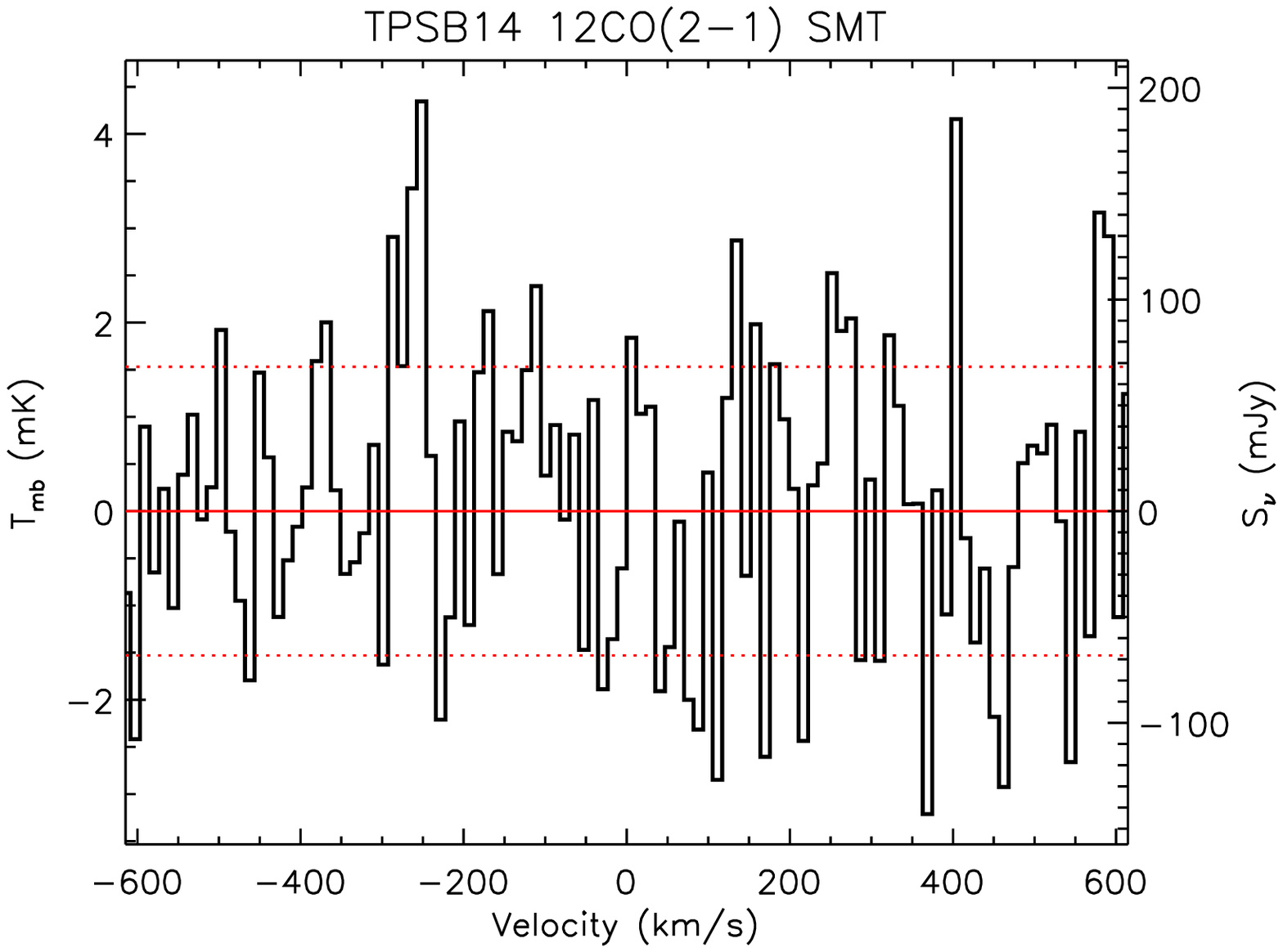}
\includegraphics[width=2.0in]{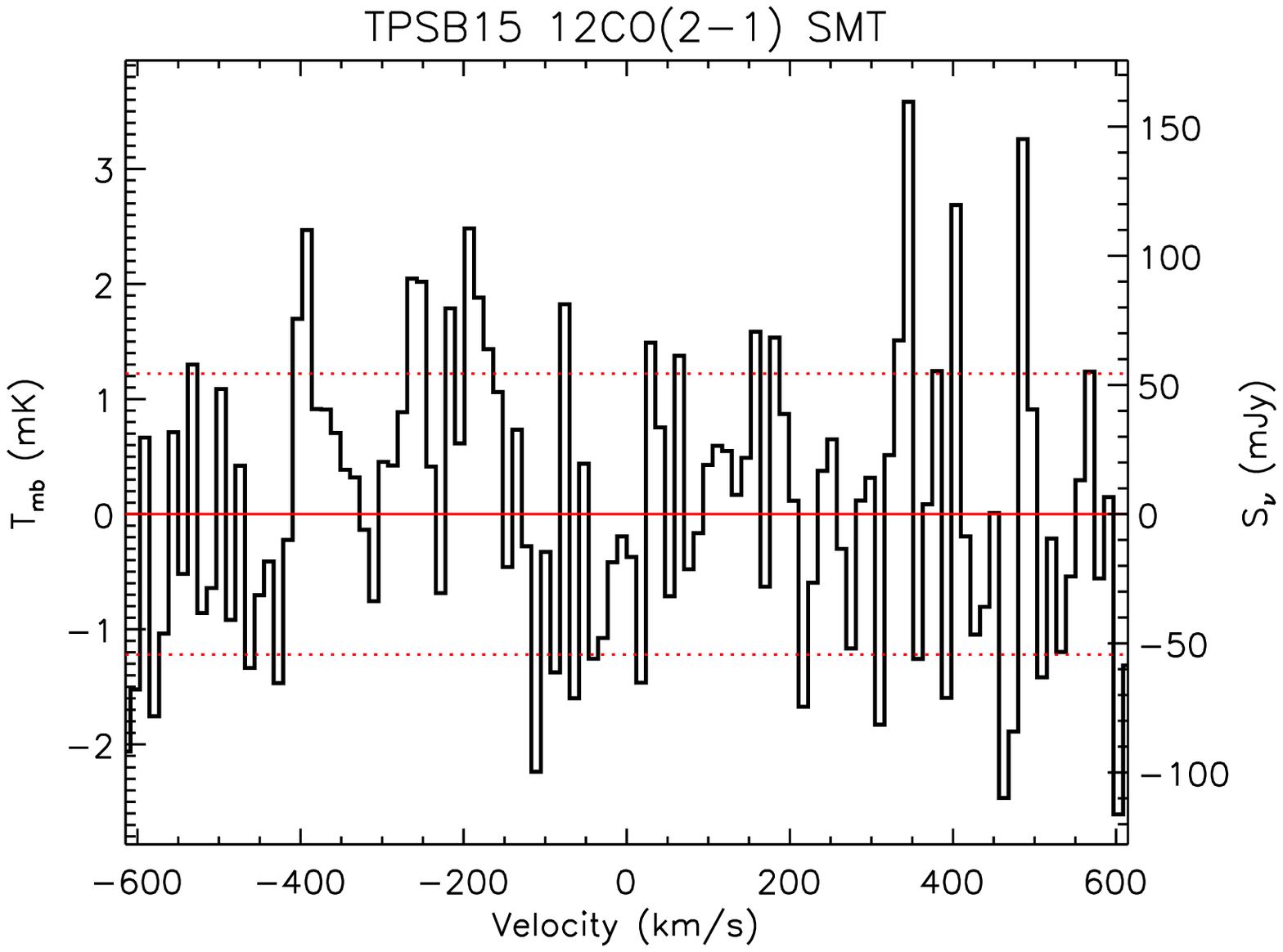}}

\subfigure{\includegraphics[width=2.0in]{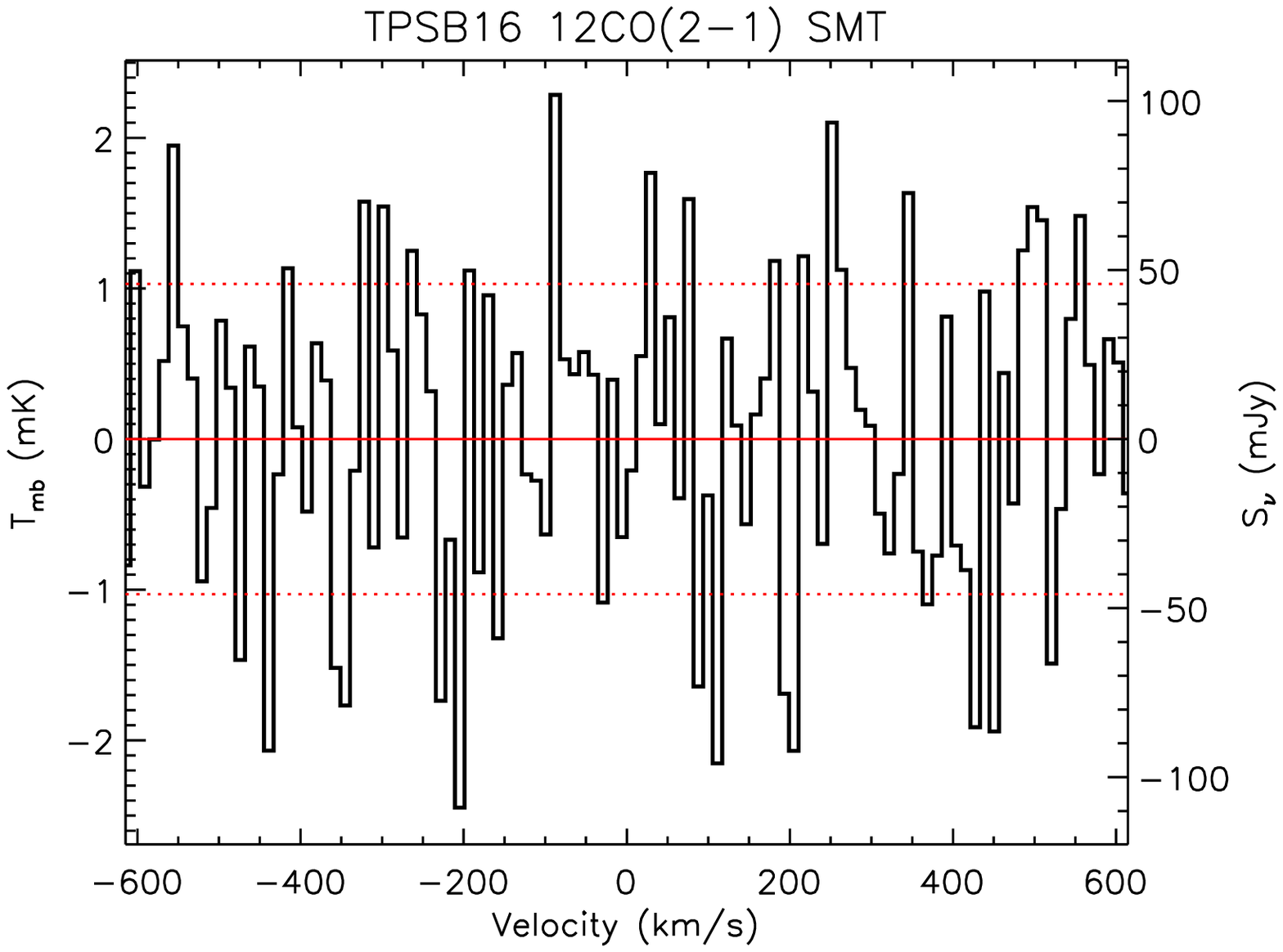}
\includegraphics[width=2.0in]{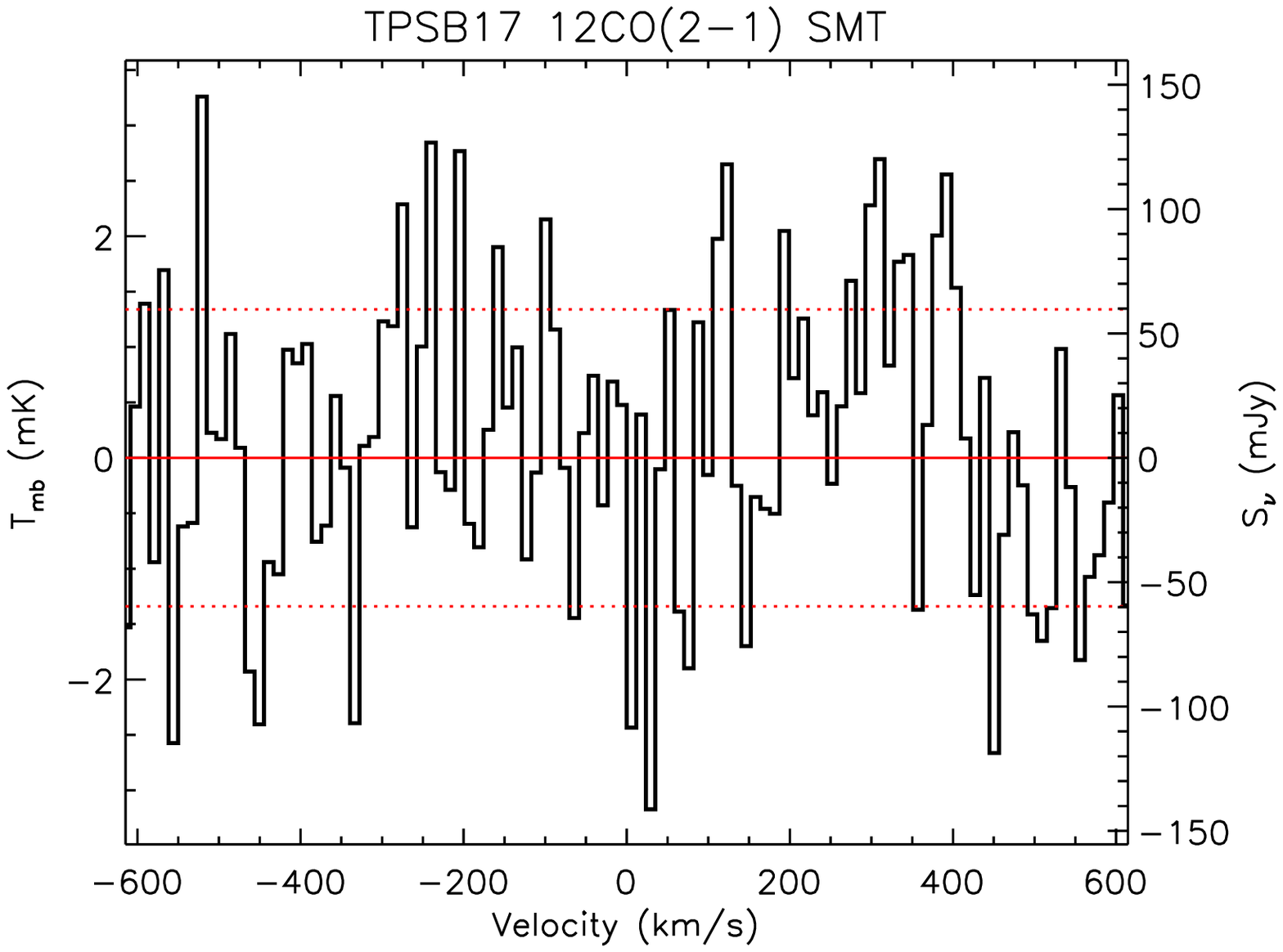}
\includegraphics[width=2.0in]{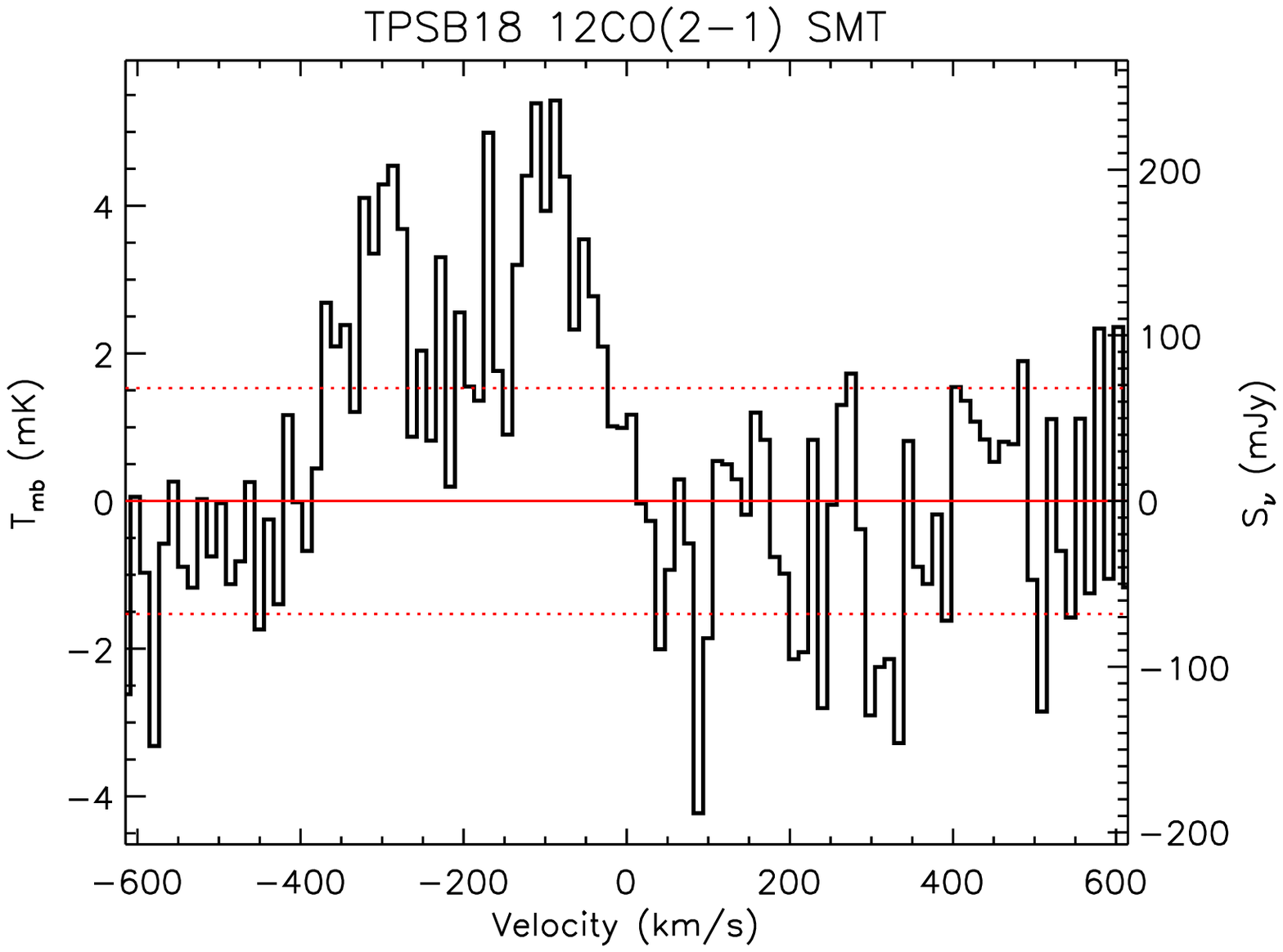}}

\subfigure{\includegraphics[width=2.0in]{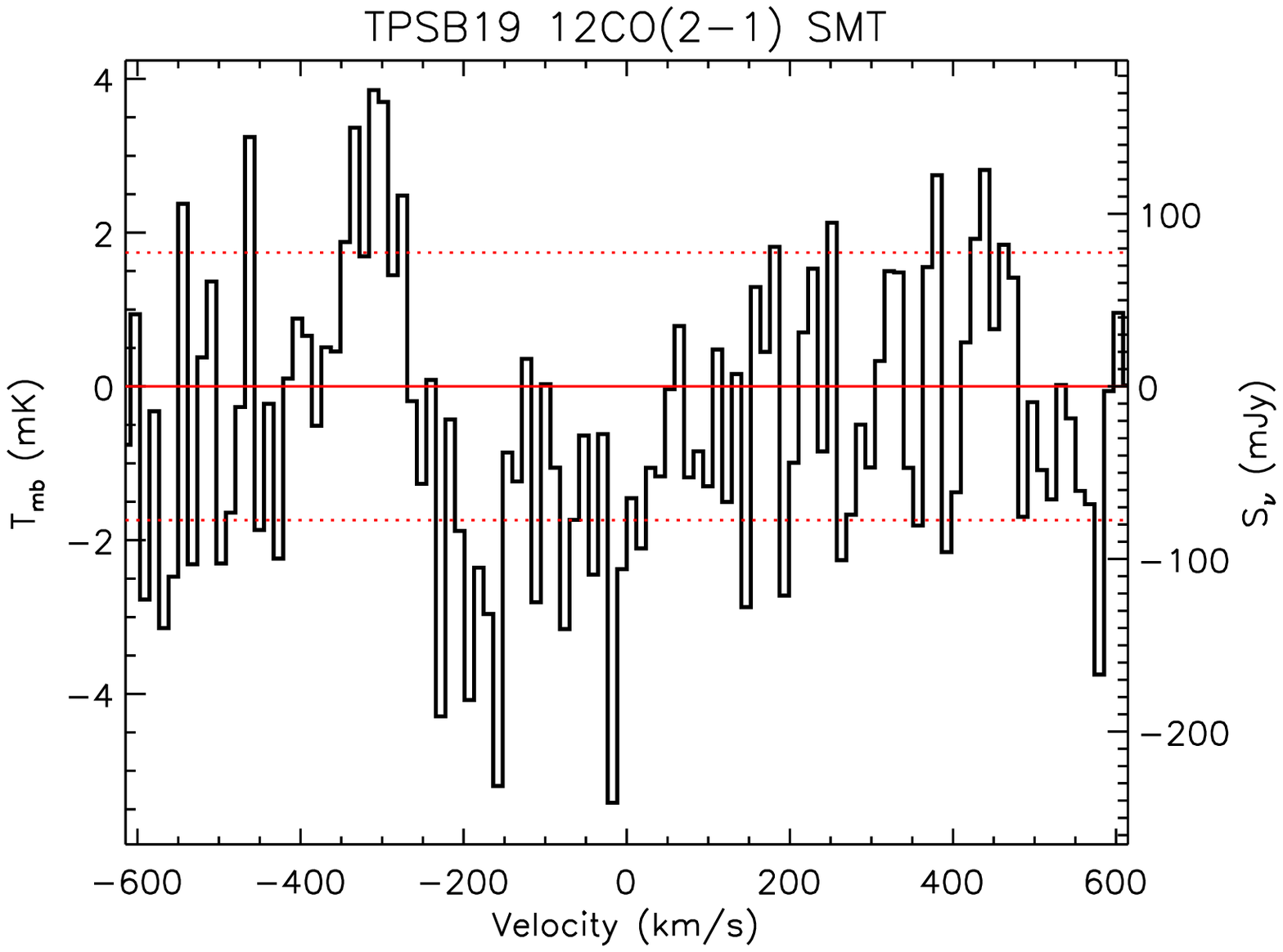}
\includegraphics[width=2.0in]{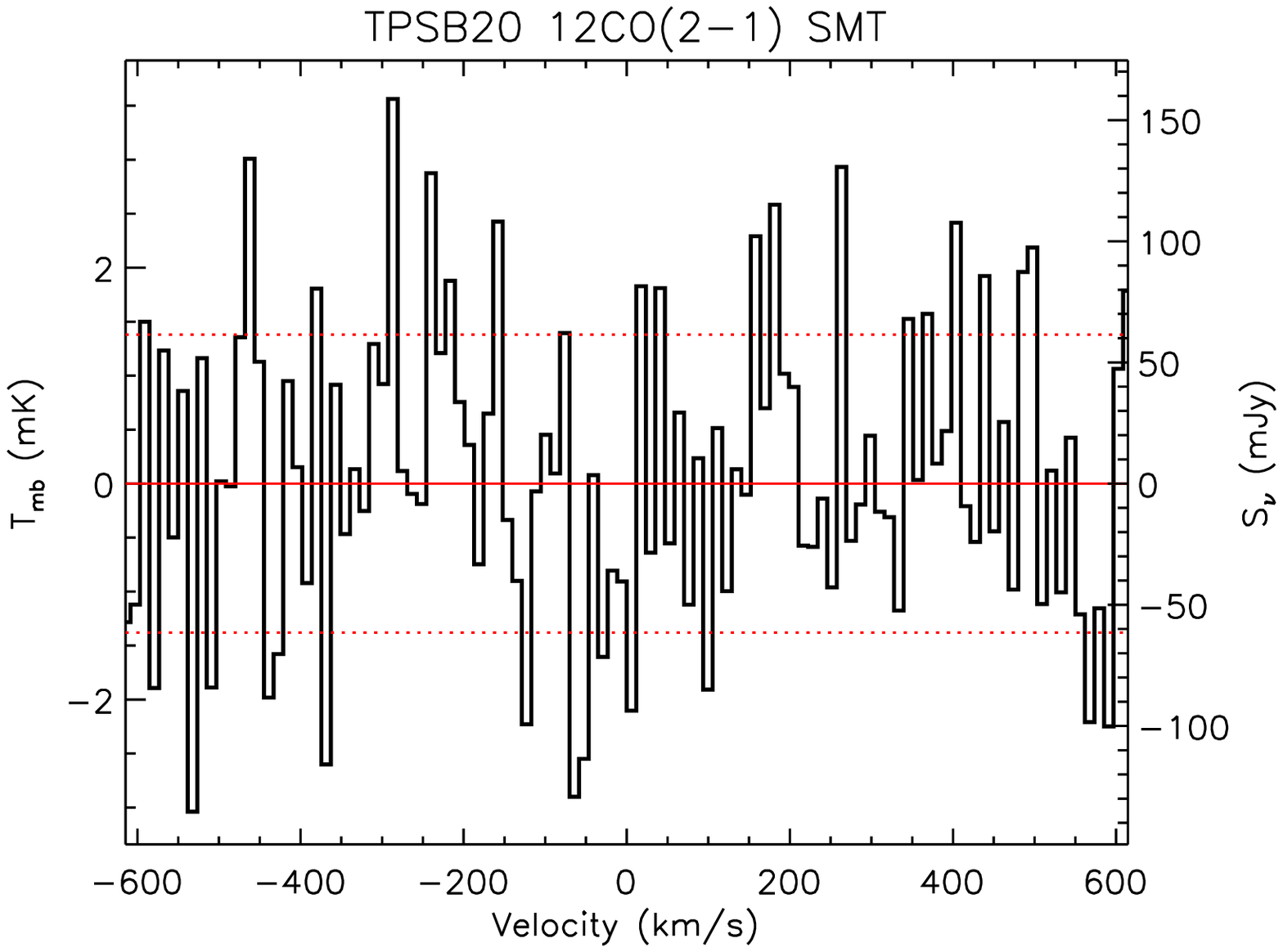}
\includegraphics[width=2.0in]{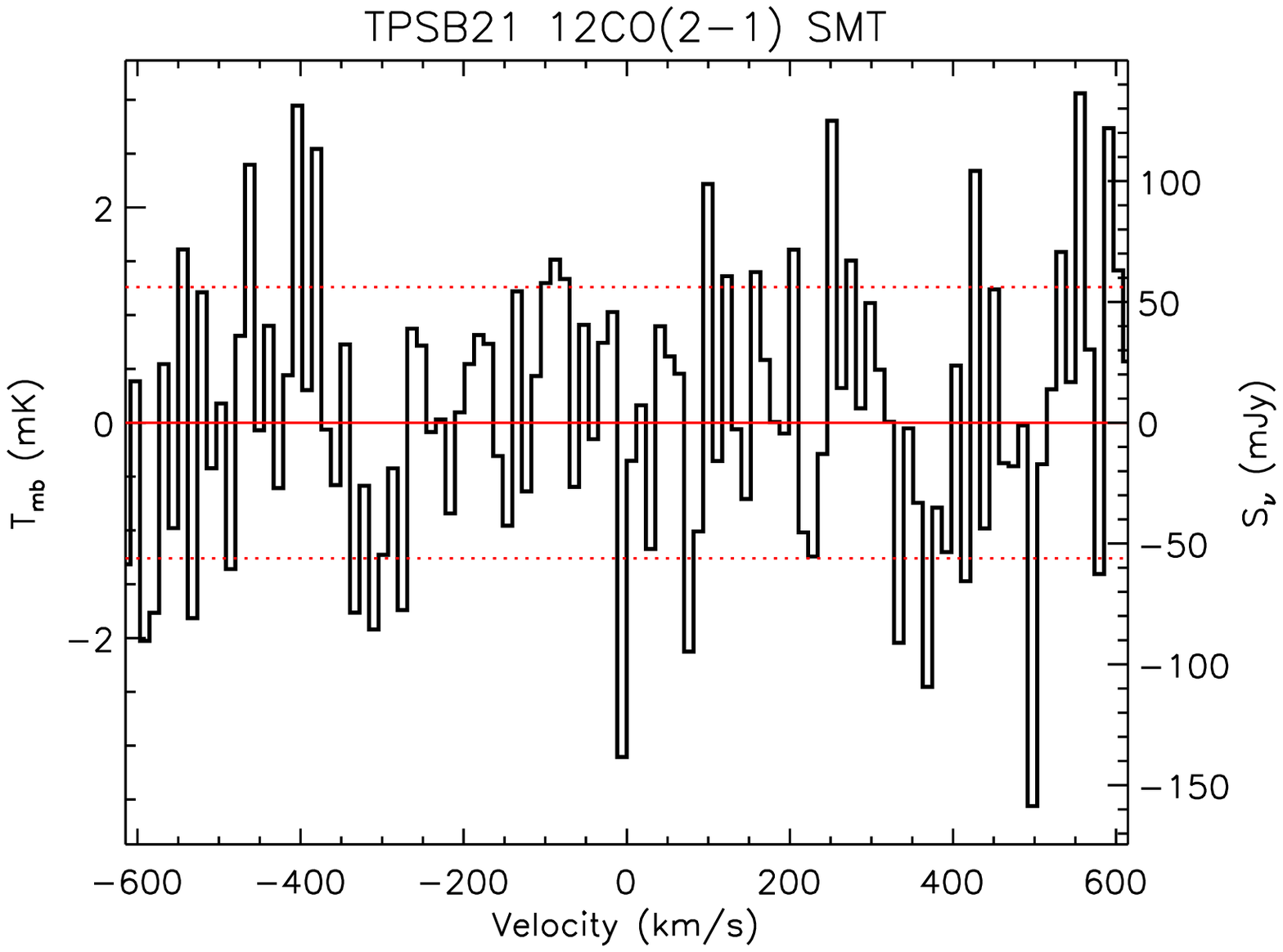}}

\subfigure{\includegraphics[width=2.0in]{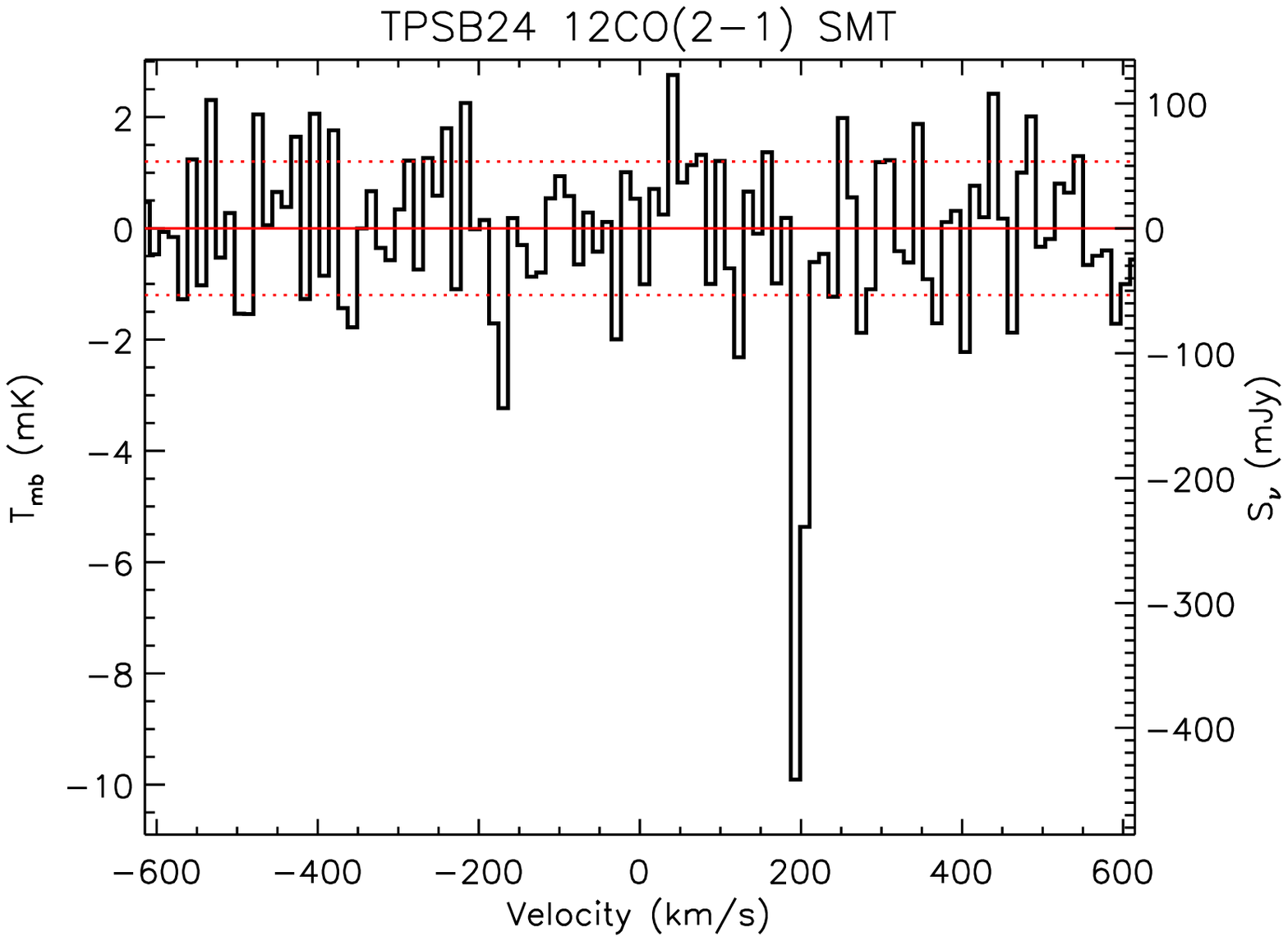}
\includegraphics[width=2.0in]{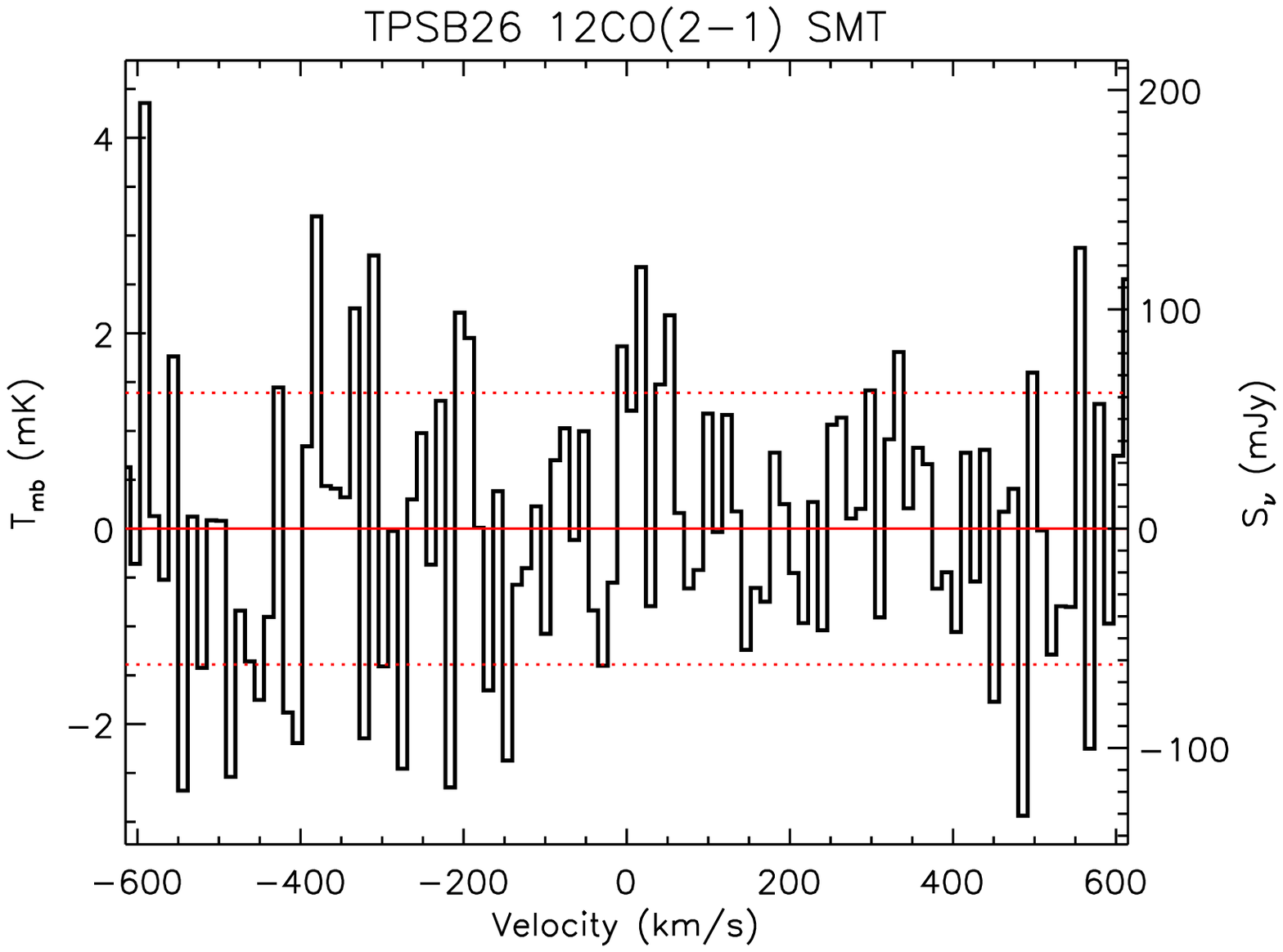}
\includegraphics[width=2.0in]{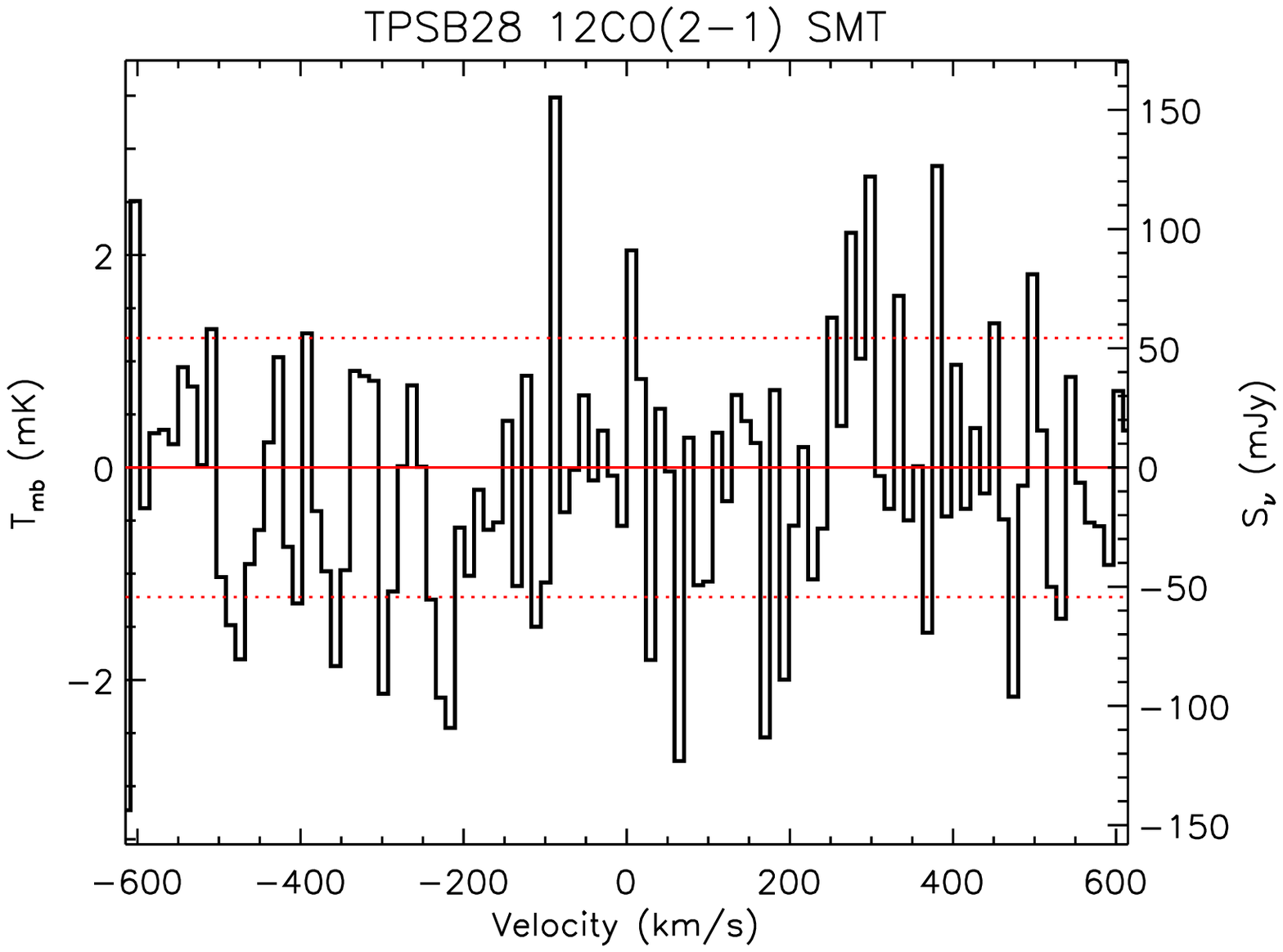}}

\caption{CO\,(2--1) spectra (continued) \label{fig:cospec2}}
\end{figure*}

\begin{figure*}
\includegraphics[width=5in]{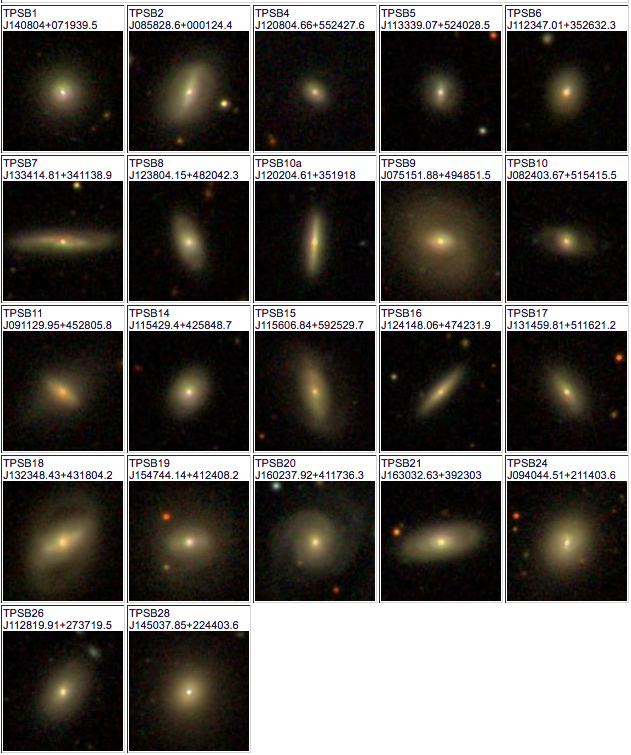}
      \caption{ SDSS cutout images of the green-valley Seyfert PSBs. They show prominent bulges. But they do not show strong merger signatures and dust extinctions, unlike some of the previously studied PSBs \citep[e.g.,][]{Alatalo+16}. This, together with their older stellar populations and red NUV-optical colors, indicates that they are  late-stage post-starbursts. \label{fig:cutouts}}
\end{figure*}

The molecular gas mass can be calculated from $L^\prime _{\rm CO}$ by assuming a CO conversion factor (mass-to-light ratio) $\alpha_{\rm CO}$, M(H$_2)= \alpha_{ \rm CO} L^\prime _{\rm CO}$. We assume an \alphacom in Milky Way  disk \citep{Bolatto+13} and $R_{21}=L^\prime _{\rm CO(2-1)}/L^\prime _{\rm CO(1-0)}=1$ . As discussed in section~\ref{sec:alphaco}, choosing a lower value of $\alpha_{ \rm CO}$ strengthens our main conclusion. The ratio $R_{21}$ is uncertain \citep{Leroy+13,Sandstrom+13}. \citet{Leroy+13} found $R_{21} \sim 0.7 \pm 0.3$ in nearby disk galaxies. We adopt the Galactic $\alpha_{\rm CO}$ value throughout the paper unless explicitly stated and all previous measurements are adjusted to our assumed value of $\alpha_{\rm CO}$. The observation are summarized in Table 1, the co-added spectra are given in Figure~\ref{fig:cospec1} \& \ref{fig:cospec2} and the SDSS cutout images of the Seyfert PSBs are shown in Figure~\ref{fig:cutouts}.

\begin{table*}
\caption{Summary of the CO observations of the new Seyfert transition PSBs. The column names are, from left to right, target name, right ascension, declination, redshift, 4000{\AA} break, dust-corrected NUV-g color, H$\delta$ absorption equivalent width, WISE flux ratio, stellar mass, molecular gas mass, detection flag.}
\begin{threeparttable}
       	\begin{tabular}{lccccccccrr}
      	\hline
         \hline
	Target & RA & Dec. &$z$&$D_n(4000)_{\rm dc}$&(NUV-g)$_{\rm dc}$ & H$\delta$ & log f$_{12}$/f$_{4.6}$ & log M$_\star$/M$_\odot$ & log M$_{\rm H_2}$/M$_\odot$   &log M$_{\rm H_2}$/M$_\star$\\
         \hline
	TPSB1   &212.01668  &      7.32762   &  0.0238    &  1.26      & 2.5      & 5.8     &       0.62     & 10.04     & 8.79    & $-1.25$\\
	TPSB2  &134.61915    &   0.02347    & 0.0285      & 1.32      & 2.8     & 5.2       &       0.59       &  10.40   &   8.64   &  $-1.76$\\
	TPSB4   &182.01943  &    55.40766    &  0.0513    &   1.40     &  3.0     & 4.0     &      0.66    & 10.30    & $<9.19$    & $<-1.11$\\
	TPSB5   &173.41286   &   52.67458     & 0.0490     &   1.40     & 3.2      & 4.9     &      0.22     & 10.19     & $<9.10$    & $<-1.09$\\
	TPSB6   &170.94591   &   35.44231    & 0.0341     &  1.42      &  2.5     & 4.9     &      0.52      & 10.29    & $<8.97$    & $<-1.32$\\
	TPSB7   &203.56173   &   34.19415     & 0.0236     &  1.31      &  2.8     & 5.5    &      0.44     & 10.15   & 8.62    & $-1.53$\\
	TPSB8   &189.51736   &   48.34506     & 0.0306     &  1.44      & 3.6      & 4.0     &      0.04     & 10.21     & $<8.81$    & $<-1.40$\\
	TPSB9   &180.51923   &   35.32168    & 0.0341    &  1.48      & 3.7     & 4.5      &      0.31   & 10.54   & $<8.90$    & $<-1.64$\\
	TPSB10 &117.96618   &  49.81432      & 0.0244    &  1.40      & 2.7     & 2.9      &      0.41     & 10.45     & 9.29    & $-1.16$\\
	TPSB10b   &126.01534    &  51.90432     & 0.0315     &  1.36      & 2.4   &  2.8      &      0.72    & 10.12     & $<9.12$  & $<-1.00$\\
	TPSB11  &137.87485    &  45.46828    & 0.0268     &  1.43      & 2.6     &  3.0      &     0.64     & 10.70     & $8.93$    & $-1.77$\\
	TPSB12  &139.49938    &  50.00218    &  0.0342     & 1.41      & 3.1      & 0.8       &      0.53    & 10.33     & $<8.94$    & $<-1.39$\\
	TPSB13  &173.16774    &  52.95040   &  0.0266     & 1.50      & 1.8      & 2.4       &      0.48    & 10.54     & $<8.68$    & $<-1.86$\\
	TPSB14  &178.62255    &  42.98021     & 0.0235      & 1.51     & 2.9      & 1.7       &      0.57     & 10.05     & $<8.71$    & $<-1.34$\\
	TPSB15  &179.02851    &  59.42492    &  0.0320     & 1.52      & 3.7     & 1.5      &      0.41     & 10.49     & $<8.85$    & $<-1.64$\\
	TPSB16  &190.45029    &  47.70888    &  0.0308     &  1.48     &  3.3     &  3.2      &      0.45   & 10.25     & $<8.81$    & $-1.44$\\
	TPSB17  &198.74928    & 51.27259     & 0.0249      & 1.48      & 3.4     &  2.8      &      0.34     & 10.02     & $<8.66$    & $<-1.36$\\
	TPSB18  &200.95187    &  43.30118    & 0.0273      & 1.28      & 3.0      & 3.3        &      0.74     & 10.68    & 9.21    & $-1.47$\\
	TPSB19  &236.93394    &  41.40230    &  0.0327     & 1.35     &  2.9     &  3.5     &       0.69    & 10.47    & $<9.00$    & $<-1.47$\\
	TPSB20  &240.65806    &  41.29344    &  0.0348     & 1.45     & 2.7      & 3.0       &       0.08     & 10.59     & $<8.97$    & $<-1.62$\\
	TPSB21  &247.63604    &  39.38420    &  0.0305     &  1.42     &  2.9     &  3.0      &       0.55     & 10.54    &  $<8.82$   &  $<-1.72$\\
	TPSB24  &145.18542    &  21.23427    &  0.0244     & 1.43      & 3.8      & 2.2       &       0.71       & 10.33     & $<8.60$    &  $<-1.73$\\
	TPSB26  &172.08298    &  27.62209   &  0.0321     &  1.42    &  3.1     &  2.8      &      0.18     & 10.31      &  $<8.89$    & $<-1.42$\\
	TPSB28  &222.65772    &  22.73433   &  0.0210     & 1.33     & 3.6     & 3.8       &      0.60     & 10.09     & $<8.47$    & $<-1.62$\\
	\hline
	\end{tabular}
	 \begin{tablenotes}
      \small
      \item  For the non-detections, the given molecular masses are  3$\sigma$ upper limits. 
          \end{tablenotes}
  \end{threeparttable}
\end{table*}

\section{Overview of Statistical Methods}\label{sec:stat}

In this section, we briefly review the statistical methods and tests used in the next section. Readers interested in more details of the methods should refer to the references provided. 

In this section, we are concerned with how to analyze data that include both CO detections and CO upper limits, and make inference that is consistent with the incomplete information at hand. In statistics, a data set is called censored when the value of a measurement is only partially known to be above or below a threshold. Survival analysis is an area of statistics that mostly deals with modeling censored data such as time to an event \citep[e.g.,][]{Klein+05,Feigelson+85,Helsel+05}. We use existing survival analysis methods to properly extract information from our censored CO observations. 

Let T denote a positive random variable representing time to an event of interest, say death. The survival function is the probability that an individual survives beyond time t, $S(t)=Pr(T \ge t)=1-F(t)$, where $F(t)$ is the cumulative distribution function (CDF). The Kaplan-Meier estimator \citep{Kaplan+58} is a non-parametric maximum likelihood estimator of the survival function, even in the presence of censoring. A non-parametric method does not require an assumption that  the data follow a specific probability distribution. Instead, the data are ranked from smallest to largest, providing information on the relative positions of each observation. In the case of ties, Kaplan-Meier method assigns the smallest rank to each observation. The censored observations are also used in calculating the ranks. Let there be a set of $\{t_k\}^n_{k=1}$ data points and of these let $t^\prime_1 < t^\prime_2 < t^\prime_3, \ldots, t^\prime_r$ be ranked, distinct uncensored values. At each time point $t^\prime_j$, we observe $d_j$, the number of deaths, $c_j$, the number of censored observations between the time $t^\prime_j$ and $t^\prime_{j-1}$ and, $n_j$, the number of individuals at risk just prior to the time $t^\prime_j$. In other words, $n_j$ is the total sample size minus those who are censored or have died before $t^\prime_j$, $n_j=n-c_j-d_j=\sum_{l \ge j} (c_l+d_l)$. The Kaplan-Meier estimator has a form : $\hat{S}(t)= \prod_{j:t^\prime_j <t}\big(1-d_j/n_j\big)$. It is a step function with jumps at times $t^\prime_j$.

To compute the empirical cumulative distribution function (ECDF), using the Kaplan-Meier method, we use the NADA package in R programing language \citep{Helsel+05,Lee+15}. We use the function \emph{cenfit} in the NADA package to compute the Kaplan-Meier estimator for the CO data \citep{Helsel+05,Lee+15}.

The survival curves or ECDF of two groups can be compared using the log-rank test, which tests the null hypothesis that two groups have the same distribution against the alternative hypothesis that two groups have different distributions \citep{Mantel+66,Cox72,Harrington+82,Martinez07}. If the data are uncensored, the log-rank test gives similar result to the Mann-Whitney test. We use the routine \emph{cendiff} in the NADA package to do the the log-rank test.

The Kendall's $\tau$ is a non-parametric correlation coefficient that can be used for testing trends in both censored and uncensored data \citep[e.g.,][]{Helsel+05}. $\tau$ ranges between -1 and 1. $\tau=0$ is no correlation, $\tau=1$ is a perfect correlation, and  $\tau=-1$ is a perfect anti-correlation.

In the uncensored  case, the slope a linear regression can be estimated using the Theil-Sen slope \citep{Sen+68}, which is the median of all pairwise slopes between two data points. This slope results in the Kendall's $\tau$ of 0 for the correlation between the residuals and the covariate, X. For censored data, the slope can be estimated using the extended version called Akritas - Theil - Sen estimator \citep{Akritas+95}. The intercept of the linear regression is the median of the residual. We use the \emph{cenken} routine in NADA package to compute the Kendall's $\tau$ correlation coefficient and fit the ATS regression line to our censored CO data \citep{Helsel+05,Lee+15}.

\section{Results}\label{sec:res} 

\begin{figure}
\includegraphics[scale=0.4,angle=270]{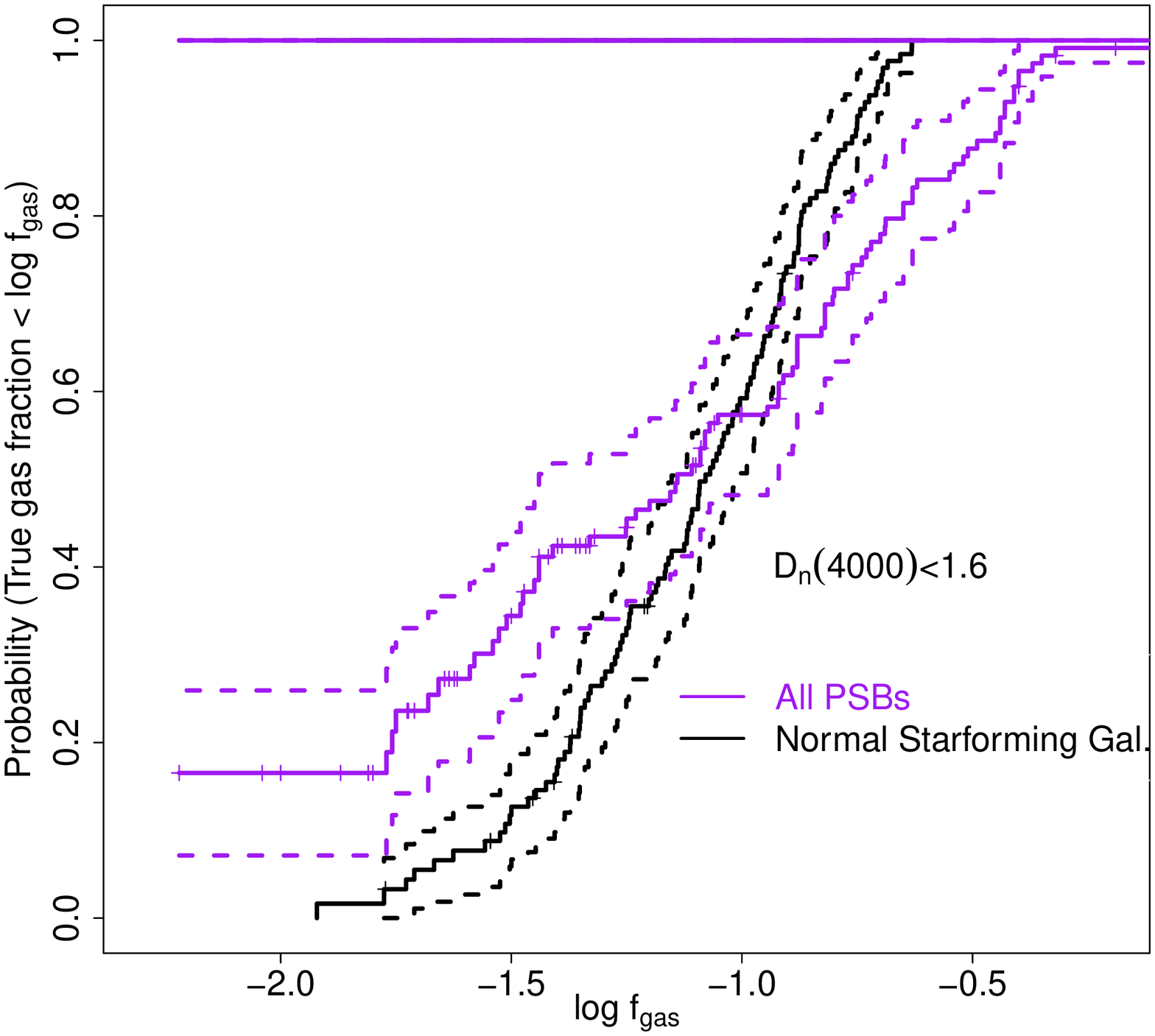}
\caption{Compares the empirical cumulative distribution function (ECDF) of molecular gas fraction, \lfgas, of PSBs and of star forming non-PSBs. The ECDF is estimated using the \citet{Kaplan+58} estimator and takes into account the upper limits, which are shown as +s. 95\% confidence curves are shown as dashed lines. The vertical axis denotes the estimated proportion of galaxies with gas fraction less than a given observed value. The distribution of gas fractions in PSBs is significantly broader (different) from that of non-PSBs. For example, the proportion of PSBs with gas fractions below $\sim 5\%$ is higher than that of starforming non-PSB galaxies. See text in section in \ref{sec:res} for a quantitative test that compare the two distributions.
\label{fig:km_psb}}
\end{figure}

\begin{figure}
\includegraphics[scale=0.4,angle=270]{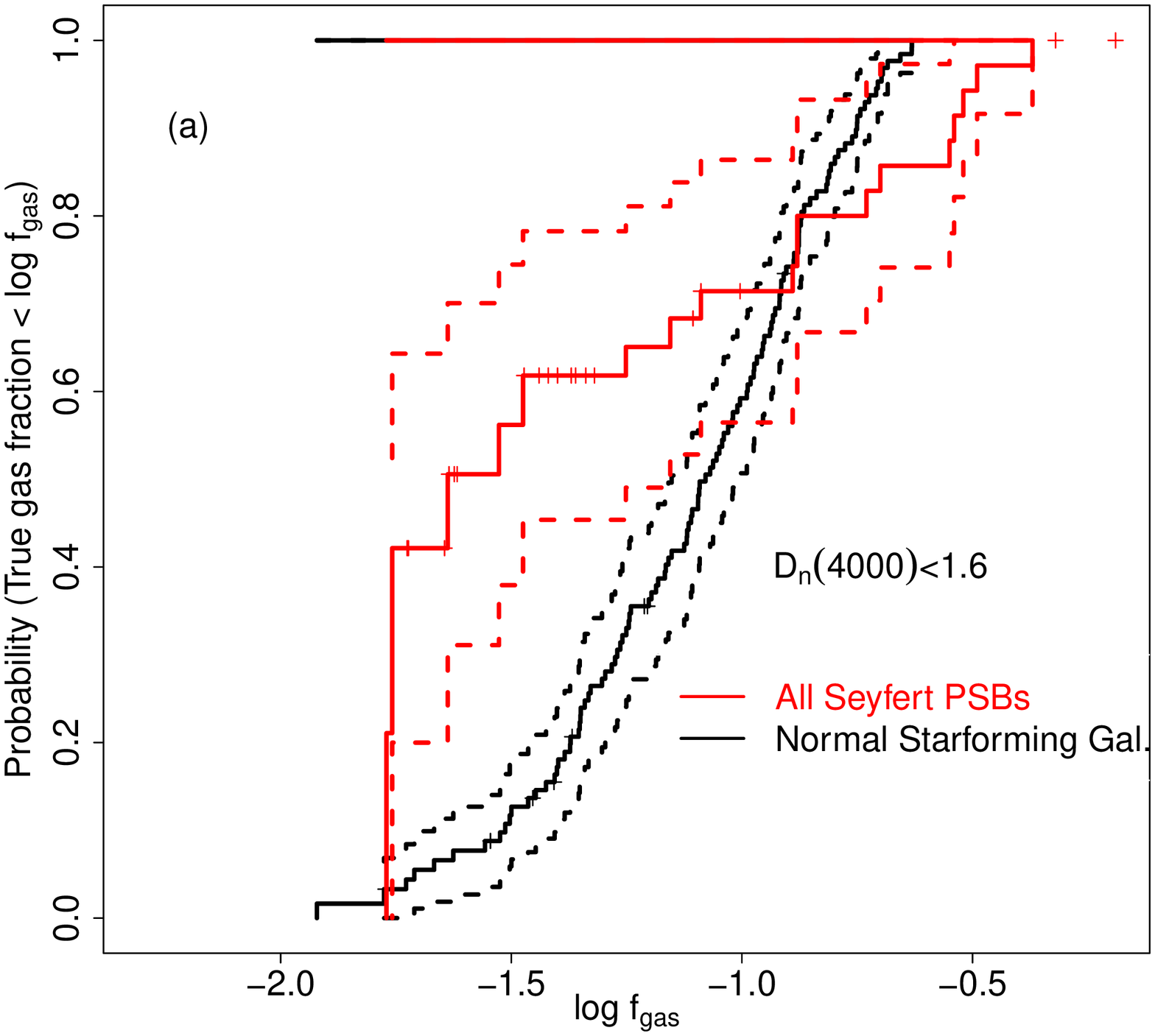}
\includegraphics[scale=0.4,angle=270]{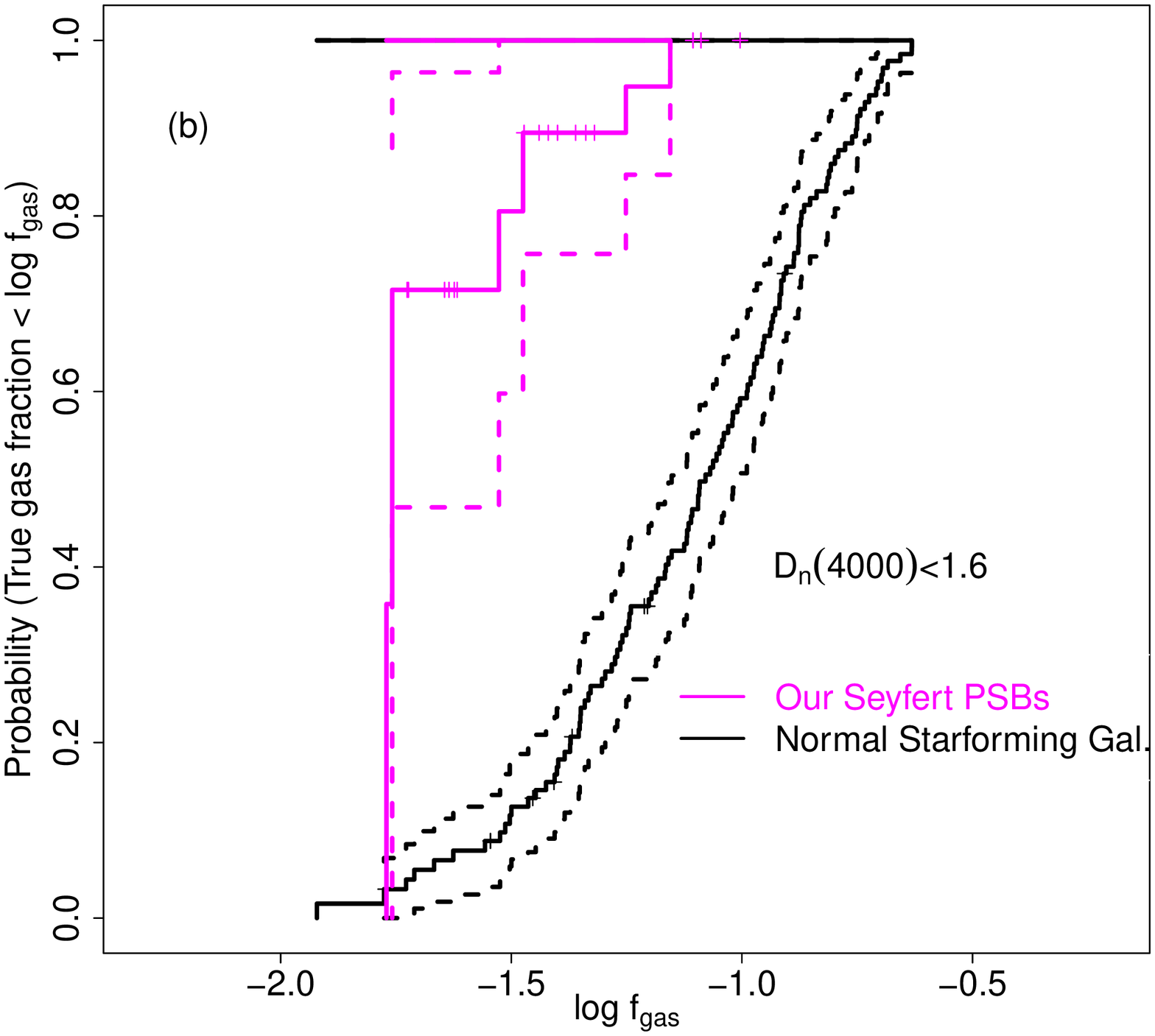}
\caption{ Panel (a) is similar to Figure~\ref{fig:km_psb} but compares the ECDF of molecular gas fraction, \lfgas, of all Seyfert PSBs and of star forming non-PSBs. The distribution gas fractions of Seyfert PSBs is significantly broader (different) than that of non-PSBs of similar $D_n(4000)$. A higher proportion of Seyfert PSBs have gas fraction fractions below $\sim 10\%$. Panel (b) only plots our new green valley Seyfert PSB sample.
\label{fig:km_sy}}
\end{figure}


Figure~\ref{fig:km_psb} compares the empirical cumulative distribution functions (ECDF) of \lfgas of normal star-forming galaxies ($D_n(4000) < 1.6$) with that of PSBs while Figure~\ref{fig:km_sy} compares the ECDF of the star-forming galaxies with Seyfert PSBs. The vertical probability axis in both plots tracks the estimated Kaplan-Meier percentiles of galaxies with gas fraction less than a given threshold \citep{Kaplan+58,Helsel+05}. The \lfgas curves for the PSBs and non-PSBs are similar at high gas fractions  but are different at lower gas fractions. The log-rank test indicates that ECDF of all PSBs or Seyfert only PSBs are significantly different from that of the young non-PSBs. When the ECDF for Seyfert PSBs is compared to that of the young non-PSBs, the test gives a $\chi^2$ of 24 on one degrees of freedom corresponding to a $p$-value of $10^{-6}$. Therefore, the null-hypothesis that the EDCFs for these two populations are the same can be rejected at a $4.8\sigma$ significance level. Similarly, the test gives a $\chi^2$ of 11.4 on one degrees of freedom, corresponding to a $p$-value of $7\times 10^{-4}$, when EDCFs of all PSBs and non-PSBs with $D_n(4000) < 1.6 $ are compared. 

Having established that there is a statistically significant difference between the gas fraction of young star-forming non-PSB galaxies and of PSBs in general, and of Seyfert PSBs in particular, in the rest of this section we explore in more detail where these trends come from. We first demonstrate that the WISE flux ratio between $12\mu$m and $4.6\mu$m is an excellent proxy for \lfgas using the COLD GASS sample of non-PSBs \citep{Saintonge+11} and then use it as a tool to constrain the gas fraction evolution in PSBs. Stellar populations younger than 0.6 Gyr dominate the $12\mu$m emission and, this ratio is known to correlate well with the specific star formation rate \citep[sSFR,][]{Donoso+12}. The AGN hot dust emission is mostly between $\sim 3-5\mu$m, some have suggested that this ratio is not appreciably affected by AGN emission and that it can trace the sSFR in AGN \citep{Donley+12,Donoso+12}. \citet{Yesuf+14} have shown that starbursts and post-starbursts form a time sequence in this ratio, with starbursts having the highest ratios and quiescent PSBs having ratios similar to those of green-valley and red-sequence galaxies.

\begin{figure}
\includegraphics[scale=0.4,angle=270]{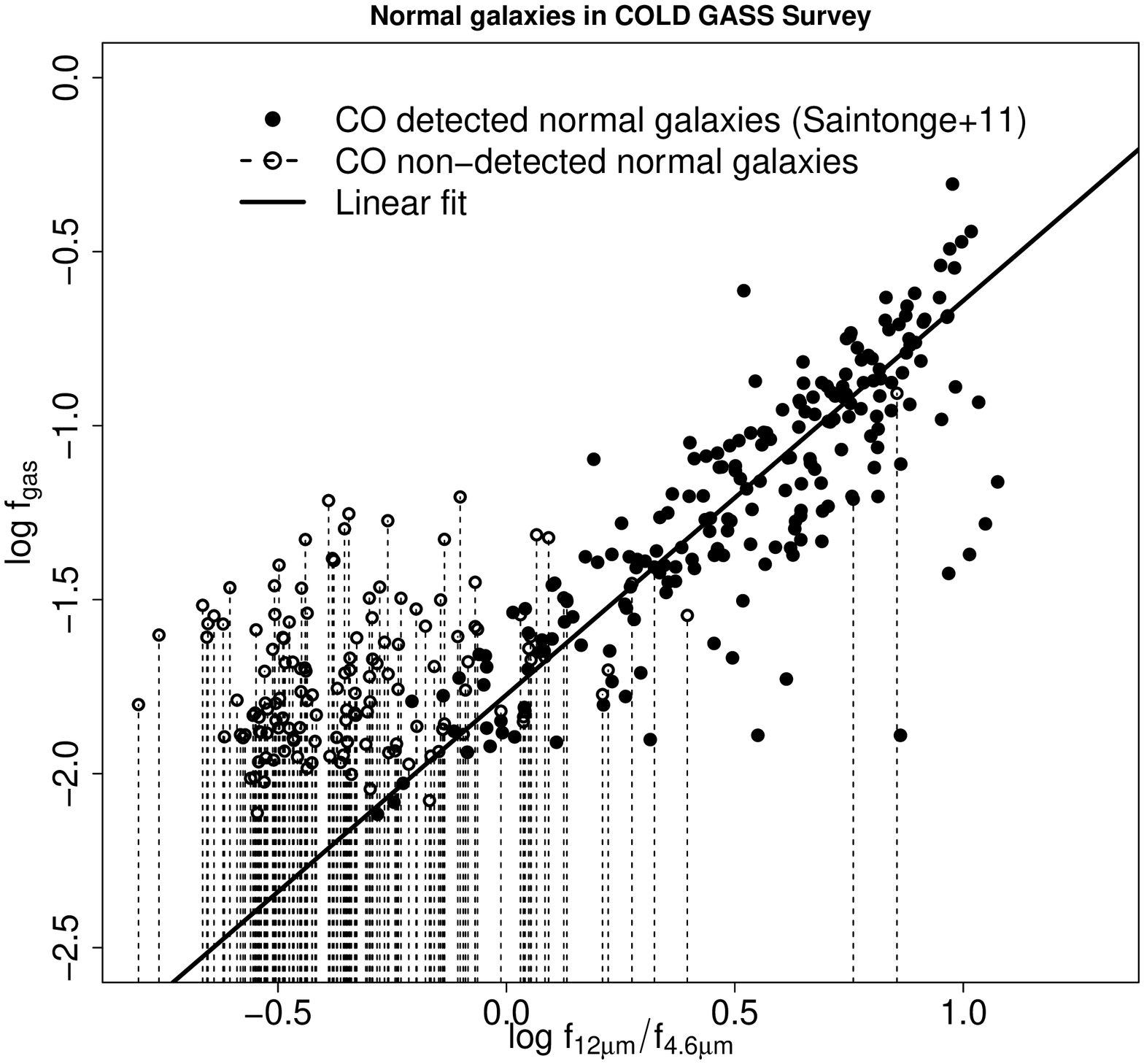}
\caption{WISE flux ratio \lfwise versus molecular gas fraction for all galaxies in COLD GASS survey \citep{Saintonge+11}. Young star-forming galaxies have high \lfwise and while old quiescent galaxies have low \lfwise. The filled circles denote galaxies with CO detections while the open circles and the dashed lines denote non-detected galaxies and their upper-limits. The black line is the linear fit to the data and it includes the upper-limits. The \lfwise is an excellent proxy for the molecular gas fraction. \label{fig:w23fg}}
\end{figure}

\begin{figure*}
\includegraphics[scale=0.7,angle=270]{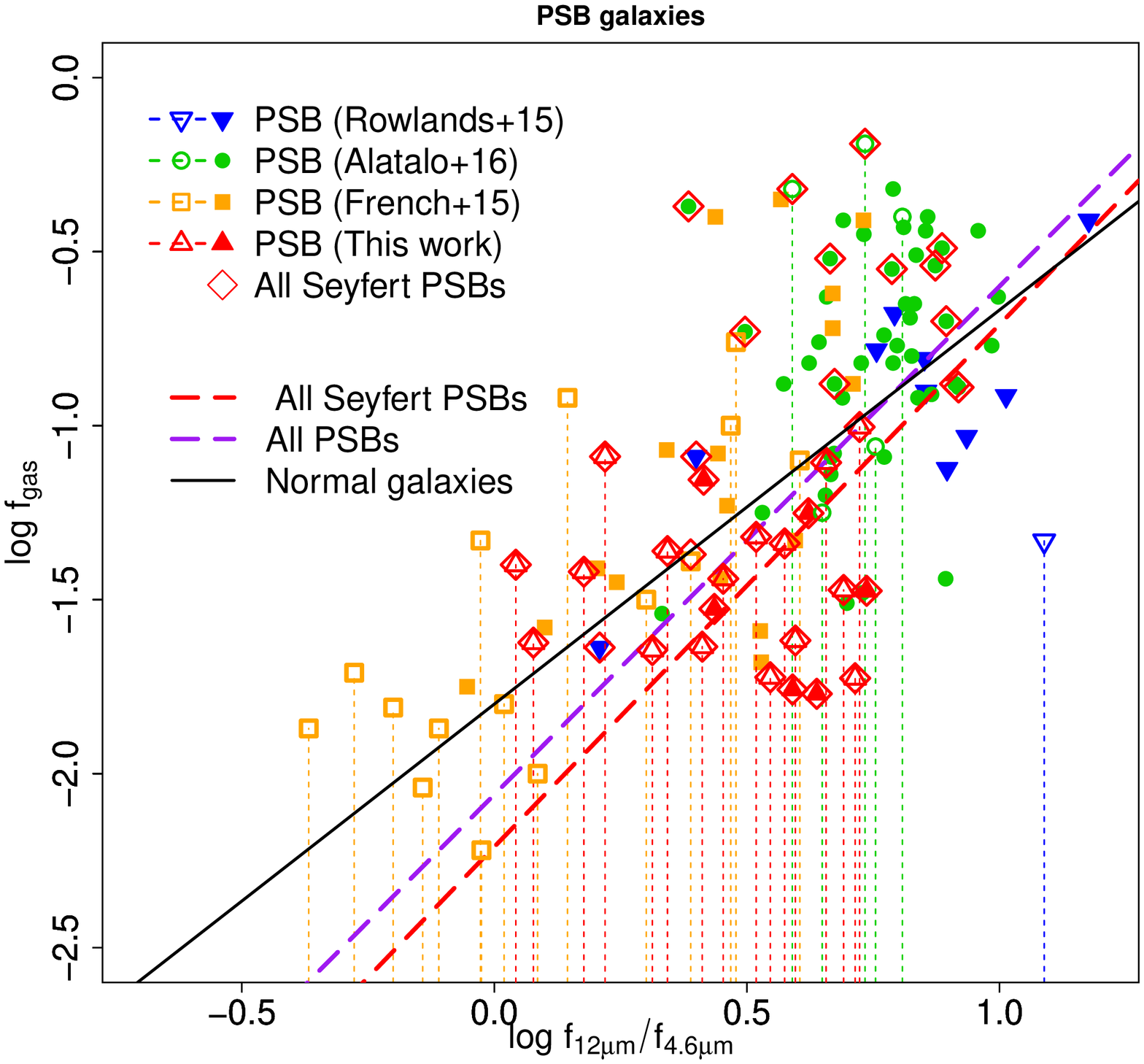}
\caption{WISE flux ratio \lfwise versus molecular gas fraction for PSBs \citep[in this work,][]{French+15,Rowlands+15,Alatalo+16}. The filled symbols denote PSBs with CO detections while the open symbols (except the purple diamonds) and the dashed lines denote non-detected PSBs and their upper-limits. The purple line is the linear fit to the Seyfert PSB data including the upper-limits while the cyan line is the fit to all PSB data. The black line is same as the best fit fine for non-PSBs in Figure~\ref{fig:w23fg}. The slope of the best fit line (the correlation) for PSBs is statistically significant (5$\sigma$). \label{fig:w23fg_psb}}
\end{figure*}

Figure~\ref{fig:w23fg} shows the WISE flux ratio \lfwise versus  \lfgas for all normal galaxies in the COLD GASS survey. As shown in the Appendix Figure~\ref{fig:w23_w12}, the general population of SDSS galaxies show bimodal \lfwise distribution that correlates with $D_n(4000)$. Star-forming galaxies have ratios between \lfwise $\sim$ 0.2 -- 1 while quiescent galaxies have ratios between \lfwise $\sim$ -0.8 -- 0.0. A tight correlation is observed between \lfwise and \lfgas in the COLD GASS sample, indicating an evolutionary process in normal galaxies in which sSFR declines as gas is used up. The slope of the best fit line is $\beta=1.13$ and its intercept is $\beta_0 = -1.8$. The Kendall's $\tau$ correlation coefficient for the fit is 0.64. The statistical significance of the correlation is more than 5$\sigma$. One can easily show that $\beta$ and $\beta_0$ are related to the logarithm of the molecular gas depletion time at a given sSFR by the relation: $\log \tau_{\rm dep}= \hat{\beta}\log {\rm sSFR}+ \hat{\beta_0}$, where  $\hat{\beta}=(\beta/\alpha-1)$,  $\hat{\beta_0}= -\beta(\alpha_0/\alpha+0.769)+\beta_0$, where $\alpha$ and $\alpha_0$ define the relation: $\log {\rm sSFR} = \alpha$(\lfwise$+0.769)+\alpha_0$. \citet{Donoso+12} computed the relation between sSFR and 4.6 - 12$\mu$m color in Vega magnitude and 0.769 is subtracted to change the color to AB magnitude. They found $\alpha = 2.5\times 0.775=1.938$ and $\alpha_0 =-12.56$ for the SDSS star-forming galaxies. Using these values with our fitted values of  $\beta$ and $\beta_0$ gives $\hat{\beta} = -0.42$ and $\hat{\beta_0} = 4.7$. In comparison, \citet{Huang+14} found $\hat{\beta}=-0.37 \pm 0.04$ and $\hat{\beta_0}=5.45 \pm 0.42$ for the COLD GASS sample. The authors used the GALEX FUV and WISE 22$\mu$m to estimate the sSFR. Our inferred depletion times are consistent with \citet{Huang+14} but there is likely a systematic offset in the two SFR estimates. 

Figure~\ref{fig:w23fg_psb} shows the relationship between the WISE flux ratio \lfwise and \lfgas for PSBs. They also show statistically significant ($5\sigma$) correlation between these two quantities. The slope of the best fit line is $\beta=1.46$ and its intercept is -2.1. Collectively,  the broad overall trend, suggests same correlation is at work: lower molecular gas fractions means lower sSFR or dust-heating. But the scatter is much larger, and looking in detail we can see that certain sub-populations of PSBs statistically differ. A major insight from this figure is that the sSFR histories of the PSBs and of Seyferts PSBs may not be as simple or well-behaved as normal galaxies. Hence, the use of standard age indicators may lead to different results depending on indicator used. This topic will be explored in future papers. The reason for the low  \lfgas of Seyferts in Figure~\ref{fig:km_sy} is consistent with low sSFR upper limits inferred from \lfwise.

Table~\ref{tbl:fgas_dist} summarizes the distribution of molecular gas fraction of our sample and of the previous samples of PSBs. Our sample has a \lfgas distribution with mean of 0.025 and dispersion 0.018. 95\% of our PSBs have gas fractions below 0.07 while 50\% have gas fractions below 0.02. The previous samples have means and dispersions of $\sim 0.1- 0.2$. It is particularly interesting that the \citet{French+15} sample, which may also contains late stage PSBs, has \lfgas distribution with mean of 0.08 and dispersion 0.12. The majority of PSBs from \citet{Rowlands+15} and \citet{Alatalo+16} samples have both high \lfwise and high molecular gas fractions in comparison to our sample. The PSBs in these two works may be precursors to the PSBs in our sample or to those in the \citet{French+15} sample. 

\section{Discussion}\label{sec:discus}

In this section, we discuss in detail how gas fractions in our sample compare with those observed in previous samples of PSBs. We will also discuss the general implication of our result and the uncertainty in the hitherto assumed CO conversion factor.

\subsection{Comparison of our sample with the previous samples of PSBs}

We study the molecular gas evolution of 116 post-starburst galaxies, of which 22 have new observations undertaken by us while the rest are compiled from the literature \citep{Rowlands+15,French+15,Alatalo+16}. The galaxies in the new sample are selected using their spectral indices and NUV-g colors, to identify them as green valley PSBs, and their emission line ratios, to identify them as Seyferts. Our sample represents about two-thirds of the Seyferts PSBs in the combined sample. The Seyfert PSBs from the previous works were serendipitously selected and are mostly in the blue-cloud. We show that \lfgas distribution of the Seyfert PSBs and of PSBs in general are significantly different from that of normal star-forming galaxies. The molecular gas fraction in the Seyfert PSBs decreases with the WISE ratios \lfwise. If there is an evolutionary sequence in Figure~\ref{fig:w23fg_psb}, from upper right to lower left, it would be plausible to conclude that our new Seyferts are fairly aged galaxies located
nearing the middle of the gas-exhaustion cycle.  This conclusion would be consistent with our selection of these galaxies based on integrated UV color as members of the green valley and well on their way to gas exhaustion. The systematically lower gas content of the new Seyferts relative to the PSBs in Figure~\ref{fig:w23fg_psb} might then be interpreted as the action of late AGNs in them in finally clearing out the gas. 

However, we call attention to an important puzzle: In Figure~\ref{fig:w23fg_psb} the \citet{French+15} sample as a whole shows a significant correlation between \lfwise and \lfgas; the PSBs with high \lfwise have high gas fractions. This is 
puzzling because the \citet{French+15} galaxies were selected to have very low H$\alpha$, which is the gold-standard for star-formation rate. Thus, the \lfwise and H$\alpha$ as SFR indicators disagree for the subset of the \citet{French+15} sample with high \lfwise. However, A-star heating is significant during the post-starburst phase, which can artificially boost IR-based SFR indicators \citep{Utomo+14, Hayward+14b}. An analysis of multiple SFR indicators, including extinction-free 1.4 GHz measurements in \S 2.5 of \citet{French+15} shows these galaxies are unlikely to have significant dust obscuration. Analyses of the near and far IR properties of this sample are the subject of a forthcoming study (Smercina et al. in prep).

\begin{table*}
\caption{Summary statistics of the distribution of gas fraction in PSBs for the new and old CO data.} \label{tbl:fgas_dist}
\begin{tabular}{lccccccc}
\hline
\hline
 Study & Mean & Std deviation &  5\% & 25\% & 50\%&75\% & 95\%\\
\hline
This work (Seyfert) & 0.025 & 0.018 &  - & 0.017 & 0.017 &0.030 & 0.070 \\
French el al (2015) & 0.081 & 0.122 &  - & - & 0.026 & 0.083 & 0.398 \\
Alatalo et al. (2016) & 0.216 & 0.214 & 0.031 & 0.083 &  0.170 & 0.288 & 0.426 \\
Rowlands et al. (2016) & 0.133 & 0.103 & 0.023 & 0.075&  0.122 & 0.165 & -  \\ 
Seyferts in all works & 0.088 & 0.116 & 0.017 &  0.017& 0.023 & 0.132 & 0.324  \\
PSBs in all works & 0.133 & 0.176 & - & 0.021 &  0.075 & 0.182 & 0.398 \\
\hline
\end{tabular}
\begin{tablenotes}
	 
      \small
      \item  The summary statistics in the table are estimated using the Kaplan-Meier method and therefore take into account the CO non-detections. The percentiles show the probability that the gas fraction of a sample is below the given value.
     Our late stage Seyfert sample has a distribution of gas fraction with the lowest mean and standard deviation compared to other samples \citep{French+15,Rowlands+15,Alatalo+16}. The \citet{French+15} sample, which may also be late stage PSBs, has a wider range of gas fraction distribution.
\end{tablenotes}
\end{table*}

The episodic lifetime of an individual AGN event is currently not well-constrained and it may range between $\sim 0.3-1$ times the effective AGN lifetime \citep[][]{Hopkins+09}. The effective AGN lifetime for the Seyfert PSBs is likely to be $\sim 2$ Gyr. This estimate uses the median $\sigma=110$\,km\,s$^{-1}$ to estimate the median black hole mass, $10^{7.4}$M$_\odot$, from the M$_{\rm BH}-\sigma$ relation \citep{Kormendy+13}, the median dust-corrected \ion{O}{iii} luminosity, $10^{7.3}$L$_\odot$, to estimate the Eddington ratio, $\lambda=0.003$, \citep{Lamastra+09}, and Figure 9 of \citet{Hopkins+09} to estimate the effective AGN lifetime from M$_{\rm BH}$ \& $\lambda$. The effective AGN lifetime for the Seyfert PSBs is longer than the quenched PSB lifetime of $\sim 1$ Gyr, which suggests that, given the above estimate of the episodic lifetime, the quenched PSBs may experience second episode of AGN activity or no AGN activity at all within their typical lifetime. Perhaps the variation in AGN activity timescale is linked to wide range in gas fraction observed in the \citet{French+15} sample, if these PSBs are truly quiescent galaxies, and AGN are linked to the removal of gas in the \citet{French+15} galaxies with low \lfwise and in our sample.

We used the \lfwise ratio as a crude first-order indicator of star formation activity in both AGN and non-AGN galaxies. Some variation is expected between the ratios of the two classes \citep{Donoso+12}. It is not clear if these variations are due to additional AGN dust heating or preferentially aged stellar populations in AGN host galaxies. Understanding the details of this variation is not important for conclusions of the current work, but future studies of the effect of AGN contamination on \lfwise will be useful.

Furthermore, it should be noted that the gas fractions in very nearby early type galaxies are found to be $\lesssim 0.1\%$ \citep[e.g.,][]{Young+11}. Many of the PSBs considered here have molecular gas fractions already consistent with detections in these early type galaxies. A full comparison with early-type galaxies is complicated by the lower redshifts and smaller physical apertures of the \citet[][ATLAS 3D]{Young+11} sample. Deeper observations of the non-detected PSBs will be necessary to study the full range in their molecular gas properties, and to compare to the deep observations of the ATLAS 3D sample of early-type galaxies.

\subsection{Implication for galaxy evolution}

It is thought that star formation quenching happens in both slow and fast modes \citep[e.g.,][]{Faber+07,Cheung+12,Barro+13,Fang+12,Dekel+14,Schawinski+14,Yesuf+14,Woo+15}. The existence of these modes, their relative importance and the physical mechanisms that drive them are still open questions in galaxy evolution. Some works have indicated that PSBs quench rapidly and may exemplify an important mode quenching for the build-up of the red sequence \citep{Kaviraj+07,Wild+09,Wong+12,Yesuf+14,Wild+16}, while others have questioned their importance \citep{DeLucia+09,Dressler+13}. 

Recent studies found large gas fractions in PSBs complicating the picture of their rapid evolution to the red sequence \citep{Rowlands+15,French+15,Alatalo+16}. However, these works have been biased against finding PSBs with Seyfert-like emission lines in the UV-optical green valley. In this work, we have observed the molecular gas properties of 22 such galaxies. The combined data presented in this work cannot rule out AGN feedback, and as a whole are consistent with a removal or destruction of molecular gas at later stages by AGN feedback or no AGN feedback at all if different subsamples of PSBs are not evolutionary connected. Better and deeper observations, especially at a later stage of PSB evolution, are needed to firmly and more directly test the merger-induced AGN feedback hypothesis. In the theoretical front, developments in cosmological simulations with AGN feedback are needed to be compared to these future observations. 

\subsection{The effect of CO conversion factor \label{sec:alphaco}}

The CO luminosity to the molecular hydrogen conversion factor is a widely acknowledged source of uncertainty in studies that use CO as a tracer of molecular hydrogen \citep{Bolatto+13}. Up to this point, we have assumed that all objects including PSBs have a conversion factor similar to the factor observed in the Milky Way disk. There is no study to date that measured the conversion factor in PSBs.  As discussed in detail in \citet{Bolatto+13}, departure from the Galactic conversion factor are both observed and expected in starburst galaxies. It has been shown that adopting the Galactic conversion factor would cause the inferred molecular gas mass to exceed the dynamical mass for the central region of a starburst galaxy \citep{Solomon+97,Downes+98}. Using galaxy merger hydrodynamic simulations, which incorporate dust and molecular line radiative transfer calculations, \citet{Narayanan+11} have shown that in merger-induced starbursts, the combined effect of increased velocity dispersion and kinetic temperature increases the velocity-integrated CO intensity, and lowers the CO conversion factor from the Galactic value by a factor of $\sim 2-10$. The authors also note that in the PSB phase, it is less trivial to simply relate the conversion factor to the gas velocity dispersion and temperature owing to varying physical conditions in the PSB galaxy. Some of their merger simulations returned to a Galactic conversion factor value quickly after the peak of the starburst, while others remained low. Furthermore, if AGN heat the bulk of molecular gas, the conversion factor could be lower from the Galactic value in PSBs with AGN. Note that a lower CO conversion factor than the assumed Galactic value will lower the gas fractions observed in PSBs compared normal galaxies.

\section{Summary \& Conclusions} \label{sec:sum}

Using the SMT, we undertook new CO\,(2--1) observations of 22 green-valley Seyfert post-starburst candidate galaxies at redshift $z =$ 0.02 -- 0.06. The sample was selected using the dust-corrected NUV-g color, H$\delta$ absorption and 4000{\AA} break to indicate a PSB signature, and the emission line ratios to indicate Seyfert activity. We analyzed our sample with previous samples of 94 PSBs. The combined sample probes a variety of stages in a possible evolutionary sequence, and spans a range of AGN properties. Our main results are:

\begin{itemize}

\item We detect molecular gas in only 6 out of 22 Seyfert PSB galaxies we targeted. Using comparable sensitivity limits, this is consistent with the detection in PSB sample by \citet{French+15} using the SMT. However, taking into account our upper limits, the mean and the dispersion of the distribution of the gas fraction in our green-valley Seyfert PSBs ($\mu = 0.025, \sigma= 0.018$) are much smaller than previous samples of Seyfert PSBs or PSBs in general \citep[$\mu \sim 0.1 - 0.2, \sigma \sim 0.1 - 0.2$,][]{French+15,Rowlands+15,Alatalo+16}. This difference may partly be explained by the evolution of molecular gas fraction with post-starburst age if our sample probes the late stage of the evolution. However, PSBs in \citet{French+15} are also have similar green-valley colors but they show a wider range of gas fraction, despite their low star-formation rate inferred from their H$\alpha$ emission. The comparison between these samples is complicated by the differing selections on H$\delta$ absorption, as weaker H$\delta$ is consistent with either a weaker starburst, or a later post-starburst age.
 
 \item The distribution of gas fraction in PSBs is significantly different from young star-forming galaxies from COLD GASS survey \citep{Saintonge+11}. PSBs are more likely to have lower gas fraction. The Seyfert PSBs have a distribution of gas fraction which is even more significantly different from that of normal star-forming galaxies of similar $D_n(4000)$. 

\item The WISE flux ratio, \lfwise, is an excellent proxy for gas fraction for both PSBs and non-PSBs. We find a statistically significant ($5\sigma$) relationship between \lfwise and \lfgas for both PSBs and normal star-forming galaxies. 
\end{itemize}

\section*{Acknowledgements}

We thank Xavier Prochaska and Ann Zabludoff for their useful comments and suggestions. H. Yesuf would like to acknowledge support by NSF grants AST-0808133 and and AST-1615730. KDF acknowledges support from NSF grant DGE-1143953, PEO, and the ARCS Phoenix Chapter and Burton Family.

The SMT is operated by the Arizona Radio Observatory (ARO), Steward Observatory, University of Arizona. We thank the operators and staff of the ARO. 

Funding for the SDSS and SDSS-II has been provided by the Alfred P. Sloan Foundation, the Participating Institutions, the National Science Foundation, the U.S. Department of Energy, the National Aeronautics and Space Administration, the Japanese Monbukagakusho, the Max Planck Society, and the Higher Education Funding Council for England. The SDSS Web site is http://www.sdss.org/.

The SDSS is managed by the Astrophysical Research Consortium for the Participating Institutions. The Participating Institutions are the American Museum of Natural History, Astrophysical Institute Potsdam, University of Basel, University of Cambridge, Case Western Reserve University, University of Chicago, Drexel University, Fermilab, the Institute for Advanced Study, the Japan Participation Group, Johns Hopkins University, the Joint Institute for Nuclear Astrophysics, the Kavli Institute for Particle Astrophysics and Cosmology, the Korean Scientist Group, the Chinese Academy of Sciences (LAMOST), Los Alamos National Laboratory, the Max-Planck-Institute for Astronomy (MPIA), the Max-Planck-Institute for Astrophysics (MPA), New Mexico State University, Ohio State University, University of Pittsburgh, University of Portsmouth, Princeton University, the United States Naval Observatory, and the University of Washington.

\bibliographystyle{mnras}
\bibliography{co_reference}

\begin{thebibliography}{}
\makeatletter
\relax
\def\mn@urlcharsother{\let\do\@makeother \do\$\do\&\do\#\do\^\do\_\do\%\do\~}
\def\mn@doi{\begingroup\mn@urlcharsother \@ifnextchar [ {\mn@doi@}
  {\mn@doi@[]}}
\def\mn@doi@[#1]#2{\def\@tempa{#1}\ifx\@tempa\@empty \href
  {http://dx.doi.org/#2} {doi:#2}\else \href {http://dx.doi.org/#2} {#1}\fi
  \endgroup}
\def\mn@eprint#1#2{\mn@eprint@#1:#2::\@nil}
\def\mn@eprint@arXiv#1{\href {http://arxiv.org/abs/#1} {{\tt arXiv:#1}}}
\def\mn@eprint@dblp#1{\href {http://dblp.uni-trier.de/rec/bibtex/#1.xml}
  {dblp:#1}}
\def\mn@eprint@#1:#2:#3:#4\@nil{\def\@tempa {#1}\def\@tempb {#2}\def\@tempc
  {#3}\ifx \@tempc \@empty \let \@tempc \@tempb \let \@tempb \@tempa \fi \ifx
  \@tempb \@empty \def\@tempb {arXiv}\fi \@ifundefined
  {mn@eprint@\@tempb}{\@tempb:\@tempc}{\expandafter \expandafter \csname
  mn@eprint@\@tempb\endcsname \expandafter{\@tempc}}}

\bibitem[\protect\citeauthoryear{{Aihara} et~al.,}{{Aihara}
  et~al.}{2011}]{Aihara+11}
{Aihara} H.,  et~al., 2011, \mn@doi [\apjs] {10.1088/0067-0049/193/2/29}, \href
  {http://adsabs.harvard.edu/abs/2011ApJS..193...29A} {193, 29}

\bibitem[\protect\citeauthoryear{Akritas, Murphy  \& LaValley}{Akritas
  et~al.}{1995}]{Akritas+95}
Akritas M.~G.,  Murphy S.~A.,   LaValley M.~P.,  1995, Journal of the American
  Statistical Association, 90, 170

\bibitem[\protect\citeauthoryear{{Alatalo} et~al.,}{{Alatalo}
  et~al.}{2011}]{Alatalo+11}
{Alatalo} K.,  et~al., 2011, \mn@doi [\apj] {10.1088/0004-637X/735/2/88}, \href
  {http://adsabs.harvard.edu/abs/2011ApJ...735...88A} {735, 88}

\bibitem[\protect\citeauthoryear{{Alatalo}, {Cales}, {Appleton}, {Kewley},
  {Lacy}, {Lisenfeld}, {Nyland}  \& {Rich}}{{Alatalo}
  et~al.}{2014}]{Alatalo+14}
{Alatalo} K.,  {Cales} S.~L.,  {Appleton} P.~N.,  {Kewley} L.~J.,  {Lacy} M.,
  {Lisenfeld} U.,  {Nyland} K.,   {Rich} J.~A.,  2014, \mn@doi [\apjl]
  {10.1088/2041-8205/794/1/L13}, \href
  {http://adsabs.harvard.edu/abs/2014ApJ...794L..13A} {794, L13}

\bibitem[\protect\citeauthoryear{{Alatalo} et~al.,}{{Alatalo}
  et~al.}{2016}]{Alatalo+16}
{Alatalo} K.,  et~al., 2016, \mn@doi [\apj] {10.3847/0004-637X/827/2/106},
  \href {http://adsabs.harvard.edu/abs/2016ApJ...827..106A} {827, 106}

\bibitem[\protect\citeauthoryear{{Assef} et~al.,}{{Assef}
  et~al.}{2013}]{Assef+13}
{Assef} R.~J.,  et~al., 2013, \mn@doi [\apj] {10.1088/0004-637X/772/1/26},
  \href {http://adsabs.harvard.edu/abs/2013ApJ...772...26A} {772, 26}

\bibitem[\protect\citeauthoryear{{Baldwin}, {Phillips}  \&
  {Terlevich}}{{Baldwin} et~al.}{1981}]{Baldwin+81}
{Baldwin} J.~A.,  {Phillips} M.~M.,   {Terlevich} R.,  1981, \mn@doi [\pasp]
  {10.1086/130766}, \href {http://adsabs.harvard.edu/abs/1981PASP...93....5B}
  {93, 5}

\bibitem[\protect\citeauthoryear{{Balogh}, {Morris}, {Yee}, {Carlberg}  \&
  {Ellingson}}{{Balogh} et~al.}{1999}]{Balogh+99}
{Balogh} M.~L.,  {Morris} S.~L.,  {Yee} H.~K.~C.,  {Carlberg} R.~G.,
  {Ellingson} E.,  1999, \mn@doi [\apj] {10.1086/308056}, \href
  {http://adsabs.harvard.edu/abs/1999ApJ...527...54B} {527, 54}

\bibitem[\protect\citeauthoryear{{Barnes} \& {Hernquist}}{{Barnes} \&
  {Hernquist}}{1991}]{Barnes+91}
{Barnes} J.~E.,  {Hernquist} L.~E.,  1991, \mn@doi [\apjl] {10.1086/185978},
  \href {http://adsabs.harvard.edu/abs/1991ApJ...370L..65B} {370, L65}

\bibitem[\protect\citeauthoryear{{Barro} et~al.,}{{Barro}
  et~al.}{2013}]{Barro+13}
{Barro} G.,  et~al., 2013, \mn@doi [\apj] {10.1088/0004-637X/765/2/104}, \href
  {http://adsabs.harvard.edu/abs/2013ApJ...765..104B} {765, 104}

\bibitem[\protect\citeauthoryear{{Bolatto}, {Wolfire}  \& {Leroy}}{{Bolatto}
  et~al.}{2013}]{Bolatto+13}
{Bolatto} A.~D.,  {Wolfire} M.,   {Leroy} A.~K.,  2013, \mn@doi [\araa]
  {10.1146/annurev-astro-082812-140944}, \href
  {http://adsabs.harvard.edu/abs/2013ARA%26A..51..207B} {51, 207}

\bibitem[\protect\citeauthoryear{{Boselli}, {Cortese}, {Boquien}, {Boissier},
  {Catinella}, {Lagos}  \& {Saintonge}}{{Boselli} et~al.}{2014}]{Boselli+14}
{Boselli} A.,  {Cortese} L.,  {Boquien} M.,  {Boissier} S.,  {Catinella} B.,
  {Lagos} C.,   {Saintonge} A.,  2014, \mn@doi [\aap]
  {10.1051/0004-6361/201322312}, \href
  {http://adsabs.harvard.edu/abs/2014A%26A...564A..66B} {564, A66}

\bibitem[\protect\citeauthoryear{{Bruzual} \& {Charlot}}{{Bruzual} \&
  {Charlot}}{2003}]{BC03}
{Bruzual} G.,  {Charlot} S.,  2003, \mn@doi [\mnras]
  {10.1046/j.1365-8711.2003.06897.x}, \href
  {http://adsabs.harvard.edu/abs/2003MNRAS.344.1000B} {344, 1000}

\bibitem[\protect\citeauthoryear{{Bruzual A.}}{{Bruzual A.}}{1983}]{Bruzual83}
{Bruzual A.} G.,  1983, \mn@doi [\apj] {10.1086/161352}, \href
  {http://adsabs.harvard.edu/abs/1983ApJ...273..105B} {273, 105}

\bibitem[\protect\citeauthoryear{{Cen}}{{Cen}}{2012}]{Cen12}
{Cen} R.,  2012, \mn@doi [\apj] {10.1088/0004-637X/755/1/28}, \href
  {http://adsabs.harvard.edu/abs/2012ApJ...755...28C} {755, 28}

\bibitem[\protect\citeauthoryear{{Charlot} \& {Fall}}{{Charlot} \&
  {Fall}}{2000}]{Charlot+00}
{Charlot} S.,  {Fall} S.~M.,  2000, \mn@doi [\apj] {10.1086/309250}, \href
  {http://adsabs.harvard.edu/abs/2000ApJ...539..718C} {539, 718}

\bibitem[\protect\citeauthoryear{{Cheung} et~al.,}{{Cheung}
  et~al.}{2012}]{Cheung+12}
{Cheung} E.,  et~al., 2012, \mn@doi [\apj] {10.1088/0004-637X/760/2/131}, \href
  {http://adsabs.harvard.edu/abs/2012ApJ...760..131C} {760, 131}

\bibitem[\protect\citeauthoryear{{Cicone} et~al.,}{{Cicone}
  et~al.}{2014}]{Cicone+14}
{Cicone} C.,  et~al., 2014, \mn@doi [\aap] {10.1051/0004-6361/201322464}, \href
  {http://adsabs.harvard.edu/abs/2014A%26A...562A..21C} {562, A21}

\bibitem[\protect\citeauthoryear{{Couch} \& {Sharples}}{{Couch} \&
  {Sharples}}{1987}]{Couch+87}
{Couch} W.~J.,  {Sharples} R.~M.,  1987, \mn@doi [\mnras]
  {10.1093/mnras/229.3.423}, \href
  {http://adsabs.harvard.edu/abs/1987MNRAS.229..423C} {229, 423}

\bibitem[\protect\citeauthoryear{{Cox}}{{Cox}}{1972}]{Cox72}
{Cox} D.~R.,  1972, JRSS, 34, 187

\bibitem[\protect\citeauthoryear{{Davis} et~al.,}{{Davis}
  et~al.}{2014}]{Davis+14}
{Davis} T.~A.,  et~al., 2014, \mn@doi [\mnras] {10.1093/mnras/stu570}, \href
  {http://adsabs.harvard.edu/abs/2014MNRAS.444.3427D} {444, 3427}

\bibitem[\protect\citeauthoryear{{De Lucia}, {Poggianti}, {Halliday},
  {Milvang-Jensen}, {Noll}, {Smail}  \& {Zaritsky}}{{De Lucia}
  et~al.}{2009}]{DeLucia+09}
{De Lucia} G.,  {Poggianti} B.~M.,  {Halliday} C.,  {Milvang-Jensen} B.,
  {Noll} S.,  {Smail} I.,   {Zaritsky} D.,  2009, \mn@doi [\mnras]
  {10.1111/j.1365-2966.2009.15435.x}, \href
  {http://adsabs.harvard.edu/abs/2009MNRAS.400...68D} {400, 68}

\bibitem[\protect\citeauthoryear{{Dekel} \& {Burkert}}{{Dekel} \&
  {Burkert}}{2014}]{Dekel+14}
{Dekel} A.,  {Burkert} A.,  2014, \mn@doi [\mnras] {10.1093/mnras/stt2331},
  \href {http://adsabs.harvard.edu/abs/2014MNRAS.438.1870D} {438, 1870}

\bibitem[\protect\citeauthoryear{{Di Matteo}, {Springel}  \& {Hernquist}}{{Di
  Matteo} et~al.}{2005}]{DiMatteo+05}
{Di Matteo} T.,  {Springel} V.,   {Hernquist} L.,  2005, \mn@doi [\nat]
  {10.1038/nature03335}, \href
  {http://adsabs.harvard.edu/abs/2005Natur.433..604D} {433, 604}

\bibitem[\protect\citeauthoryear{{Donley} et~al.,}{{Donley}
  et~al.}{2012}]{Donley+12}
{Donley} J.~L.,  et~al., 2012, \mn@doi [\apj] {10.1088/0004-637X/748/2/142},
  \href {http://adsabs.harvard.edu/abs/2012ApJ...748..142D} {748, 142}

\bibitem[\protect\citeauthoryear{{Donoso} et~al.,}{{Donoso}
  et~al.}{2012}]{Donoso+12}
{Donoso} E.,  et~al., 2012, \mn@doi [\apj] {10.1088/0004-637X/748/2/80}, \href
  {http://adsabs.harvard.edu/abs/2012ApJ...748...80D} {748, 80}

\bibitem[\protect\citeauthoryear{{Downes} \& {Solomon}}{{Downes} \&
  {Solomon}}{1998}]{Downes+98}
{Downes} D.,  {Solomon} P.~M.,  1998, \mn@doi [\apj] {10.1086/306339}, \href
  {http://adsabs.harvard.edu/abs/1998ApJ...507..615D} {507, 615}

\bibitem[\protect\citeauthoryear{{Dressler} \& {Gunn}}{{Dressler} \&
  {Gunn}}{1983}]{Dressler+83}
{Dressler} A.,  {Gunn} J.~E.,  1983, \mn@doi [\apj] {10.1086/161093}, \href
  {http://adsabs.harvard.edu/abs/1983ApJ...270....7D} {270, 7}

\bibitem[\protect\citeauthoryear{{Dressler}, {Oemler}, {Poggianti}, {Gladders},
  {Abramson}  \& {Vulcani}}{{Dressler} et~al.}{2013}]{Dressler+13}
{Dressler} A.,  {Oemler} Jr. A.,  {Poggianti} B.~M.,  {Gladders} M.~D.,
  {Abramson} L.,   {Vulcani} B.,  2013, \mn@doi [\apj]
  {10.1088/0004-637X/770/1/62}, \href
  {http://adsabs.harvard.edu/abs/2013ApJ...770...62D} {770, 62}

\bibitem[\protect\citeauthoryear{{Fabello}, {Kauffmann}, {Catinella},
  {Giovanelli}, {Haynes}, {Heckman}  \& {Schiminovich}}{{Fabello}
  et~al.}{2011}]{Fabello+11}
{Fabello} S.,  {Kauffmann} G.,  {Catinella} B.,  {Giovanelli} R.,  {Haynes}
  M.~P.,  {Heckman} T.~M.,   {Schiminovich} D.,  2011, \mn@doi [\mnras]
  {10.1111/j.1365-2966.2011.18825.x}, \href
  {http://adsabs.harvard.edu/abs/2011MNRAS.416.1739F} {416, 1739}

\bibitem[\protect\citeauthoryear{{Faber} et~al.,}{{Faber}
  et~al.}{2007}]{Faber+07}
{Faber} S.~M.,  et~al., 2007, \mn@doi [\apj] {10.1086/519294}, \href
  {http://adsabs.harvard.edu/abs/2007ApJ...665..265F} {665, 265}

\bibitem[\protect\citeauthoryear{{Fang}, {Faber}, {Salim}, {Graves}  \&
  {Rich}}{{Fang} et~al.}{2012}]{Fang+12}
{Fang} J.~J.,  {Faber} S.~M.,  {Salim} S.,  {Graves} G.~J.,   {Rich} R.~M.,
  2012, \mn@doi [\apj] {10.1088/0004-637X/761/1/23}, \href
  {http://adsabs.harvard.edu/abs/2012ApJ...761...23F} {761, 23}

\bibitem[\protect\citeauthoryear{{Feigelson} \& {Nelson}}{{Feigelson} \&
  {Nelson}}{1985}]{Feigelson+85}
{Feigelson} E.~D.,  {Nelson} P.~I.,  1985, \mn@doi [\apj] {10.1086/163225},
  \href {http://adsabs.harvard.edu/abs/1985ApJ...293..192F} {293, 192}

\bibitem[\protect\citeauthoryear{{Feruglio}, {Maiolino}, {Piconcelli}, {Menci},
  {Aussel}, {Lamastra}  \& {Fiore}}{{Feruglio} et~al.}{2010}]{Feruglio+10}
{Feruglio} C.,  {Maiolino} R.,  {Piconcelli} E.,  {Menci} N.,  {Aussel} H.,
  {Lamastra} A.,   {Fiore} F.,  2010, \mn@doi [\aap]
  {10.1051/0004-6361/201015164}, \href
  {http://adsabs.harvard.edu/abs/2010A%26A...518L.155F} {518, L155}

\bibitem[\protect\citeauthoryear{{Fischer} et~al.,}{{Fischer}
  et~al.}{2010}]{Fischer+10}
{Fischer} J.,  et~al., 2010, \mn@doi [\aap] {10.1051/0004-6361/201014676},
  \href {http://adsabs.harvard.edu/abs/2010A%26A...518L..41F} {518, L41}

\bibitem[\protect\citeauthoryear{{French}, {Yang}, {Zabludoff}, {Narayanan},
  {Shirley}, {Walter}, {Smith}  \& {Tremonti}}{{French}
  et~al.}{2015}]{French+15}
{French} K.~D.,  {Yang} Y.,  {Zabludoff} A.,  {Narayanan} D.,  {Shirley} Y.,
  {Walter} F.,  {Smith} J.-D.,   {Tremonti} C.~A.,  2015, \mn@doi [\apj]
  {10.1088/0004-637X/801/1/1}, \href
  {http://adsabs.harvard.edu/abs/2015ApJ...801....1F} {801, 1}

\bibitem[\protect\citeauthoryear{{Garc{\'{\i}}a-Burillo}
  et~al.,}{{Garc{\'{\i}}a-Burillo} et~al.}{2014}]{Garcia-Burillo+14}
{Garc{\'{\i}}a-Burillo} S.,  et~al., 2014, \mn@doi [\aap]
  {10.1051/0004-6361/201423843}, \href
  {http://adsabs.harvard.edu/abs/2014A%26A...567A.125G} {567, A125}

\bibitem[\protect\citeauthoryear{{Ger{\'e}b}, {Morganti}, {Oosterloo},
  {Hoppmann}  \& {Staveley-Smith}}{{Ger{\'e}b} et~al.}{2015}]{Gereb+15}
{Ger{\'e}b} K.,  {Morganti} R.,  {Oosterloo} T.~A.,  {Hoppmann} L.,
  {Staveley-Smith} L.,  2015, \mn@doi [\aap] {10.1051/0004-6361/201424810},
  \href {http://adsabs.harvard.edu/abs/2015A%26A...580A..43G} {580, A43}

\bibitem[\protect\citeauthoryear{{Gladders}, {L{\'o}pez-Cruz}, {Yee}  \&
  {Kodama}}{{Gladders} et~al.}{1998}]{Gladders+98}
{Gladders} M.~D.,  {L{\'o}pez-Cruz} O.,  {Yee} H.~K.~C.,   {Kodama} T.,  1998,
  \mn@doi [\apj] {10.1086/305858}, \href
  {http://adsabs.harvard.edu/abs/1998ApJ...501..571G} {501, 571}

\bibitem[\protect\citeauthoryear{{Halsel}}{{Halsel}}{2012}]{Helsel+05}
{Halsel} D.,  2012, Statistics for Censored Environmental Data, 2 edn.
Wiley

\bibitem[\protect\citeauthoryear{Harrington \& Fleming}{Harrington \&
  Fleming}{1982}]{Harrington+82}
Harrington D.~P.,  Fleming T.~R.,  1982, Biometrika, 69, 553

\bibitem[\protect\citeauthoryear{{Hayward} et~al.,}{{Hayward}
  et~al.}{2014}]{Hayward+14b}
{Hayward} C.~C.,  et~al., 2014, \mn@doi [\mnras] {10.1093/mnras/stu1843}, \href
  {http://adsabs.harvard.edu/abs/2014MNRAS.445.1598H} {445, 1598}

\bibitem[\protect\citeauthoryear{Hopkins \& Hernquist}{Hopkins \&
  Hernquist}{2009}]{Hopkins+09}
Hopkins P.~F.,  Hernquist L.,  2009, The Astrophysical Journal, 698, 1550

\bibitem[\protect\citeauthoryear{{Hopkins}, {Hernquist}, {Cox}, {Di Matteo},
  {Robertson}  \& {Springel}}{{Hopkins} et~al.}{2006}]{Hopkins+06}
{Hopkins} P.~F.,  {Hernquist} L.,  {Cox} T.~J.,  {Di Matteo} T.,  {Robertson}
  B.,   {Springel} V.,  2006, \mn@doi [\apjs] {10.1086/499298}, \href
  {http://adsabs.harvard.edu/abs/2006ApJS..163....1H} {163, 1}

\bibitem[\protect\citeauthoryear{{Hopkins}, {Hernquist}, {Cox}  \& {Kere{\v
  s}}}{{Hopkins} et~al.}{2008}]{Hopkins+08}
{Hopkins} P.~F.,  {Hernquist} L.,  {Cox} T.~J.,   {Kere{\v s}} D.,  2008,
  \mn@doi [\apjs] {10.1086/524362}, \href
  {http://adsabs.harvard.edu/abs/2008ApJS..175..356H} {175, 356}

\bibitem[\protect\citeauthoryear{{Huang} \& {Kauffmann}}{{Huang} \&
  {Kauffmann}}{2014}]{Huang+14}
{Huang} M.-L.,  {Kauffmann} G.,  2014, \mn@doi [\mnras]
  {10.1093/mnras/stu1232}, \href
  {http://adsabs.harvard.edu/abs/2014MNRAS.443.1329H} {443, 1329}

\bibitem[\protect\citeauthoryear{{Kaplan} \& {Meier}}{{Kaplan} \&
  {Meier}}{1958}]{Kaplan+58}
{Kaplan} E.~L.,  {Meier} P.~M.,  1958, JASA, 53

\bibitem[\protect\citeauthoryear{{Kauffmann} et~al.,}{{Kauffmann}
  et~al.}{2003}]{kauffmann03c}
{Kauffmann} G.,  et~al., 2003, \mn@doi [\mnras]
  {10.1111/j.1365-2966.2003.07154.x}, \href
  {http://adsabs.harvard.edu/abs/2003MNRAS.346.1055K} {346, 1055}

\bibitem[\protect\citeauthoryear{{Kaviraj}, {Kirkby}, {Silk}  \&
  {Sarzi}}{{Kaviraj} et~al.}{2007}]{Kaviraj+07}
{Kaviraj} S.,  {Kirkby} L.~A.,  {Silk} J.,   {Sarzi} M.,  2007, \mn@doi
  [\mnras] {10.1111/j.1365-2966.2007.12475.x}, \href
  {http://adsabs.harvard.edu/abs/2007MNRAS.382..960K} {382, 960}

\bibitem[\protect\citeauthoryear{{Kewley}, {Dopita}, {Sutherland}, {Heisler}
  \& {Trevena}}{{Kewley} et~al.}{2001}]{Kewley+01}
{Kewley} L.~J.,  {Dopita} M.~A.,  {Sutherland} R.~S.,  {Heisler} C.~A.,
  {Trevena} J.,  2001, \mn@doi [\apj] {10.1086/321545}, \href
  {http://adsabs.harvard.edu/abs/2001ApJ...556..121K} {556, 121}

\bibitem[\protect\citeauthoryear{Klein \& Moeschberger}{Klein \&
  Moeschberger}{2005}]{Klein+05}
Klein J.~P.,  Moeschberger M.~L.,  2005, Survival analysis: techniques for
  censored and truncated data.
Springer Science \& Business Media

\bibitem[\protect\citeauthoryear{{Kormendy}}{{Kormendy}}{2013}]{Kormendy+13}
{Kormendy} J.,  2013, in {Thomas} D.,  {Pasquali} A.,   {Ferreras} I.,  eds,
  IAU Symposium Vol. 295, The Intriguing Life of Massive Galaxies. pp 241--256,
  \mn@doi{10.1017/S174392131300495X}

\bibitem[\protect\citeauthoryear{{Lamastra}, {Bianchi}, {Matt}, {Perola},
  {Barcons}  \& {Carrera}}{{Lamastra} et~al.}{2009}]{Lamastra+09}
{Lamastra} A.,  {Bianchi} S.,  {Matt} G.,  {Perola} G.~C.,  {Barcons} X.,
  {Carrera} F.~J.,  2009, \mn@doi [\aap] {10.1051/0004-6361/200912023}, \href
  {http://adsabs.harvard.edu/abs/2009A%26A...504...73L} {504, 73}

\bibitem[\protect\citeauthoryear{Lee \& Lee}{Lee \& Lee}{2015}]{Lee+15}
Lee L.,  Lee M.~L.,  2015, Package NADA

\bibitem[\protect\citeauthoryear{{Leroy} et~al.,}{{Leroy}
  et~al.}{2013}]{Leroy+13}
{Leroy} A.~K.,  et~al., 2013, \mn@doi [\aj] {10.1088/0004-6256/146/2/19}, \href
  {http://adsabs.harvard.edu/abs/2013AJ....146...19L} {146, 19}

\bibitem[\protect\citeauthoryear{{Mantel}}{{Mantel}}{1966}]{Mantel+66}
{Mantel} N.,  1966, Cancer chemotherapy reports. Part 1, 50, 163

\bibitem[\protect\citeauthoryear{{Martin} et~al.,}{{Martin}
  et~al.}{2005}]{Martin+05}
{Martin} D.~C.,  et~al., 2005, \mn@doi [\apjl] {10.1086/426387}, \href
  {http://adsabs.harvard.edu/abs/2005ApJ...619L...1M} {619, L1}

\bibitem[\protect\citeauthoryear{Martinez}{Martinez}{2007}]{Martinez07}
Martinez R. L. M.~C.,  2007, Diagnostics for choosing between Log-rank and
  Wilcoxon tests.
ProQuest

\bibitem[\protect\citeauthoryear{{Murray}, {Quataert}  \& {Thompson}}{{Murray}
  et~al.}{2005}]{Murray+05}
{Murray} N.,  {Quataert} E.,   {Thompson} T.~A.,  2005, \mn@doi [\apj]
  {10.1086/426067}, \href {http://adsabs.harvard.edu/abs/2005ApJ...618..569M}
  {618, 569}

\bibitem[\protect\citeauthoryear{{Narayanan} et~al.,}{{Narayanan}
  et~al.}{2008}]{Narayanan+08}
{Narayanan} D.,  et~al., 2008, \mn@doi [\apjs] {10.1086/533500}, \href
  {http://adsabs.harvard.edu/abs/2008ApJS..176..331N} {176, 331}

\bibitem[\protect\citeauthoryear{{Narayanan}, {Krumholz}, {Ostriker}  \&
  {Hernquist}}{{Narayanan} et~al.}{2011}]{Narayanan+11}
{Narayanan} D.,  {Krumholz} M.,  {Ostriker} E.~C.,   {Hernquist} L.,  2011,
  \mn@doi [\mnras] {10.1111/j.1365-2966.2011.19516.x}, \href
  {http://adsabs.harvard.edu/abs/2011MNRAS.418..664N} {418, 664}

\bibitem[\protect\citeauthoryear{{Pawlik}, {Wild}, {Walcher}, {Johansson},
  {Villforth}, {Rowlands}, {Mendez-Abreu}  \& {Hewlett}}{{Pawlik}
  et~al.}{2016}]{Pawlik+16}
{Pawlik} M.~M.,  {Wild} V.,  {Walcher} C.~J.,  {Johansson} P.~H.,  {Villforth}
  C.,  {Rowlands} K.,  {Mendez-Abreu} J.,   {Hewlett} T.,  2016, \mn@doi
  [\mnras] {10.1093/mnras/stv2878}, \href
  {http://adsabs.harvard.edu/abs/2016MNRAS.456.3032P} {456, 3032}

\bibitem[\protect\citeauthoryear{{Poggianti}, {Smail}, {Dressler}, {Couch},
  {Barger}, {Butcher}, {Ellis}  \& {Oemler}}{{Poggianti}
  et~al.}{1999}]{Poggianti+99}
{Poggianti} B.~M.,  {Smail} I.,  {Dressler} A.,  {Couch} W.~J.,  {Barger}
  A.~J.,  {Butcher} H.,  {Ellis} R.~S.,   {Oemler} Jr. A.,  1999, \mn@doi
  [\apj] {10.1086/307322}, \href
  {http://adsabs.harvard.edu/abs/1999ApJ...518..576P} {518, 576}

\bibitem[\protect\citeauthoryear{{Rowlands}, {Wild}, {Nesvadba}, {Sibthorpe},
  {Mortier}, {Lehnert}  \& {da Cunha}}{{Rowlands} et~al.}{2015}]{Rowlands+15}
{Rowlands} K.,  {Wild} V.,  {Nesvadba} N.,  {Sibthorpe} B.,  {Mortier} A.,
  {Lehnert} M.,   {da Cunha} E.,  2015, \mn@doi [\mnras]
  {10.1093/mnras/stu2714}, \href
  {http://adsabs.harvard.edu/abs/2015MNRAS.448..258R} {448, 258}

\bibitem[\protect\citeauthoryear{{Rupke}, {Veilleux}  \& {Sanders}}{{Rupke}
  et~al.}{2002}]{Rupke+02}
{Rupke} D.~S.,  {Veilleux} S.,   {Sanders} D.~B.,  2002, \mn@doi [\apj]
  {10.1086/339789}, \href {http://adsabs.harvard.edu/abs/2002ApJ...570..588R}
  {570, 588}

\bibitem[\protect\citeauthoryear{{Saintonge} et~al.,}{{Saintonge}
  et~al.}{2011}]{Saintonge+11}
{Saintonge} A.,  et~al., 2011, \mn@doi [\mnras]
  {10.1111/j.1365-2966.2011.18677.x}, \href
  {http://adsabs.harvard.edu/abs/2011MNRAS.415...32S} {415, 32}

\bibitem[\protect\citeauthoryear{{Saintonge} et~al.,}{{Saintonge}
  et~al.}{2012}]{Saintonge+12}
{Saintonge} A.,  et~al., 2012, \mn@doi [\apj] {10.1088/0004-637X/758/2/73},
  \href {http://adsabs.harvard.edu/abs/2012ApJ...758...73S} {758, 73}

\bibitem[\protect\citeauthoryear{{Sanders}, {Soifer}, {Elias}, {Madore},
  {Matthews}, {Neugebauer}  \& {Scoville}}{{Sanders} et~al.}{1988}]{Sanders+88}
{Sanders} D.~B.,  {Soifer} B.~T.,  {Elias} J.~H.,  {Madore} B.~F.,  {Matthews}
  K.,  {Neugebauer} G.,   {Scoville} N.~Z.,  1988, \mn@doi [\apj]
  {10.1086/165983}, \href {http://adsabs.harvard.edu/abs/1988ApJ...325...74S}
  {325, 74}

\bibitem[\protect\citeauthoryear{{Sandstrom} et~al.,}{{Sandstrom}
  et~al.}{2013}]{Sandstrom+13}
{Sandstrom} K.~M.,  et~al., 2013, \mn@doi [\apj] {10.1088/0004-637X/777/1/5},
  \href {http://adsabs.harvard.edu/abs/2013ApJ...777....5S} {777, 5}

\bibitem[\protect\citeauthoryear{{Scannapieco} \& {Oh}}{{Scannapieco} \&
  {Oh}}{2004}]{Scannapieco+04}
{Scannapieco} E.,  {Oh} S.~P.,  2004, \mn@doi [\apj] {10.1086/386542}, \href
  {http://adsabs.harvard.edu/abs/2004ApJ...608...62S} {608, 62}

\bibitem[\protect\citeauthoryear{{Schawinski}, {Thomas}, {Sarzi}, {Maraston},
  {Kaviraj}, {Joo}, {Yi}  \& {Silk}}{{Schawinski} et~al.}{2007}]{Schawinski+07}
{Schawinski} K.,  {Thomas} D.,  {Sarzi} M.,  {Maraston} C.,  {Kaviraj} S.,
  {Joo} S.-J.,  {Yi} S.~K.,   {Silk} J.,  2007, \mn@doi [\mnras]
  {10.1111/j.1365-2966.2007.12487.x}, \href
  {http://adsabs.harvard.edu/abs/2007MNRAS.382.1415S} {382, 1415}

\bibitem[\protect\citeauthoryear{{Schawinski} et~al.,}{{Schawinski}
  et~al.}{2009}]{Schawinski+09}
{Schawinski} K.,  et~al., 2009, \mn@doi [\apj] {10.1088/0004-637X/690/2/1672},
  \href {http://adsabs.harvard.edu/abs/2009ApJ...690.1672S} {690, 1672}

\bibitem[\protect\citeauthoryear{{Schawinski} et~al.,}{{Schawinski}
  et~al.}{2014}]{Schawinski+14}
{Schawinski} K.,  et~al., 2014, \mn@doi [\mnras] {10.1093/mnras/stu327}, \href
  {http://adsabs.harvard.edu/abs/2014MNRAS.440..889S} {440, 889}

\bibitem[\protect\citeauthoryear{{Sen}}{{Sen}}{1968}]{Sen+68}
{Sen} P.~K.,  1968, JASA, 63, 1379

\bibitem[\protect\citeauthoryear{{Silk} \& {Nusser}}{{Silk} \&
  {Nusser}}{2010}]{Silk+10}
{Silk} J.,  {Nusser} A.,  2010, \mn@doi [\apj] {10.1088/0004-637X/725/1/556},
  \href {http://adsabs.harvard.edu/abs/2010ApJ...725..556S} {725, 556}

\bibitem[\protect\citeauthoryear{{Silk} \& {Rees}}{{Silk} \&
  {Rees}}{1998}]{Silk+98}
{Silk} J.,  {Rees} M.~J.,  1998, \aap, \href
  {http://adsabs.harvard.edu/abs/1998A%26A...331L...1S} {331, L1}

\bibitem[\protect\citeauthoryear{{Snyder}, {Cox}, {Hayward}, {Hernquist}  \&
  {Jonsson}}{{Snyder} et~al.}{2011}]{Snyder+11}
{Snyder} G.~F.,  {Cox} T.~J.,  {Hayward} C.~C.,  {Hernquist} L.,   {Jonsson}
  P.,  2011, \mn@doi [\apj] {10.1088/0004-637X/741/2/77}, \href
  {http://adsabs.harvard.edu/abs/2011ApJ...741...77S} {741, 77}

\bibitem[\protect\citeauthoryear{{Solomon}, {Downes}, {Radford}  \&
  {Barrett}}{{Solomon} et~al.}{1997}]{Solomon+97}
{Solomon} P.~M.,  {Downes} D.,  {Radford} S.~J.~E.,   {Barrett} J.~W.,  1997,
  \apj, \href {http://adsabs.harvard.edu/abs/1997ApJ...478..144S} {478, 144}

\bibitem[\protect\citeauthoryear{{Spoon} et~al.,}{{Spoon}
  et~al.}{2013}]{Spoon+13}
{Spoon} H.~W.~W.,  et~al., 2013, \mn@doi [\apj] {10.1088/0004-637X/775/2/127},
  \href {http://adsabs.harvard.edu/abs/2013ApJ...775..127S} {775, 127}

\bibitem[\protect\citeauthoryear{{Springel}, {Di Matteo}  \&
  {Hernquist}}{{Springel} et~al.}{2005}]{springel05b}
{Springel} V.,  {Di Matteo} T.,   {Hernquist} L.,  2005, \mn@doi [\mnras]
  {10.1111/j.1365-2966.2005.09238.x}, \href
  {http://adsabs.harvard.edu/abs/2005MNRAS.361..776S} {361, 776}

\bibitem[\protect\citeauthoryear{{Sturm} et~al.,}{{Sturm}
  et~al.}{2011}]{Sturm+11}
{Sturm} E.,  et~al., 2011, \mn@doi [\apjl] {10.1088/2041-8205/733/1/L16}, \href
  {http://adsabs.harvard.edu/abs/2011ApJ...733L..16S} {733, L16}

\bibitem[\protect\citeauthoryear{{Sun}, {Greene}, {Zakamska}  \&
  {Nesvadba}}{{Sun} et~al.}{2014}]{Sun+14}
{Sun} A.-L.,  {Greene} J.~E.,  {Zakamska} N.~L.,   {Nesvadba} N.~P.~H.,  2014,
  \mn@doi [\apj] {10.1088/0004-637X/790/2/160}, \href
  {http://adsabs.harvard.edu/abs/2014ApJ...790..160S} {790, 160}

\bibitem[\protect\citeauthoryear{{Toomre} \& {Toomre}}{{Toomre} \&
  {Toomre}}{1972}]{Toomre+72}
{Toomre} A.,  {Toomre} J.,  1972, \mn@doi [\apj] {10.1086/151823}, \href
  {http://adsabs.harvard.edu/abs/1972ApJ...178..623T} {178, 623}

\bibitem[\protect\citeauthoryear{{Utomo}, {Kriek}, {Labb{\'e}}, {Conroy}  \&
  {Fumagalli}}{{Utomo} et~al.}{2014}]{Utomo+14}
{Utomo} D.,  {Kriek} M.,  {Labb{\'e}} I.,  {Conroy} C.,   {Fumagalli} M.,
  2014, \mn@doi [\apjl] {10.1088/2041-8205/783/2/L30}, \href
  {http://adsabs.harvard.edu/abs/2014ApJ...783L..30U} {783, L30}

\bibitem[\protect\citeauthoryear{{Veilleux} et~al.,}{{Veilleux}
  et~al.}{2013}]{Veilleux+13}
{Veilleux} S.,  et~al., 2013, \mn@doi [\apj] {10.1088/0004-637X/776/1/27},
  \href {http://adsabs.harvard.edu/abs/2013ApJ...776...27V} {776, 27}

\bibitem[\protect\citeauthoryear{{Wild}, {Walcher}, {Johansson}, {Tresse},
  {Charlot}, {Pollo}, {Le F{\`e}vre}  \& {de Ravel}}{{Wild}
  et~al.}{2009}]{Wild+09}
{Wild} V.,  {Walcher} C.~J.,  {Johansson} P.~H.,  {Tresse} L.,  {Charlot} S.,
  {Pollo} A.,  {Le F{\`e}vre} O.,   {de Ravel} L.,  2009, \mn@doi [\mnras]
  {10.1111/j.1365-2966.2009.14537.x}, \href
  {http://adsabs.harvard.edu/abs/2009MNRAS.395..144W} {395, 144}

\bibitem[\protect\citeauthoryear{{Wild}, {Heckman}  \& {Charlot}}{{Wild}
  et~al.}{2010}]{Wild+10}
{Wild} V.,  {Heckman} T.,   {Charlot} S.,  2010, \mn@doi [\mnras]
  {10.1111/j.1365-2966.2010.16536.x}, \href
  {http://adsabs.harvard.edu/abs/2010MNRAS.405..933W} {405, 933}

\bibitem[\protect\citeauthoryear{{Wild}, {Charlot}, {Brinchmann}, {Heckman},
  {Vince}, {Pacifici}  \& {Chevallard}}{{Wild} et~al.}{2011}]{Wild+11}
{Wild} V.,  {Charlot} S.,  {Brinchmann} J.,  {Heckman} T.,  {Vince} O.,
  {Pacifici} C.,   {Chevallard} J.,  2011, \mn@doi [\mnras]
  {10.1111/j.1365-2966.2011.19367.x}, \href
  {http://adsabs.harvard.edu/abs/2011MNRAS.417.1760W} {417, 1760}

\bibitem[\protect\citeauthoryear{{Wild}, {Almaini}, {Dunlop}, {Simpson},
  {Rowlands}, {Bowler}, {Maltby}  \& {McLure}}{{Wild} et~al.}{2016}]{Wild+16}
{Wild} V.,  {Almaini} O.,  {Dunlop} J.,  {Simpson} C.,  {Rowlands} K.,
  {Bowler} R.,  {Maltby} D.,   {McLure} R.,  2016, \mn@doi [\mnras]
  {10.1093/mnras/stw1996}, \href
  {http://adsabs.harvard.edu/abs/2016MNRAS.463..832W} {463, 832}

\bibitem[\protect\citeauthoryear{{Wong} et~al.,}{{Wong} et~al.}{2012}]{Wong+12}
{Wong} O.~I.,  et~al., 2012, \mn@doi [\mnras]
  {10.1111/j.1365-2966.2011.20159.x}, \href
  {http://adsabs.harvard.edu/abs/2012MNRAS.420.1684W} {420, 1684}

\bibitem[\protect\citeauthoryear{{Woo}, {Dekel}, {Faber}  \& {Koo}}{{Woo}
  et~al.}{2015}]{Woo+15}
{Woo} J.,  {Dekel} A.,  {Faber} S.~M.,   {Koo} D.~C.,  2015, \mn@doi [\mnras]
  {10.1093/mnras/stu2755}, \href
  {http://adsabs.harvard.edu/abs/2015MNRAS.448..237W} {448, 237}

\bibitem[\protect\citeauthoryear{{Wright} et~al.,}{{Wright}
  et~al.}{2010}]{Wright+10}
{Wright} E.~L.,  et~al., 2010, \mn@doi [\aj] {10.1088/0004-6256/140/6/1868},
  \href {http://adsabs.harvard.edu/abs/2010AJ....140.1868W} {140, 1868}

\bibitem[\protect\citeauthoryear{{Yesuf}, {Faber}, {Trump}, {Koo}, {Fang},
  {Liu}, {Wild}  \& {Hayward}}{{Yesuf} et~al.}{2014}]{Yesuf+14}
{Yesuf} H.~M.,  {Faber} S.~M.,  {Trump} J.~R.,  {Koo} D.~C.,  {Fang} J.~J.,
  {Liu} F.~S.,  {Wild} V.,   {Hayward} C.~C.,  2014, \mn@doi [\apj]
  {10.1088/0004-637X/792/2/84}, \href
  {http://adsabs.harvard.edu/abs/2014ApJ...792...84Y} {792, 84}

\bibitem[\protect\citeauthoryear{{Young} et~al.,}{{Young}
  et~al.}{2011}]{Young+11}
{Young} L.~M.,  et~al., 2011, \mn@doi [\mnras]
  {10.1111/j.1365-2966.2011.18561.x}, \href
  {http://adsabs.harvard.edu/abs/2011MNRAS.414..940Y} {414, 940}

\bibitem[\protect\citeauthoryear{{Zabludoff}, {Zaritsky}, {Lin}, {Tucker},
  {Hashimoto}, {Shectman}, {Oemler}  \& {Kirshner}}{{Zabludoff}
  et~al.}{1996}]{Zabludoff+96}
{Zabludoff} A.~I.,  {Zaritsky} D.,  {Lin} H.,  {Tucker} D.,  {Hashimoto} Y.,
  {Shectman} S.~A.,  {Oemler} A.,   {Kirshner} R.~P.,  1996, \mn@doi [\apj]
  {10.1086/177495}, \href {http://adsabs.harvard.edu/abs/1996ApJ...466..104Z}
  {466, 104}

\bibitem[\protect\citeauthoryear{{Zubovas} \& {King}}{{Zubovas} \&
  {King}}{2012}]{Zubovas+12}
{Zubovas} K.,  {King} A.,  2012, \mn@doi [\apjl] {10.1088/2041-8205/745/2/L34},
  \href {http://adsabs.harvard.edu/abs/2012ApJ...745L..34Z} {745, L34}

\makeatother
\end{thebibliography}

\section{Appendix}\label{sec:app}

This section presents materials that support the analysis in the main sections of the paper.

\subsection{Evolution of H$\delta$ equivalent width after a starburst}

\begin{figure}
\includegraphics[scale=0.45]{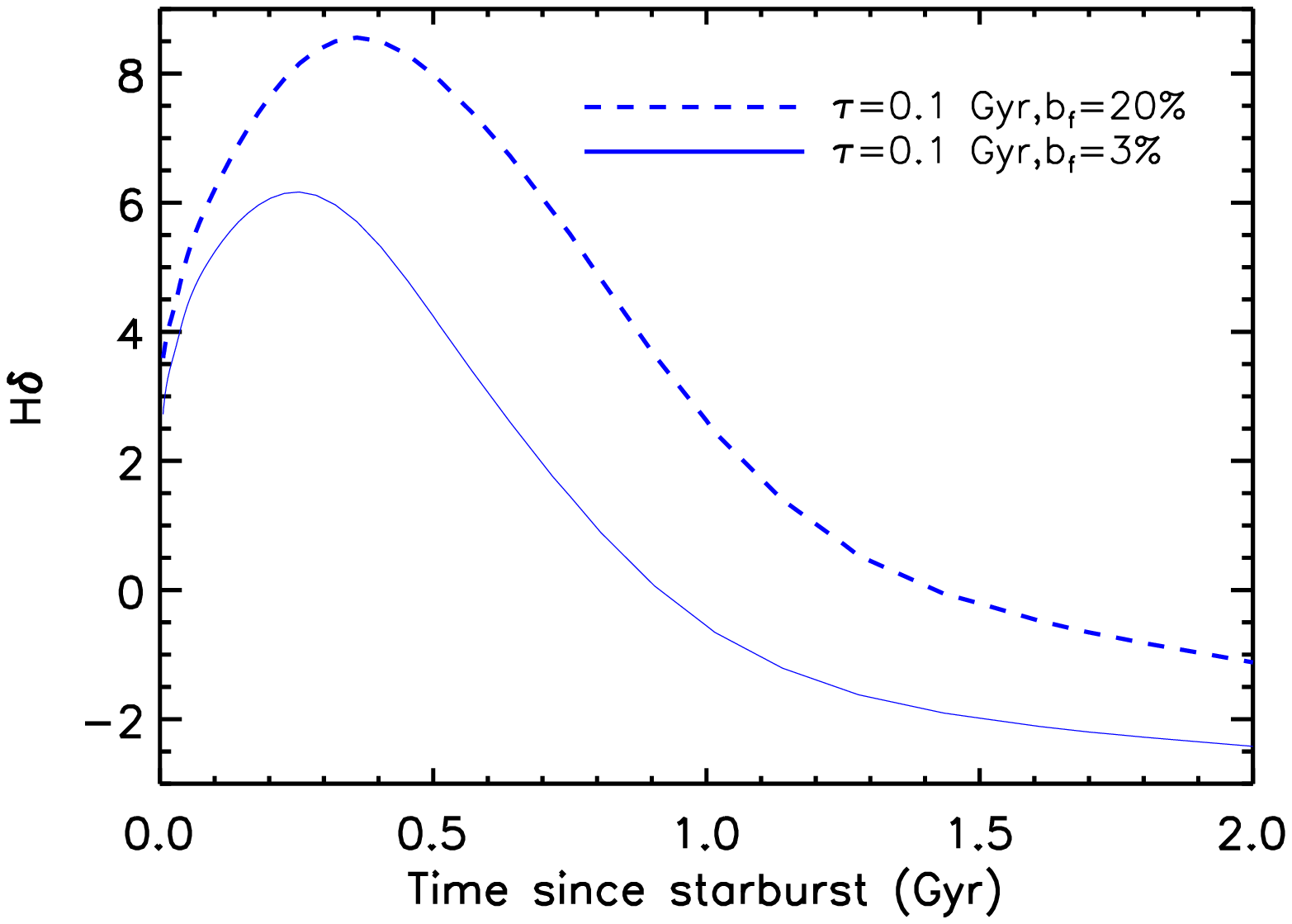}
\includegraphics[scale=0.45]{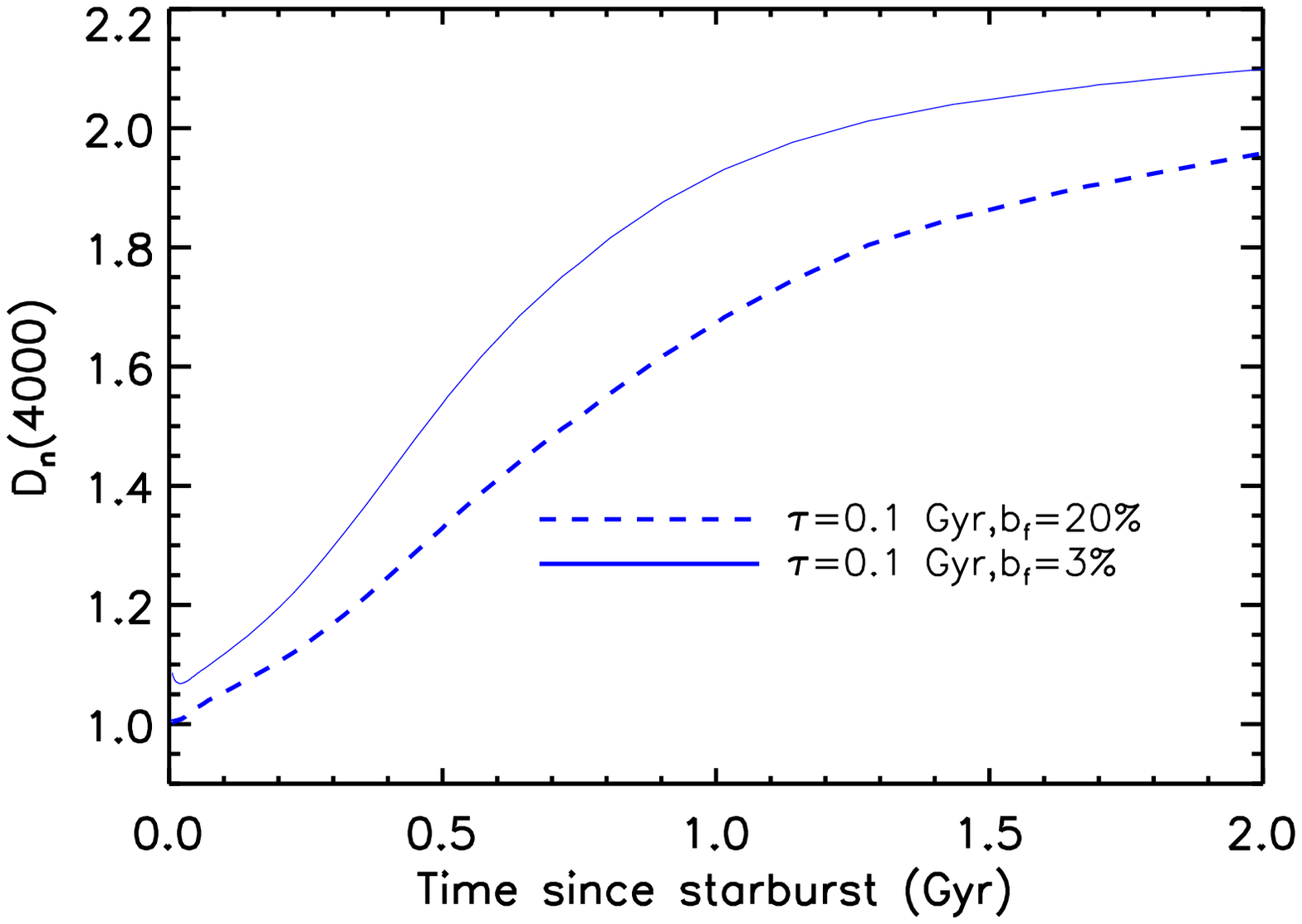}
\caption{The evolution the H$\delta$ absorption and $D_n(4000)$ with time since the starburst for the \citet{BC03} burst model tracks with a star formation timescale, $\tau = 0.1$ Gyr, and a burst mass fraction ($b_f$) 3\% or 20\%. \label{fig:t_bc03}}
\end{figure}

Figure~\ref{fig:t_bc03} shows the evolution of H$\delta$ equivalent width for \citet{BC03} starburst models, with a star formation timescale, $\tau = 0.1$ Gyr, and a burst mass fraction ($b_f$) $3\%$ or $20\%$. As detailed in \citet{Yesuf+14}. The H$\delta$ is dependent on the magnitude of the burst and there is generally a degeneracy between the burst mass fraction and the burst timescale. The figure, demonstrates with particular models above, that late stage PSBs can have H$\delta < 3${\AA}, which is lower than the threshold often used to define PSBs.

\subsection{A comparison of our PSB sample with previous samples}

\begin{figure}
\includegraphics[scale=0.4,angle=270]{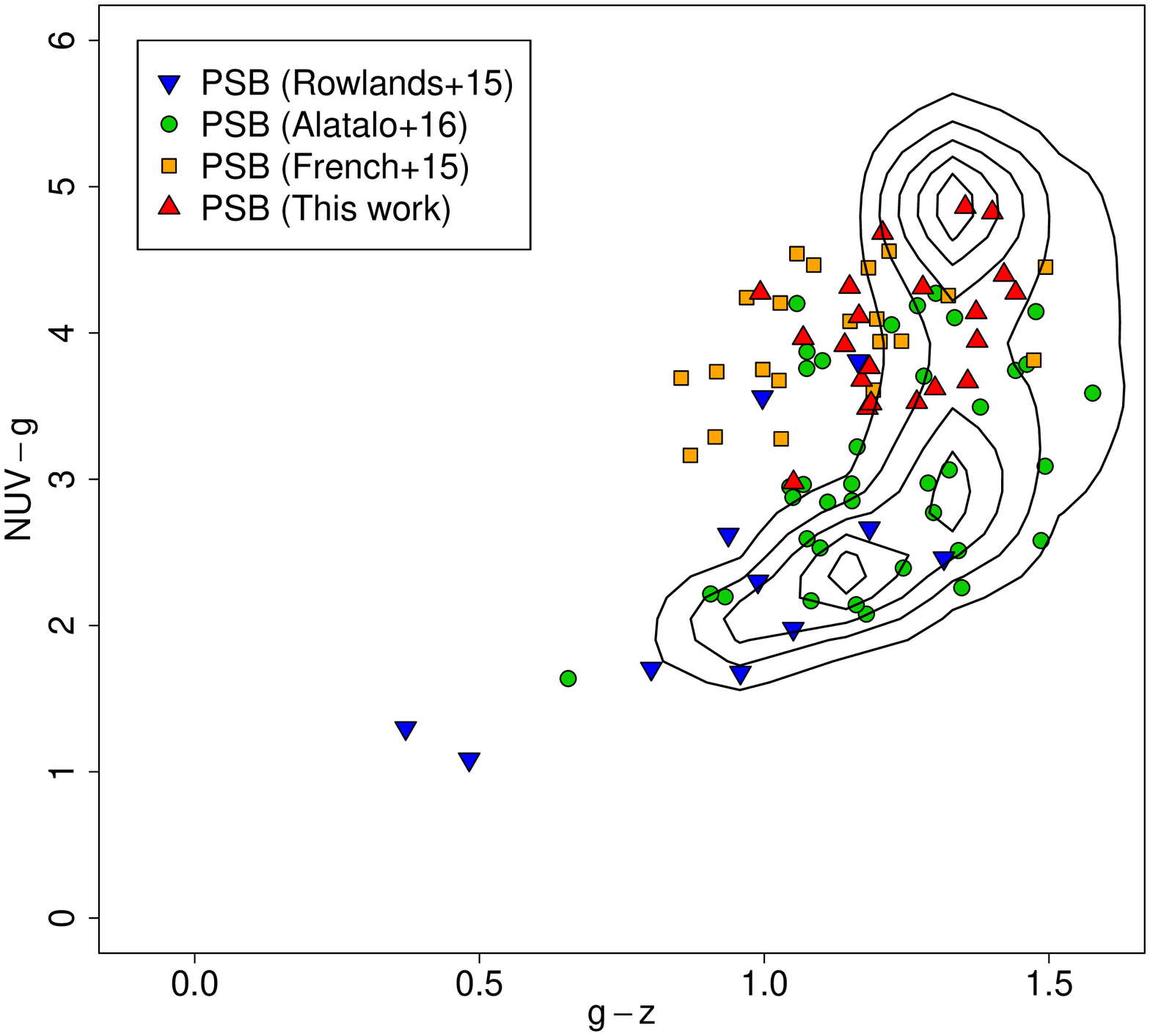}
\includegraphics[scale=0.4,angle=270]{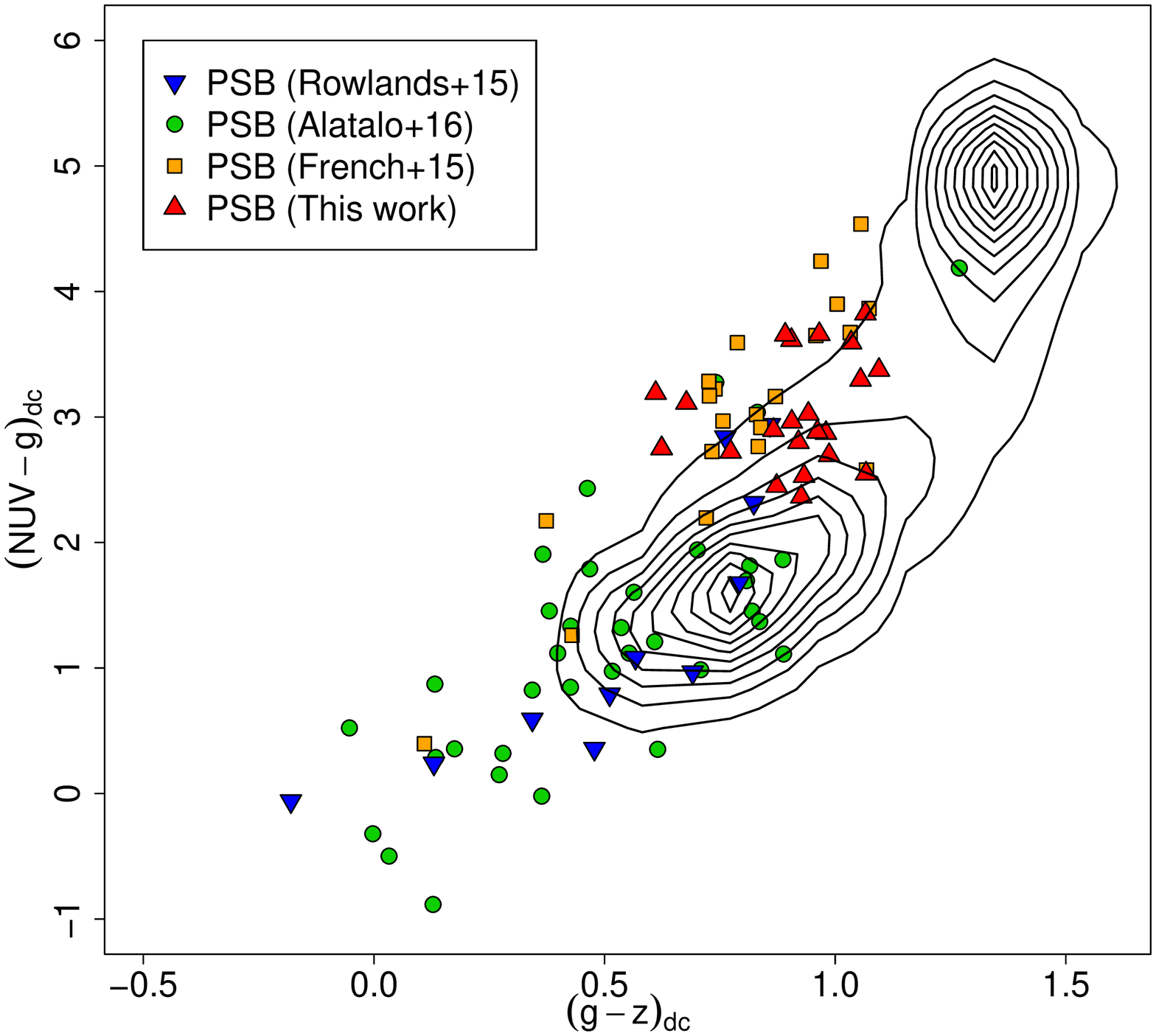}
\caption{The top panel shows the two color plot of NUV-g and g-z before the dust correction and the bottom panel shows the two color plot after the dust correction. The colored points are PSBs  \citep[in this work,][]{French+15,Rowlands+15,Alatalo+16}. The contours represent the number density of SDSS galaxies at redshift $z=0.02-0.06$ with stellar mass M$=10^{10}-10^{11}$ M$_\odot$. The positions of our Seyfert PSBs in the diagrams indicate that they are not dusty star-forming galaxies rather they are transitional galaxies in green valley. \label{fig:nuvgz}}
\end{figure}

\begin{figure}
\includegraphics[scale=0.4,angle=270]{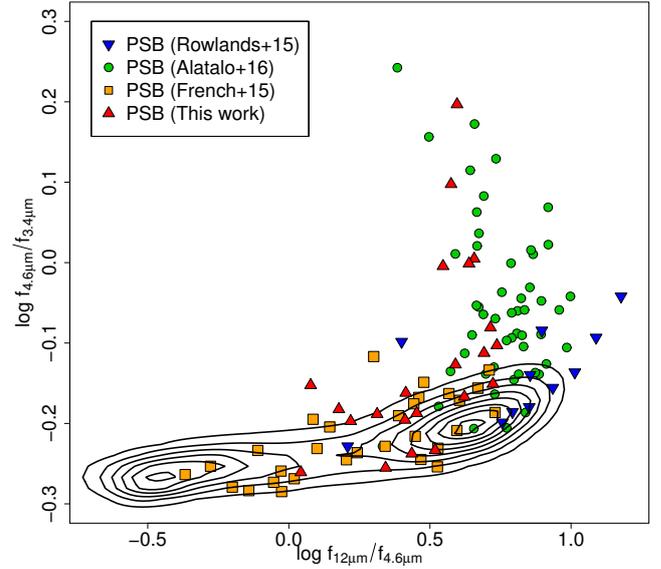}
\caption[The WISE two color diagram for the different PSB samples]{The WISE two color diagram. \lfwise is indicator of sSFR and while the $\log f_\mathrm{4.6}/f_\mathrm{3.4}\,$ is indictor hot dust emission by AGN or starburst or both. The contours denote the distribution of bi-model galaxy population at $z =0.02-0.06$ and log M (M$_\odot) = 10-11$. The blue-cloud is the density cloud with peak at \lfwise $\sim 0.6$ while the red-sequence is the density cloud with peak at \lfwise $\sim -0.45$. PSBs show a wide range in \lfwise and ones with \lfwiseh $> -0.06$ may be AGN.  \label{fig:w23_w12}}
\end{figure}

This section present additional figures that compare the stellar population of our Seyfert PSBs with previous samples of PSBs \citep{French+15,Rowlands+15,Alatalo+16}. Figure~\ref{fig:nuvgz} shows the NUV-g versus g-z colors before and after the dust correction. Before the dust-correction, starforming galaxies form a diagonal track which stretches from blue to red colors. The red end of this track is inhabited by dusty galaxies. The quiescent galaxies form a separate concentration above the dusty starforming galaxies. The locations our Seyfert PSBs in the  two color diagrams indicate that they are not dusty star-forming galaxies but are transitional galaxies in green valley. The dust-corrected diagram in addition shows that stellar ages of our PSBs and the PSBs in \citet{French+15} are comparable and the PSBs in \citet{Rowlands+15} and \citet{Alatalo+16} are younger than our sample. Figure~\ref{fig:w23_w12} depicts \lfwise versus \lfwiseh \citep{Wright+10}. Stellar populations younger than 0.6 Gyr dominate the 12\,$\mu$m emission and \lfwise is known to correlate well with sSFR \citep{Donoso+12} and \lfwiseh is sensitive to hot dust emission from AGN. Normal galaxies form a tight bi-modal sequence with some vertical scatter. Starforming galaxies are located at high \lfwise, while quiescent galaxies are located at low \lfwise. 

\begin{figure}
\includegraphics[scale=0.4,angle=270]{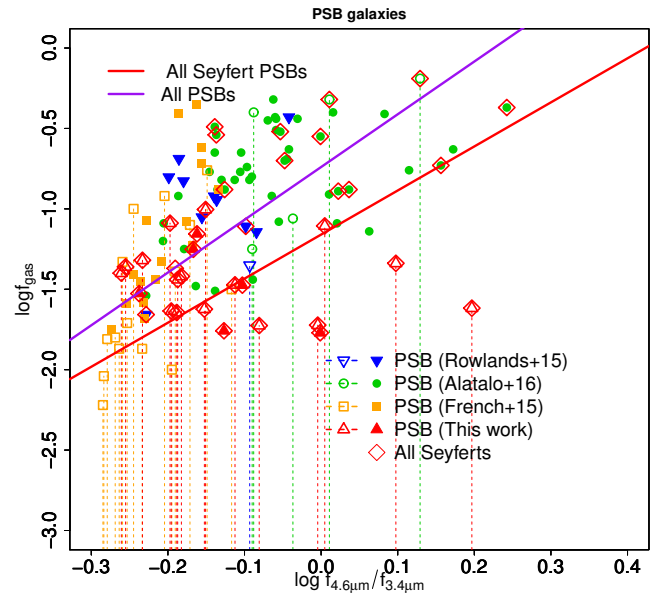}
\caption{WISE flux ratio \lfwiseh versus molecular gas fraction for PSBs \citep[in this work,][]{French+15,Rowlands+15,Alatalo+16}. The observed statistically significant trend of decreasing gas fraction with \lfwiseh is consistent with stellar dust heating process that decreases with the starburst age and it is inconsistent with AGN feedback that happens close to the starburst phase. \label{fig:w12fg_psb}}
\end{figure}

Figure~\ref{fig:w12fg_psb} depicts the WISE flux ratio \lfwiseh versus \lfgas for PSBs. The \lfwiseh $> -0.06$ criterion can identify hot dust emission from AGN albeit only with $\sim 50\%$ reliability \citep{Assef+13}. Anti-correlation between \lfwiseh and \lfgas may naively expected if AGN feedback destroys the molecular gas by heating it. Such trend is not seen in the data. There is instead somewhat significant ($\sim 2\,\sigma$) correlation between \lfwiseh and \lfgas for the BPT-identified Seyfert PSBs.  The current data cannot rule out AGN feedback, and the observed trend is consistent with delayed AGN feedback, which is inefficient at earlier times when the molecular gas fraction is above $\sim10$\%. Future study with large number of PSBs with \lfwiseh $> -0.06$ may constrain this model better. It should be noted that correlation between \lfwiseh and \lfgas for all PSBs in the combined sample is highly significant (>$5\sigma$) and it is likely tracing stellar dust heating process that declines with the starburst age.

\subsection{CO luminosities of Seyfert PSBs}
Table~\ref{tbl:co_lum} presents the observed CO luminosities of our new Seyfert PSB sample or 3 $\sigma$ upper limits in the case of no detections. Note that TPSB2 and TPSB23 are the same object, which is observed twice and we have combined the two observations in the analysis in presented in the main text of the paper. 


\begin{table}
\caption{CO\,(2--1) line luminosities and FWHM from fitting a Gaussian to the lines.}
\begin{tabular}{lcccc}
\hline
\hline
Target & RA & Dec. & L$^{\prime}_{\rm CO} $& FWHM  \\
 & (degree) & (degree) &  (10$^{7}$ K km s$^{-1}$ pc$^{2}$)  &  (km s$^{-1})$\\
\hline
TPSB1   &  212.01667   &    7.3276444    &  $14.17  \pm  2.74$    &    $109.8  \pm 13.8$ \\
TPSB2   &  134.61917   &   0.02346944   &  $21.5222  \pm   4.95$   &  $234.8 \pm  54.0$\\
TPSB4   &  182.01942   &   55.407672     &  $< 35.36$              &      \ldots    \\
TPSB5   &  173.41283  &    52.674611     &  $<29.27$               &      \ldots     \\
TPSB6   &  170.94588   &    35.442308    &  $<21.61$               &       \ldots    \\
TPSB7   &   203.56175  &    34.194147    & $9.59   \pm  2.55$      &   $201.7 \pm  48.3$\\
TPSB8   &   189.51733  &     48.345097   &  $<14.97$               &     \ldots   \\
TPSB9   &   117.96617  &    49.814314    &  $44.81 \pm 11.92$    &   $137.8  \pm  22.2$ \\
TPSB10 &   180.51921  &     35.321681   &  $<17.77$               &      \ldots  \\
TPSB11  &  137.87483  &    45.468278    & $19.60 \pm  4.72$      &   $ 370.2  \pm 169.2$ \\
TPSB12  &  139.49937  &     50.002175   & $<20.09$                &       \ldots  \\
TPSB13  &  173.16771  &     52.950400   & $<11.08$                &       \ldots  \\
TPSB14  &  178.62254  &     42.980203   & $<11.86$                &       \ldots  \\
TPSB15  &  179.02850   &    59.424919   & $<16.27$                &       \ldots  \\
TPSB16  &  190.45025   &    47.708878   & $<14.73$                &       \ldots  \\
TPSB17  &  198.74925  &    51.272583    & $<10.41$                &       \ldots  \\
TPSB18  &   200.95183  &    43.301181   & $37.94 \pm  5.70$      &     $436.3 \pm 125.8$\\
TPSB19  &   236.93392  &    41.402294   & $<23.06$                &       \ldots  \\
TPSB20  &   240.65804  &    41.293433   & $<21.51$                &       \ldots  \\
TPSB21  &   247.63600  &    39.384192   & $<15.28$                &       \ldots  \\
TPSB23  &   134.61913  &    0.02346944 & $15.20 \pm  4.18$      &     $126.1 \pm   24.0$ \\
TPSB24  &  145.18546   &   21.234358    & $<9.23$                  &      \ldots  \\
TPSB26  &   172.08300  &     27.622097   & $<17.80$                &    \ldots   \\
TPSB28  &   222.65771  &    22.734336    & $<6.82$                  &     \ldots  \\
\hline
\end{tabular}
\label{tbl:co_lum}
\end{table}


\bsp	
\label{lastpage}
\end{document}